\documentclass[11pt]{article}
\pdfoutput=1

\usepackage{a4wide}
\usepackage[fleqn]{amsmath}
\usepackage{fourier} % math & rm
\usepackage[scaled=0.875]{helvet} % ss
\usepackage{tikz}
\usetikzlibrary{matrix}

\tikzset{
  pretableaumatrix/.style={
    ampersand replacement=\&,
    matrix of math nodes,
    outer sep=1mm,
    inner sep=0mm,
    anchor=center,
    row sep={between borders,-\pgflinewidth},
    column sep={between borders,-\pgflinewidth},
    dottedentry/.style={densely dotted},
    dashedentry/.style={densely dashed},
    spaceentry/.style={draw=none,execute at begin node=\null},
  },
  pretableaunode/.style={
    font=\small,
    draw=gray,
    sharp corners,
    rectangle,
    anchor=base,
    text height=3.75mm,
    text depth=1.25mm,
    minimum height=5mm,
    minimum width=5mm,
    inner sep=0mm,
    outer sep=0mm,
    doublewidth/.style={minimum width=10mm},
    footnotesize/.style={font=\footnotesize},
    scriptsize/.style={font=\scriptsize},
  },
  tableaumatrix/.style={
    pretableaumatrix,
    every node/.append style={
      pretableaunode,
    },
  },
  medtableaumatrix/.style={
    pretableaumatrix,
    every node/.append style={
      pretableaunode,
      font=\footnotesize,
      text height=2.75mm,
      text depth=.75mm,
      minimum height=3.5mm,
      minimum width=3.5mm
    },
  },
  smalltableaumatrix/.style={
    pretableaumatrix,
    every node/.append style={
      pretableaunode,
      font=\scriptsize,
      text height=1.85mm,
      text depth=.15mm,
      minimum height=2.5mm,
      minimum width=2.5mm,
    },
  },
  tinytableaumatrix/.style={
    pretableaumatrix,
    every node/.append style={
      pretableaunode,
      font=\tiny,
      text height=1.25mm,
      text depth=.15mm,
      minimum height=1.75mm,
      minimum width=1.75mm
    },
  },
  tableau/.style={
    baseline=-1.25mm,
    every matrix/.style={tableaumatrix},
  },
  medtableau/.style={
    baseline=-1.25mm,
    every matrix/.style={medtableaumatrix},
  },
  smalltableau/.style={
    baseline=-1.25mm,
    every matrix/.style={smalltableaumatrix},
  },
  preshapetableaumatrix/.style={
    pretableaumatrix,
    execute at end cell={\strut},
    every node/.append style={
      draw=black,
      anchor=base,
      inner sep=0mm,
      outer sep=0mm,
    },
    shadedentry/.style={fill=gray},
    darkshadedentry/.style={fill=darkgray},
  },
  medshapetableaumatrix/.style={
    preshapetableaumatrix,
    every node/.append style={
      text height=2.75mm,
      text depth=.75mm,
      minimum height=3.5mm,
      minimum width=3.5mm
    },
  },
  shapetableaumatrix/.style={
    ampersand replacement=\&,
    matrix of math nodes,
    outer sep=0mm,
    inner sep=0mm,
    anchor=base,
    row sep={between borders,-\pgflinewidth},
    column sep={between borders,-\pgflinewidth},
    execute at begin cell={\strut},
    every node/.append style={draw,anchor=base,text height=1mm,text depth=.5mm,minimum size=1.5mm,inner sep=0mm,outer sep=0mm},
  },
  shapetableau/.style={
    every matrix/.style={shapetableaumatrix},
  },
  topalign/.style={
    every matrix/.append style={name=maintableau,anchor=maintableau-1-1.base},
    baseline,
  },
}

\newcommand*\smalltableau[2][]{\tikz[smalltableau,#1]\matrix{#2};}
\RequirePackage{amsmath,amssymb,amsxtra,amsthm}

\RequirePackage{mathrsfs}
\RequirePackage{setspace}
\RequirePackage{array}
\RequirePackage{booktabs}

\RequirePackage{colortbl}
\RequirePackage{xcolor}
\colorlet{titlerowcolor}{gray!15}
\usepackage[square,numbers]{natbib}

\definecolor{blue3}{RGB}{31,119,180}
\definecolor{red3}{RGB}{214,39,40}
\definecolor{orange3}{RGB}{255,127,14}
\definecolor{green3}{RGB}{44,160,44}

\usepackage{colortbl}
\definecolor{lightgreen}{cmyk}{0.2, 0, 0.2, 0.2}
\definecolor{lightgray}{cmyk}{0.1,0.2,0,0.1}
\definecolor{lightgray2}{cmyk}{0.1,0.1,0,0.1}

\usepackage{hyperref}
\hypersetup{colorlinks=true,linkcolor=teal,citecolor=orange3,urlcolor=green3,pdfencoding=auto}

\numberwithin{equation}{section}
\numberwithin{table}{section}
\numberwithin{figure}{section}

\author{
  \begin{minipage}{1.00\linewidth}
    \vspace{1cm}
    \begin{center}
      \begin{small}
        \textbf{Carlo Angelantonj$^1$, Cezar Condeescu$^2$, Emilian Dudas$^3$ and Giorgio Leone$^1$}
        \\ \vspace{1cm}
        ${}^1$ {\em Dipartimento di Fisica, Universit\`a di Torino and INFN Sezione di Torino}
        \\
        {\em Via Pietro Giuria 1, 10125 Torino, Italy}
        \\
        ${}^2$ {\em Department of Theoretical Physics, ``Horia Hulubei'' National Institute of Physics and Nuclear Engineering, P.O. Box MG-6, Magurele - Bucharest, 077125, Jud. Ilfov, Romania}
        \\
        ${}^3$ {\em CPHT Ecole Polytechnique, CNRS, IP Paris, 91128 Palaiseau, France}
     \end{small}
    \end{center}
    \vspace{1cm}
  \end{minipage}
}

\date{}

\title{\vspace{3cm}
  \begin{huge} \textbf{Rigid Vacua with Brane Supersymmetry Breaking} 	
  \end{huge}
  \\ \vspace{.7cm}
}

\begin{document}

\begin{titlepage}
  \maketitle
  \thispagestyle{empty}

  \vspace{-14cm}
  \begin{flushright}
%{\bf    \today}
   \end{flushright}
\begin{flushright}
 CPHT-RR015.032024
\end{flushright}

  \vspace{11cm}

  \begin{center}
    \textsc{Abstract}\\
  \end{center}
We construct new string vacua featuring Brane Supersymmetry Breaking, based on $T^4/\mathbb{Z}_N$ orientifolds with $\text{O}5_+$ planes and $\overline{\text{D}5}$ branes. Differently from the original construction, in these vacua the cancellation of twisted R-R charges of D-branes and orientifold planes is achieved in a non-trivial way, and results in rigid configurations with few open-string moduli, which highly restrict the possible deformations of the model. The breaking of supersymmetry and, consequently, the non-vanishing untwisted and twisted NS-NS tadpoles, generates a rich potential both for the dilaton and the blown-up moduli.  

\noindent
We also uncover the stringy origin of the $J$ form entering the gauge kinetic functions in the low-energy effective action, and display its relation to the NS-NS tadpoles for the scalars in the tensor multiplets. As a result, the $J$ form can be consistently identified also when supersymmetry is broken, thus solving an embarrassing puzzle related to its very existence. We also discuss the unitarity constraints for one-dimensional defects in these vacua, where the Ka$\check{\text{c}}$-Moody algebra for D9 and $\overline{\text{D}5}$ can be realised both in the left-moving and right-moving sectors.

\vfill

{\small
\begin{itemize}
\item[E-mail:] {\tt carlo.angelantonj@unito.it}
\\
{\tt ccezar@theory.nipne.ro}
\\
{\tt emilian.dudas@polytechnique.edu}
\\
{\tt giorgio.leone@unito.it}

\end{itemize}
}
%\vfill

\end{titlepage}

\setstretch{1.1}

%%%%%%%%%%%%%%%%%%%%%%%%%
%\setcounter{section}{18}
%%%%%%%%%%%%%%%%%%%%%%%%%

{		\hypersetup{linkcolor=black}
		\tableofcontents	}

\section{Introduction}

Orientifold vacua \cite{Sagnotti:1987tw, Pradisi:1988xd, Horava:1989vt, Bianchi:1990yu, Bianchi:1990tb, Bianchi:1991eu, Dudas:2000bn, Angelantonj:2002ct}
allow for interesting constructions where supersymmetry, exact in the closed string sector, is explicitly broken on the D-branes, while still leading to a classically stable vacuum free of tachyonic excitations. The simplest instance of such class of vacua is given by the Sugimoto model \cite{Sugimoto:1999tx} in $D=10$, where $\overline{\text{D}9}$ branes support a $\text{USp} (32)$ gauge group and chiral fermions in the anti-symmetric (reducible) $496$-dimensional representation. The consistency of the construction relies on the presence of a singlet fermion, $\boldsymbol{496} = \boldsymbol{495} + \boldsymbol{1}$, which plays the role of the Volkov-Akulov field \cite{Volkov:1972jx}, and a non-vanishing dilaton potential ascribed to uncancelled NS-NS tadpoles. These two ingredients guarantee that supersymmetry is non-linearly realised on the D-branes and the massless gravitino, still present in the closed-string spectrum, couples to a conserved current \cite{Dudas:2000nv}. At the microscopic level, the vacuum in \cite{Sugimoto:1999tx} involves $\text{O}9_+$ planes with positive tension and charge preserving half of the 32 supercharges of the parent type IIB theory. The cancellation of R-R tadpoles, and of irreducible gravitational anomalies, then requires the introduction of 32 anti-branes which would preserve the complementary supercharges. 
The interaction of  $\text{O}9_+$ planes and $\overline{\text{D}9}$ branes is thus the source for supersymmetry breaking in the open-string sector where bosons and fermions come in different representations of the Chan-Paton (CP) gauge group, and leaves behind an uncancelled NS-NS tension. Notice that the action of the world-sheet parity $\varOmega$ in the closed string spectrum is compatible both with the presence of $\text{O}9_-$ and $\text{O}9_+$ planes, and thus, in this case, the breaking of supersymmetry is an optional choice, since the same closed-string spectrum can be tied to both the open sector of type I superstring and the $\overline{\text{D}9}$ branes of Sugimoto.
This ambiguity is not a peculiarity of $D=10$, but is also present in lower dimensions, where one has always the freedom of flipping the tension and charge of \emph{all} orientifold planes, therefore breaking supersymmetry on the various (anti-)D-branes. 

A more interesting possibility occurs in lower dimensions, whenever orientifold planes of different dimensionality are present in the vacuum. In these cases, one has the option of flipping the tension and charge of just one set of O-planes. This choice also breaks  supersymmetry in the open-string sector, but is now incompatible with a supersymmetric vacuum, contrary to the construction in \cite{Sugimoto:1999tx}. The prototype example is given by a variation of the $T^4/\mathbb{Z}_2$ compactifications, named  Brane Supersymmetry Breaking (BSB) \cite{Antoniadis:1999xk}. The orientifold projection now involves the world-sheet parity  $\varOmega$ combined with an automorphism $\sigma$ of K3 which reverts  the twisted cohomology. Under the combined action $\varOmega\sigma$ each fixed point  yields a tensor multiplet  instead of the standard hypermultiplet of \cite{Bianchi:1990yu, Gimon:1996rq}, so that the closed-string sector comprises an ${\mathscr N} = (1,0) $ supergravity multiplet coupled to four hypermultiplets and seventeen tensor multiplets. The change of projection induced by $\sigma$ in the twisted sector implies that the BSB vacuum involves $\text{O}9_-$ planes together with sixteen $\text{O}5_+$ ones. The cancellation of the (untwisted) R-R charge then calls the introduction of 32 D9 and $\overline{\text{D}5}$ branes, and supersymmetry is automatically broken in the open-string sector due to the interaction of the $\overline{\text{D}5}$ branes with the D9 ones and the orientifold planes. A runaway dilaton potential is generated due to the uncancelled tensions of  $\text{O}5_+$ planes and $\overline{\text{D}5}$ branes, while the open-string excitations always involve singlet fermions which play the role of Volkov-Akulov fields, and ensure that supersymmetry is non-linearly realised also in this six-dimensional vacuum \cite{Pradisi:2001yv}. Although the change of orientifold projection implies that each fixed point supports a (non-dynamical) twisted R-R six-form potential, the O-planes of  \cite{Antoniadis:1999xk}, and of similar constructions \cite{Aldazabal:1999jr, Angelantonj:1999jh, Angelantonj:1999ms} in six and four dimensions, do not carry twisted charges, and pairs of fractional D9 and $\overline{\text{D}5}$ branes can be freely moved across the compactification manifold without affecting the twisted tadpole conditions.

Actually, this possibility of deforming the vacuum configurations by freely moving the D-branes is not a universal feature, and  depends on the simple choice of cancelling the  twisted tadpoles made in \cite{Antoniadis:1999xk, Aldazabal:1999jr, Angelantonj:1999jh, Angelantonj:1999ms}. In this paper, we provide more examples of BSB constructions, in six and four dimensions, where the vanishing of twisted charges occurs in a non-trivial way and the vacua are somehow rigid. To this end, we revisit the $T^4 /\mathbb{Z}_2$ (with and without $B$-field background) and $T^6/\mathbb{Z}_2 \times \mathbb{Z}_2$ vacua in six and four dimensions, and show how new interesting configurations can be obtained if the twisted charge of the fractional D9 branes is cancelled against that of $\overline{\text{D}5}$'s. Similarly, in the (new) $T^4/\mathbb{Z}_4$ and $T^4/\mathbb{Z}_6$ cases, the orientifold planes located on the $\mathbb{Z}_4$ and $\mathbb{Z}_6$ fixed points carry a non-trivial twisted charge which must be cancelled by fractional branes. Since twisted tadpoles must be neutralised locally, the new vacua are rigid in that the deformation modes associated to charged open-string moduli, which are usually present in more standard constructions, are now absent. Na\"\i vely moving branes around the compactification manifold would indeed result in an anomalous theory.  Moreover, the non-trivial cancellation of twisted R-R charges of O-planes and D-branes leaves its imprint in uncancelled twisted NS-NS tadpoles which, together with the ubiquitous non-vanishing dilaton tadpole, induce a richer tree-level scalar potential for the dilaton and the blown-up moduli. The analysis of their dynamics, which goes beyond the scope of this paper, is of great importance to understand the fate of such vacua and is expected to have interesting applications to cosmology and the development of the Swampland program \cite{Vafa:2005ui} (see \cite{Brennan:2017rbf, Palti:2019pca, vanBeest:2021lhn,  Agmon:2022thq} for a review).

In the case of ${\mathscr N}=(1,0)$ supersymmetric vacua in six dimensions, the low energy effective action (LEEA) involves a K\"ahler $J$ form which determines the metric on the moduli space of scalar fields in the tensor multiplets and the gauge kinetic functions. Supersymmetry relates it to the Wess-Zumino counterterms induced by the generalised Green-Schwarz-Sagnotti mechanism \cite{Green:1984sg, Sagnotti:1992qw}, and therefore it is linked to the data of the residual anomaly polynomial. When supersymmetry is broken, the na\"\i ve construction of the $J$ form in terms of the 't Hooft anomaly coefficients induces ghost-like couplings for scalars and gauge fields in the LEEA, which are clearly inconsistent and led to the conclusion that $J$ could not be properly defined in BSB vacua \cite{Angelantonj:2020pyr}.  A deeper scrutiny of the stringy origin of the various couplings in the LEEA actually reveals that the connection of the $J$ form with the anomaly polynomial is accidental, and holds only when supersymmetry is exact. In fact, the scalars in the tensor multiplets originate from the NS-NS sector of the closed strings, and therefore their coupling with the gauge fields living on the D-branes is encoded in the NS-NS tadpoles and not in the R-R ones, which instead determine the structure of the anomaly polynomial. The two  tadpoles are equal when supersymmetry is exact, but are crucially different in BSB vacua.  As a result, using the data from the NS-NS tadpoles one can construct the K\"ahler $J$ form also when supersymmetry is non-linearly realised \cite{Pradisi:2001yv},
thus solving the puzzle raised in \cite{Angelantonj:2020pyr}. 

Although these non-supersymmetric models correspond to \emph{bona fide} string vacua, and therefore are believed  to provide consistent quantum gravity theories, we continue our study \cite{Angelantonj:2020pyr}  of the unitarity constraints imposed on the one-dimensional defects, in an attempt to extend the work of \cite{Kim:2019vuc}\footnote{For other bottom-up studies of supersymmetric vacua, see  e.g. \cite{Lee:2019skh, Kim:2019ths, Katz:2020ewz, Tarazi:2021duw, Morrison:2021wuv}.} to the case when supersymmetry is broken. Such an extension is actually required also when supersymmetry is exact, since the D1 and $\text{D}5'$ defects in vacua with D9 and D5 (anti-)branes may admit a description in terms of gauge instantons, a situation which goes beyond the hypothesis of \cite{Kim:2019vuc}. A detailed study of the light excitations living on such defects and the structure of the anomaly inflow shows a common pattern. As expected, the coefficients of the induced gauge anomalies do not have a well defined sign, which suggests that the associated Ka$\check{\text{c}}$-Moody algebras be realised both in the left-moving and right-moving sectors of the underlying CFT on the two-dimensional world-volume of the defect. Moreover,  both chiral fermions and non-chiral bosons are charged with respect to the bulk gauge groups, which may hinder the realisation of the Ka$\check{\text{c}}$-Moody algebra in the infra-red (IR).  In all cases, we confirm the presence of string defects with null charge vectors, put forward in \cite{Angelantonj:2020pyr}.   It would be interesting to extract from this top-down analysis model-independent features which could provide general bottom-up tools to further shape the landscape of consistent vacua.

The paper is organised as follows. In Section \ref{Sec:BSBZ2} we review the construction of the BSB vacuum without \cite{Antoniadis:1999xk} and with \cite{Angelantonj:1999jh} a background $B_{ab}$ field and discuss alternative ways of cancelling R-R tadpoles which correspond to $\overline{\text{D}5}$ branes evenly distributed among the sixteen fixed points. We also discuss possible deformations via magnetic fields and/or brane recombinations. In Sections \ref{Sec:BSBZ4} and \ref{Sec:BSBZ6} we build new BSB vacua based on the $T^4/\mathbb{Z}_4$ and $T^4/\mathbb{Z}_6$ orbifolds, respectively, which now involve fractional orientifold planes and rigid configurations of D-branes. In Section \ref{Sec:Anomaly} we discuss the structure of the anomaly polynomials for these BSB vacua and their relation to the R-R tadpoles. We also show how the K\"ahler $J$ form, describing the gauge coupling constants and the moduli space of the scalars in the tensor multiplets, can be properly defined also when supersymmetry is broken and is linked to the structure of NS-NS tadpoles, which are no-longer connected by supersymmetry to the R-R ones. In Section \ref{Sec:defects} we introduce defects in the six-dimensional vacua, realised as D1 branes or $\text{D}5'$ branes wrapping the compact space, and analyse the unitarity constraints of their CFT's. In Section \ref{Sec:4d} we build new rigid BSB vacua in four-dimensions based on the $T^6/\mathbb{Z}_2 \times \mathbb{Z}_2$ orbifold with discrete torsion. Finally, the Appendices \ref{App:BSBZN}, \ref{App:ZNDefects} and \ref{App:Anomaly} collect the generic expressions for the one-loop partition functions for the vacuum configurations and the defects, and useful information on the structure of the anomaly polynomials, respectively.

\section{The $T^4/\mathbb{Z}_2$ orbifold}
\label{Sec:BSBZ2}

\subsection{The prototype vacuum}
\label{SSec:Z2prot}

The first instance of a vacuum with BSB was constructed in \cite{Antoniadis:1999xk}, and emerged from the $T^4/\mathbb{Z}_2$ orientifold. The standard world-sheet parity $\varOmega$ was dressed with the automorphism $\sigma$ of K3 which acts non-trivially on the twisted two-cycles. As a result, out of the type IIB ${\mathscr N}=(2,0)$ tensor multiplets living on the sixteen fixed points, an ${\mathscr N}=(1,0)$ tensor multiplet survives the $\varOmega\sigma$ projection, instead of the standard hypermultiplets, as can be deduced from the Klein bottle amplitude\footnote{Here, and in the following, we display only the contributions which are instrumental for the discussion. The full amplitudes, as well as the definition of the characters, can be found in Appendix \ref{App:BSBZN}} 
\begin{equation}
{\mathscr K} = \tfrac{1}{2} (P+W) \, \tau_{0,0} - 16 \, \tau_{1,0}\,.
\end{equation}
The minus sign for the twisted contributions, induced by the action of $\sigma$, has dramatic consequences on the structure of the orientifold planes, as can be appreciated by looking at the transverse channel amplitude
\begin{equation} \label{tildekleinZ2}
\tilde{\mathscr K} = \frac{2^{-5}}{2} \left( - 2^5\,\sqrt{v} + \frac{2^5}{\sqrt{v}}\right)^2 \tau_{0,0} + \ldots\,,
\end{equation}
with $v$ the volume of the $T^4$. The coefficient of $\tau_{0,0}$ determines the one-point function of the graviton and (non-dynamical) R-R ten-form potential with the orientifold planes. Therefore, the relative minus sign in the terms inside the bracket implies that the O9 and O5 planes carry opposite tension and charge. Conventionally, the BSB vacuum involves $\text{O}9_-$, with negative charge and tension, and $\text{O}5_+$ planes, with positive charge and tension. 

Although these two objects preserve the same  supercharges, they require the introduction of D9 and $\overline{\text{D}5}$ branes, which results in an explicit breaking of supersymmetry in the open-string sector. In the original construction of \cite{Antoniadis:1999xk}, all $\overline{\text{D}5}$'s were localised on a single fixed point, so that the twisted tadpole conditions
\begin{equation}
N_1 =0\,, \qquad N_1 + 4 D_1 =0\,,
\label{twZ2tadpoles}
\end{equation}
together with the untwisted ones, yield a CP gauge group 
\begin{equation} \label{CPZ2proto}
G_\text{CP} = \text{SO}(16)\times \text{SO}(16) \Big|_{\text{D}9} \times \text{USp}(16)\times \text{USp}(16) \Big|_{\overline{\text{D}5}}
\end{equation}
with  the would-be gauginos in the representations
\begin{equation}
(\boldsymbol{120}, \boldsymbol{1};\boldsymbol{1},\boldsymbol{1}) + (\boldsymbol{1}, \boldsymbol{120};\boldsymbol{1},\boldsymbol{1}) + (\boldsymbol{1},\boldsymbol{1}; \boldsymbol{120}, \boldsymbol{1}) + (\boldsymbol{1},\boldsymbol{1}; \boldsymbol{1}, \boldsymbol{120})\,,
\end{equation}
which no longer correspond to the adjoint representation in the case of $\overline{\text{D}5}$, since the adjoint of $\text{USp}(16)$ is symmetric and  136-dimen\-sio\-nal. Supersymmetry is thus explicitly broken due to the interaction between the $\overline{\text{D}5}$ branes and the $\text{O}5_+$ planes. The antisymmetric representation of a symplectic gauge group is reducible and contains a singlet which is supposed to play the role of the Volkov-Akulov field \cite{Pradisi:2001yv}, similarly to what happens \cite{Dudas:2000nv} in the Sugimoto model \cite{Sugimoto:1999tx}, so that supersymmetry is, actually, non-linearly realised. The rest of the spectrum comprises hypermultiplets in the bi-fundamental representations $(\boldsymbol{16}, \boldsymbol{16};\boldsymbol{1},\boldsymbol{1}) +(\boldsymbol{1}, \boldsymbol{1};\boldsymbol{16},\boldsymbol{16})$, two scalars in the $(\boldsymbol{16}, \boldsymbol{1};\boldsymbol{1},\boldsymbol{16}) +(\boldsymbol{1}, \boldsymbol{16};\boldsymbol{16},\boldsymbol{1})$ and a left-handed (LH) symplectic Majorana-Weyl (sMW) fermion in the $(\boldsymbol{16}, \boldsymbol{1};\boldsymbol{16},\boldsymbol{1}) +(\boldsymbol{1}, \boldsymbol{16};\boldsymbol{1},\boldsymbol{16})$ representations. This symplectic Majorana-Weyl fermion is a peculiar feature of six-dimensional vacua where the fundamental Weyl fermion, a pseudo-real spinor of $\text{SU}^*(4)$, can be subjected to an additional Majorana condition, if this is supplemented by the conjugation in a pseudo-real representation. In this case, this is indeed possible, since the sMW fermions are valued in the fundamental representation of a symplectic gauge group.

The action of the $\mathbb{Z}_2$ orbifold on the CP charges, $N_1 = n_1 - n_2$ and $D_1 = d_1 - d_2$, implies that this vacuum involves different types of fractional branes. In fact, from the twisted tadpoles \eqref{twZ2tadpoles} we read that both D9 and $\overline{\text{D}5}$ branes carry charges with respect to the twisted six-form potentials localised at the fixed points and, in particular, the $n_1$ and $n_2$ branes have opposite charge, and similarly for the $d_1$ and $d_2$ ones.  The choice implicitly made in \cite{Antoniadis:1999xk} was to cancel the twisted charges among the D9  and $\overline{\text{D}5}$ branes, independently. In this vacuum,  ``bound states'' of D9 branes, and/or $\overline{\text{D}5}$ ones,  with vanishing twisted charge can then be moved in the bulk without affecting the cancellation of twisted tadpoles. This has the effect of breaking the CP gauge group, and culminates in the maximal breaking $\text{SO}(16)^2\times \text{USp} (16) ^2\to \text{SO}(16)\times \text{USp} (16)$ when the scalar fields in the bi-fundamental representations $(\boldsymbol{16}, \boldsymbol{16};\boldsymbol{1},\boldsymbol{1}) +(\boldsymbol{1}, \boldsymbol{1};\boldsymbol{16},\boldsymbol{16})$ acquire non-trivial \emph{vev}'s, and all branes are moved away from the fixed points or have non-trivial Wilson lines\footnote{In the following, with an abuse of terminology, we often refer to generic Wilson lines on the D9 branes as moving them in the bulk.}. Suitable magnetic fields on the D9 branes  can further deform the model \cite{Angelantonj:2000hi} inducing, for instance, the maximal recombination $ \text{SO}(16)\times \text{USp} (16) \to \text{U} (8)$ in the bulk. This Higgsing of the gauge group involves vev's for the scalars stretched between the D9 and $\overline{\text{D}5}$ branes and, as such, does not admit a full world-sheet description.  

Notice, that  this way of cancelling the twisted R-R charges, $N_1=0$ and $D_1=0$ implies that also the twisted NS-NS tadpoles,
\begin{equation} \label{NSNStadpolesZ2}
N_1 =0 \,, \qquad N_1 - 4 D_1 =0\,,
\end{equation}
vanish automatically, so that only the untwisted NS-NS conditions are not satisfied. As a result, the induced scalar potential  
\begin{equation}
V (\phi) = 64 \, e^{-\phi}\,, \label{DilPot}
\end{equation}
depends only on the ten-dimensional dilaton $\phi$. 

\subsection{An almost rigid variation}
\label{SSec:Z2rigid}

Although the choice of localising all $\overline{\text{D}5}$ branes on a single fixed point is quite natural because of its simplicity, it does not exhausts all possibilities. It turns out that interesting new vacua can be obtained by distributing them over all sixteen fixed points. In this way, the twisted R-R tadpoles 
\begin{equation}
N_1 + 4 D_{(i),1} = 0 \,, \qquad i=1,\ldots , 16\,,
\end{equation}
together with the standard untwisted ones, admit the unique solution
\begin{equation} \label{CPZ2}
G_\text{CP} = \text{SO} (12) \times \text{SO} (20) \Big|_{\text{D}9}\times \text{USp} (2)^{16}\Big|_{\overline{\text{D}5}}\,.
\end{equation}
Aside from gauge bosons in the adjoint representation of $G_\text{CP}$, the open-string massless spectrum now comprises  would-be LH MW gauginos in the representation $(\boldsymbol{66}, \boldsymbol{1}; \boldsymbol{1}_i) +  (\boldsymbol{1} , \boldsymbol{190}  ;\boldsymbol{1}_i) + \sum_i (\boldsymbol{1} , \boldsymbol{1}  ;\boldsymbol{1}_i) $, hypermultiplets in the $(\boldsymbol{12}, \boldsymbol{20}; \boldsymbol{1}_i) $, two scalars in the representation $\sum_i (\boldsymbol{1} , \boldsymbol{20}  ;\boldsymbol{2}_i)$ and a LH sMW fermion in the representation $\sum_i (\boldsymbol{12} , \boldsymbol{1}  ;\boldsymbol{2}_i)$.

This model is quite rigid since the $ \overline{\text{D}5}$ branes cannot be moved away from the fixed points, and their twisted charge must be locally cancelled by the D9 branes\footnote{This is not the first construction where the twisted charges are cancelled among branes of different types. Similar possibilities were considered, for instance, in \cite{Aldazabal:1999nu} for supersymmetric vacua in $D=4$.}. As a result, only a subset of space-filling branes can be moved in the bulk, resulting in $G_\text{CP} = \text{SO} (12) \Big|_\text{D9 bulk} \times SO(8)\Big|_\text{D9} \times \text{USp} (2)^{16}\Big|_{\overline{\text{D}5}}$. Also the magnetisation of this vacuum is quite subtle as can be seen from the T-dual representation in terms of intersecting branes. In this picture, the orientifold planes wrap the \emph{bulk} cycle
\begin{equation} \label{eq:O9intersectingbranes}
{\boldsymbol \Pi}_O = 16 \, ( {\boldsymbol \pi}_1 - {\boldsymbol \pi}_2 )\,,
\end{equation}
while the T-dualised D9's  are associated to the cycles
\begin{equation} \label{eq:D9intersectingbranes}
{\boldsymbol \Pi}_+ = \tfrac{1}{2} {\boldsymbol \pi}_1 + \tfrac{1}{2} \left( {\boldsymbol e}_{11} + {\boldsymbol e}_{12} + {\boldsymbol e}_{21} + {\boldsymbol e}_{22} \right)\,,
\qquad
{\boldsymbol \Pi}_- = \tfrac{1}{2} {\boldsymbol \pi}_1 - \tfrac{1}{2} \left( {\boldsymbol e}_{11} + {\boldsymbol e}_{12} + {\boldsymbol e}_{21} + {\boldsymbol e}_{22} \right)\,.
\end{equation}
while the T-dualised  $\overline{\text{D}5}$'s wrap the cycles
\begin{equation} \label{eq:D5intersectingbranes}
\begin{split}
\bar{\boldsymbol \Pi}_1 &= - \tfrac{1}{2} {\boldsymbol \pi}_2 + \tfrac{1}{2} \left( {\boldsymbol e}_{11} + {\boldsymbol e}_{13} + {\boldsymbol e}_{31} + {\boldsymbol e}_{33} \right),
\\
\bar{\boldsymbol \Pi}_3 &= - \tfrac{1}{2} {\boldsymbol \pi}_2 + \tfrac{1}{2} \left( {\boldsymbol e}_{11} + {\boldsymbol e}_{13} - {\boldsymbol e}_{31} - {\boldsymbol e}_{33} \right),
\\
\bar{\boldsymbol \Pi}_5 &= - \tfrac{1}{2} {\boldsymbol \pi}_2 + \tfrac{1}{2} \left( {\boldsymbol e}_{12} + {\boldsymbol e}_{14} + {\boldsymbol e}_{32} + {\boldsymbol e}_{34} \right),
\\
\bar{\boldsymbol \Pi}_7 &= - \tfrac{1}{2} {\boldsymbol \pi}_2 + \tfrac{1}{2} \left( {\boldsymbol e}_{12} + {\boldsymbol e}_{14} - {\boldsymbol e}_{32} - {\boldsymbol e}_{34} \right),
\\
\bar{\boldsymbol \Pi}_9 &= - \tfrac{1}{2} {\boldsymbol \pi}_2 + \tfrac{1}{2} \left( {\boldsymbol e}_{21} + {\boldsymbol e}_{23} + {\boldsymbol e}_{41} + {\boldsymbol e}_{43} \right),
\\
\bar{\boldsymbol \Pi}_{11} &= - \tfrac{1}{2} {\boldsymbol \pi}_2 + \tfrac{1}{2} \left( {\boldsymbol e}_{21} + {\boldsymbol e}_{23} - {\boldsymbol e}_{41} - {\boldsymbol e}_{43} \right),
\\
\bar{\boldsymbol \Pi}_{13} &= - \tfrac{1}{2} {\boldsymbol \pi}_2 + \tfrac{1}{2} \left( {\boldsymbol e}_{22} + {\boldsymbol e}_{24} + {\boldsymbol e}_{42} + {\boldsymbol e}_{44} \right),
\\
\bar{\boldsymbol \Pi}_{15} &= - \tfrac{1}{2} {\boldsymbol \pi}_2 + \tfrac{1}{2} \left( {\boldsymbol e}_{22} + {\boldsymbol e}_{24} - {\boldsymbol e}_{42} - {\boldsymbol e}_{44} \right),
\end{split}
\qquad
\begin{split}
\bar{\boldsymbol \Pi}_2 &= - \tfrac{1}{2} {\boldsymbol \pi}_2 + \tfrac{1}{2} \left( {\boldsymbol e}_{11} - {\boldsymbol e}_{13} - {\boldsymbol e}_{31} + {\boldsymbol e}_{33} \right),
\\
\bar{\boldsymbol \Pi}_4 &= - \tfrac{1}{2} {\boldsymbol \pi}_2 + \tfrac{1}{2} \left( {\boldsymbol e}_{11} - {\boldsymbol e}_{13} + {\boldsymbol e}_{31} - {\boldsymbol e}_{33} \right),
\\
\bar{\boldsymbol \Pi}_6 &= - \tfrac{1}{2} {\boldsymbol \pi}_2 + \tfrac{1}{2} \left( {\boldsymbol e}_{12} - {\boldsymbol e}_{14} - {\boldsymbol e}_{32} + {\boldsymbol e}_{34} \right),
\\
\bar{\boldsymbol \Pi}_8 &= - \tfrac{1}{2} {\boldsymbol \pi}_2 + \tfrac{1}{2} \left( {\boldsymbol e}_{12} - {\boldsymbol e}_{14} + {\boldsymbol e}_{32} - {\boldsymbol e}_{34} \right),
\\
\bar{\boldsymbol \Pi}_{10} &= - \tfrac{1}{2} {\boldsymbol \pi}_2 + \tfrac{1}{2} \left( {\boldsymbol e}_{21} - {\boldsymbol e}_{23} - {\boldsymbol e}_{41} + {\boldsymbol e}_{43} \right),
\\
\bar{\boldsymbol \Pi}_{12} &= - \tfrac{1}{2} {\boldsymbol \pi}_2 + \tfrac{1}{2} \left( {\boldsymbol e}_{21} - {\boldsymbol e}_{23} + {\boldsymbol e}_{41} - {\boldsymbol e}_{43} \right),
\\
\bar{\boldsymbol \Pi}_{14} &= - \tfrac{1}{2} {\boldsymbol \pi}_2 + \tfrac{1}{2} \left( {\boldsymbol e}_{22} - {\boldsymbol e}_{24} - {\boldsymbol e}_{42} + {\boldsymbol e}_{44} \right),
\\
\bar{\boldsymbol \Pi}_{16} &= - \tfrac{1}{2} {\boldsymbol \pi}_2 + \tfrac{1}{2} \left( {\boldsymbol e}_{22} - {\boldsymbol e}_{24} + {\boldsymbol e}_{42} - {\boldsymbol e}_{44} \right).
\end{split}
\end{equation}
Here ${\boldsymbol \pi}_1 = {\boldsymbol a}_1 \otimes {\boldsymbol a}_2$ and ${\boldsymbol \pi}_2 = {\boldsymbol b}_1 \otimes {\boldsymbol b}_2$ are the \emph{horizontal} and \emph{vertical} two-cycles of the $T^2 \times T^2$, while ${\boldsymbol e}_{ij}$ are the sixteen collapsed cycles associated to the blown-up singularities, with intersecting matrix ${\boldsymbol e}_{ij} \circ {\boldsymbol e}_{kl} = -2 \delta_{ik} \delta_{jl}$. The untwisted (associated to the $\boldsymbol{\pi}_a$ cycles) and twisted (associated to the collapsed cycles $\boldsymbol{e}_{ij}$) R-R tadpole conditions are now summarised in
\begin{equation}
n_1\,  {\boldsymbol \Pi}_+ + n_2\,  {\boldsymbol \Pi}_- + \sum_{i=1}^{16} d_i \, \bar{\boldsymbol \Pi}_i = {\boldsymbol \Pi}_O\,,
\end{equation}
and admit the simple solution $n_1 = 12$, $n_2 =20$ and $d_i =2$. Clearly, twelve pairs of D9 branes can be moved in the bulk without affecting the tadpoles, while the $\overline{\text{D}5}$ branes can only be recombined at once together with all the remaining D9's. At this point, one can move again all D7 and $\overline{\text{D}7}'$ branes to wrap the cycles 
\begin{equation}
\begin{split}
{\boldsymbol \Pi}_+ &= \tfrac{1}{2} {\boldsymbol \pi}_1 + \tfrac{1}{2} \left( {\boldsymbol e}_{11} + {\boldsymbol e}_{12} + {\boldsymbol e}_{21} + {\boldsymbol e}_{22} \right)\,, \\
{\boldsymbol \Pi}_- &= \tfrac{1}{2} {\boldsymbol \pi}_1 - \tfrac{1}{2} \left( {\boldsymbol e}_{11} + {\boldsymbol e}_{12} + {\boldsymbol e}_{21} + {\boldsymbol e}_{22} \right)\, ,
\end{split}
\qquad
\begin{split}
 \bar{\boldsymbol  \Pi}'_+ &= -\tfrac{1}{2} {\boldsymbol \pi}_2 + \tfrac{1}{2} \left( {\boldsymbol e}_{11} + {\boldsymbol e}_{13} + {\boldsymbol e}_{31} + {\boldsymbol e}_{33} \right)\,, \\
\bar{\boldsymbol  \Pi}'_- &= -\tfrac{1}{2} {\boldsymbol \pi}_2 - \tfrac{1}{2} \left( {\boldsymbol e}_{11} + {\boldsymbol e}_{13} + {\boldsymbol e}_{31} + {\boldsymbol e}_{33} \right)\, ,
\end{split}
\end{equation}
which is nothing but the T-dual description of the original BSB model \cite{Antoniadis:1999xk} with gauge group $\text{SO} (16)^2 \times \text{USp} (16)^2$. In this sense, the almost rigid brane configuration is continuously connected to the original vacuum, although via non-trivial Higgsing.

In the almost rigid vacuum, the twisted NS-NS tadpoles 
\begin{equation}
N_1 - 4 D_{(i),1} = -16 \,, \qquad i=1,\ldots , 16\,,
\end{equation}
cannot be cancelled, resulting in a richer scalar potential
\begin{equation}
V (\phi , \xi_i ) = 64 \, e^{-\phi} + 16\, e^{-\phi}\, \sum_{i=1}^{16} \xi_i + O (\xi^2)\,,
\label{Z2rigidDilPot}
\end{equation}
for the ten-dimensional dilaton $\phi$ and the scalars $\xi_i$ in the tensor multiplets associated to the fixed points. This writing of $V$ is purely schematic and the complete knowledge of the numerical coefficients would require a canonical normalisation of the twisted fields. This goes beyond the scope of this work. Also, one expects terms involving higher powers of the twisted scalars, but these cannot be directly extracted from the tadpole conditions. A control of the full potential would be very interesting since a non-trivial vacuum for the $\xi$ fields would imply a "spontaneous" resolution of the orbifold singularities. 

\subsection{Adding a background $B$ field}
\label{SSec:Z2Bab}

A similar picture arises if a background for the Kalb-Ramond field is turned on \cite{Bianchi:1991eu, Angelantonj:1999jh, Kakushadze:1998bw} or, in the intersecting brane picture, a skew torus is considered \cite{Angelantonj:1999xf}. In these cases, the defects have interesting properties which will be discussed in Section \ref{Sec:defects}.
Although the quantum fluctuations of $B_{ab}$ or the off-diagonal components of $g_{ab}$ are projected away, a non-trivial (quantised) background is still allowed, and results in a rank reduction for the CP gauge group. In this $\mathbb{Z}_2$ orientifold, the element  $\varOmega\sigma$ selects the states with $p_L=p_R$, thus leaving the KK momenta unaffected, while $\varOmega\sigma g$ selects the states with $p_L=-p_R$, thus requiring that the winding modes satisfy the condition 
\begin{equation} \label{evenwindings}
    \frac{2}{\alpha'} B_{ab} \, n^b= 2 m_a \, .
\end{equation}
This, in turn, implies that both $\text{O}5_+$ and $\text{O}5_-$ planes are present, and their number $n_\pm$ is related to the rank $b$ of the background $B$ field \cite{Angelantonj:1999jh, Kakushadze:1998bw}, 
\begin{equation}
n_\pm= 2^3 (1\pm 2^{-b/2} ) \,.
\end{equation}
The direct-channel Klein-bottle amplitude thus reads
\begin{equation}
{\mathscr K} = \tfrac{1}{2} (P+W (B) ) \, \tau_{0,0} - (n_+ - n_-) \, \tau_{1,0}\,,
\end{equation}
and the closed string massless spectrum now comprises $1+n_+$ tensor multiplets and $4+n_-$ hypermultiplets, aside from the ${\mathscr N}=(1,0)$ supergravity multiplet. The presence of O5 planes of different type is reflected in the transverse channel Klein-bottle amplitude
\begin{equation}
    \tilde{\mathscr K} = \frac{2^{-5}}{2} \left( - 2^5\, \sqrt{v} + \frac{2^{5-b/2}}{\sqrt{v}}\right)^2 \tau_{0,0} + \ldots 
    \label{trkleinB}
\end{equation}
where the factor $2^{5-b/2}$ counts the left-over tension and charge of the O5 planes, which is still positive.
As in the standard case without $B$-field background, D9 and $\overline{\text{D}5}$ branes are needed to cancel the R-R charge of the orientifold planes, so that supersymmetry will be broken in the open-string sector. The full annulus and M\"obius strip amplitudes can be found in \cite{Angelantonj:1999jh}, together with the simple solution where all $\overline{\text{D}5}$ branes are localised at the same fixed point and the gauge group is $G_\text{CP} = \text{SO} (2^{4-b/2} ) \times  \text{SO} (2^{4-b/2} ) \Big|_{\text{D}9} \times  \text{USp} (2^{4-b/2} ) \times  \text{USp} (2^{4-b/2} ) \Big|_{\overline{\text{D}5}}$. Again, pairs of D9 and $\overline{\text{D}5}$ branes can be moved in the bulk without affecting the twisted tadpoles, and the maximum breaking of the gauge group is $\text{SO} (2^{4-b/2} ) \times \text{USp} (2^{4-b/2} )$ when all branes are moved away from the fixed points. 

Also in this case, however, one has the option of distributing the $\overline{\text{D}5}$ branes among the various fixed points, and thus cancel the twisted R-R charge of the D9's against that of the $\overline{\text{D}5}$'s,
\begin{equation}
    \begin{split}
        & N_0= \sum_{i=1}^{2^{4-b}} D_{(i),0} = 2^{5- \frac{b}{2}}
        \\
        & N_1 +  2^{2-\frac{b}{2}} D_{(i),1}=0 \, , \qquad i=1, \ldots, 2^{4-b} \, .
    \end{split}
\end{equation}
A simple solution of this type is $n_1 = 2^{4-b/2}+4$, $n_2 = 2^{4-b/2}-4$, $d_1 =0$ and $d^i_2 = 2^{1+b/2}$, which corresponds to the gauge group
\begin{equation}
G_\text{CP} = \text{SO} (2^{4-b/2}+4) \times \text{SO} ( 2^{4-b/2}-4 ) \Big|_{\text{D}9} \times \prod_{i=1}^{2^{4-b}} \text{USp} (2^{1+b/2}) \Big|_{\overline{\text{D}5}} \,. \label{CPBgauge}
\end{equation}
The massless spectrum then comprises  a LH MW fermion in the representations
\begin{equation}
    ( \smalltableau{ \null \\ \null \\}\, , 1 ; 1) + ( 1, \smalltableau{ \null \\ \null \\} \,; 1 ) + \sum_i ( 1, 1; \smalltableau{ \null \\ \null \\}_i )  \, ,
\end{equation}
four scalars and a RH MW fermion in the representation
\begin{equation}
    ( \smalltableau{ \null \\} \,, \smalltableau{ \null \\} \,; 1 )  \, ,
\end{equation}
$2^{\frac{b}{2}}$ scalars in  the representations
\begin{equation}
    \sum_i ( \smalltableau{ \null \\}\,, 1; \smalltableau{ \null \\}_i)  \, ,
\end{equation}
and, finally, $2^{\frac{b}{2}-1}$ LH sMW fermions in the representations
\begin{equation}
\sum_i ( 1,  \smalltableau{ \null \\}\, ; \smalltableau{ \null \\}_i)  \,.
\end{equation}
Also in this case, the NS-NS twisted tadpoles are un-cancelled and induce additional terms in the scalar potential of the form
\begin{equation}
V (\phi , \xi_i ) = 2^{6-b}\, e^{-\phi} + 16 \, e^{-\phi}\, \sum_{i=1}^{2^{4-b}} \xi_i + O (\xi^2)\,.
\label{Z2BDilPot}
\end{equation} 
with $\xi_i$ the scalars in the twisted tensor multiplets, and $\phi$ the ten-dimensional dilaton. Again, the numerical coefficients are purely symbolic and should be carefully determined after the twisted fields are canonically normalised. 

As in the case without $B$-field, some D9 branes can be moved in the bulk while the $\overline{\text{D}5}$ branes cannot be displaced, unless all D9's and $\overline{\text{D}5}$'s are recombined.

In this example, the D9 and $\overline{\text{D}5}$ branes both support real CP charges, and is due to the interaction of the branes with $\text{O}_+$ planes. In \cite{Angelantonj:1999jh} it is was also presented the alternative construction with complex CP charges for both D9's and $\overline{\text{D}5}$'s, associated to unitary gauge group and $\text{O}_+$ planes. Actually, it is also possible to have the hybrid situation where D9 branes support complex CP charges while those of the  $\overline{\text{D}5}$'s are still real, or \emph{vice versa}, thus amending what stated in \cite{Angelantonj:1999jh}. In this case, the D9 branes do not carry any twisted charge, so that, aside from trivial brane displacements, the unique solution has CP gauge group
\begin{equation}
G_\text{CP} = \text{U} (2^{4-b/2})\Big|_{\text{D}9} \times \text{USp} (2^{4-b/2} )\times \text{USp} (2^{4-b/2} ) \, \Big|_{\overline{\text{D}5}}\,,
\label{UxUSpZ2}
\end{equation}
with LH MW fermions in the  $(\smalltableau{  \null \\}\times \overline{\smalltableau{  \null \\}}\, ;1,1)+ (1; \smalltableau{  \null \\ \null \\ }\, ,1)+  (1; 1, \smalltableau{  \null \\ \null \\ }\, )$ representation, four scalars and a RH MW fermion in the $(\smalltableau{  \null \& \null \\} +  \overline{\smalltableau{  \null \&\null\\}}\, ;1,1) + (1; \smalltableau{  \null \\ }\, , \smalltableau{  \null \\ }\,)$ representation, and $2^{b/2-1}$ copies of two scalars and a LH sMW fermion in the representation $ ( \smalltableau{  \null \\ }\, + \overline{\smalltableau{  \null \\ }}\, ; \smalltableau{  \null \\ }\, , 1) +  ( \smalltableau{  \null \\ }\, + \overline{\smalltableau{  \null \\ }}\,; 1, \smalltableau{  \null \\ }\, )$. The twisted NS-NS tadpoles vanish in this vacuum, so that the scalar potential only depends on the ten-dimensional dilaton, as in eq. \eqref{DilPot}.

\section{The $T^4/\mathbb{Z}_4$ orbifold}
\label{Sec:BSBZ4}

K3 compactifications actually admit the orbifold limits \cite{Dixon:1986jc} $T^4/\mathbb{Z}_N$, with $N=2,3,4,6$, and it is thus natural to extend the BSB construction to the other choices of $N$. Since BSB is rooted in the presence of two O-planes of different type, the $\mathbb{Z}_3$ case is not of interest for us since it does not have an order-two element, and therefore only the O9 plane is present. Clearly, one has the freedom to flip its tension and charge and break supersymmetry in the open sector as in the ten-dimensional  Sugimoto vacuum \cite{Sugimoto:1999tx}, but this is not what we are concerned with in this work. Therefore, we shall limit the discussions to the cases $N=4$ and $N=6$, starting from the former.

For simplicity, we consider a factorised $T^4 = T^2 \times T^2$, with complex coordinates $(z_1,z_2)$. The $\mathbb{Z}_4$ group acts as a 90-degree rotation on each $T^2$, which is then a symmetry only for a squared torus with fixed complex structure $U=i$. Compatibility with supersymmetry requires an equal or opposite rotation on the two $T^2$'s. Our choice is
\begin{equation}
g:\quad (z_1 , z_2) \to  (i z_1 , - i z_2)\,,
\end{equation}
where $g$ is the generator of $\mathbb{Z}_4$. 

When discussing strings on orbifolds, the configuration of fixed points is rather important since it gives multiplicities of the twisted states and determines the structure of the associated Hilbert spaces. The action of $g$ has four fixed points with coordinates
\begin{equation}
\zeta_1 = \left(0,0\right)\,, \qquad \zeta_2 = \left(0,\tfrac{1}{\sqrt{2}} e^{i\pi/4}\right)\,, \qquad \zeta_3 = \left(\tfrac{1}{\sqrt{2}} e^{i\pi/4},0\right)\,, \qquad \zeta_4 = \left(\tfrac{1}{\sqrt{2}} e^{i\pi/4} , \tfrac{1}{\sqrt{2}} e^{i\pi/4} \right)\,,
\end{equation}
which are also fixed under the action of $g^3$. Therefore the Hilbert spaces associated to the $g$-twisted and $g^3$-twisted sectors come in four copies and each involves a full $\mathbb{Z}_4$ projector. Different is the case for the $g^2$-twisted sector, since $g^2$ generates a $\mathbb{Z}_2$ subgroup which admits the sixteen fixed points of Section \ref{Sec:BSBZ2}. Only four of them are also fixed under the action of $\mathbb{Z}_4$, while the remaining twelve form six $\mathbb{Z}_4$ doublets. The Hilbert space associated to the $g^2$-twisted sector is then richer and comprises four copies of spaces involving a full $\mathbb{Z}_4$ projector  and six copies of spaces involving just a $\mathbb{Z}_2$ projector.

\subsection{The closed-string sector}
\label{Sec:Z4closed}

We have all the ingredients to study the $\mathbb{Z}_4$ orientifold with BSB. The starting point is, as usual, the type IIB torus amplitude. Insisting on a geometrical action of the $\mathbb{Z}_4$ rotations, which treats symmetrically the left and right moving string coordinates, the $g$ twisted sector from the left-movers is naturally coupled to the $g^3$ one from the right-movers, and \emph{vice versa}, resulting in the charge conjugation modular invariant partition function. The complete structure of the amplitude is given in Appendix \ref{App:BSBZN}. Here we just give the contribution of the massless characters which are also defined and listed in the Appendix, 
\begin{equation}
\begin{split}
{\mathscr Z}_\text{IIB} &= \tau_{0,0} \bar \tau_{0,0} + \tau_{0,1}\bar\tau_{0,3} + \tau_{0,3}\bar \tau_{0,1} 
\\
&+ 4 \, \tau_{1,0} \bar\tau_{3,0}  + 4 \, \tau_{3,0} \bar\tau_{1,0} 
\\
&+4 \, \tau_{2,0} \bar\tau_{2,0} + 6 \, \tau_{2,0} \bar\tau_{2,0} \,.
\end{split}
\label{torusZ4}
\end{equation}
The first line encodes the contribution from the untwisted sector, which comprises the ${\mathscr N}=(2,0)$ supergravity multiplet and three tensor multiplets. The second line describes the $g$ and $g^3$ twisted sectors, which indeed come in four copies associated to the four $\mathbb{Z}_4$ fixed points, and  give additional eight tensor multiplets. Finally, the third line encodes the contribution of the $g^2$ twisted sector which yields four tensor multiplets from the four $\mathbb{Z}_4$ fixed points together with six additional tensors from the six doubles of $\mathbb{Z}_2$ fixed points. Altogether, the type IIB compactification on $T^4/\mathbb{Z}_4$ gives the unique anomaly-free spectrum with a gravitational multiplet coupled to twenty-one tensor multiplets. 

Following \cite{Antoniadis:1999xk}, the orientifold projection combines the standard world-sheet parity $\varOmega$ with an inner automorphism $\sigma$ of K3, so that the two-cycles associated to the fixed points in the $g^2$ twisted sector are odd under $\sigma$. The full Klein bottle amplitude is given in Appendix \ref{App:BSBZN}, and, if one restricts to the massless characters, reads
\begin{equation}
{\mathscr K} = \tau_{0,0}  - (4+6)\,  \tau_{2,0}  \,.
\end{equation}
At this stage, the only difference with the supersymmetric $\mathbb{Z}_4$ orientifold \cite{Dabholkar:1996pc, Gimon:1996ay} is in the extra minus sign in front of the twisted characters, ascribed to the action of $\sigma$. The projected closed-string spectrum then comprises a gravitational multiplet with ${\mathscr N} = (1,0)$ supersymmetry coupled to six hypermultiplets and fifteen tensor multiplets, out of which ten come from the $g^2$ twisted sector, replacing the ten hypermultiplets of the supersymmetric vacuum. 

The presence of the $\sigma$ automorphism is reflected in the nature of the orientifold planes present in this construction. This can be easily seen from the transverse channel Klein bottle amplitude, whose massless contributions read
\begin{equation}
\tilde{\mathscr K} =  \frac{2^{-5}}{4} \left\{ \left(- 2^5 \, \sqrt{v} + \frac{2^5}{\sqrt{v}} \right)^2 \, \tau_{0,0}  + \left( -2^5\, \sqrt{v} - \frac{2^5}{\sqrt{v}} \right)^2 (\tau_{0,1}+\tau_{0,3} ) + 4\times 2^{10}\, \tau_{2,0}  \right\}\,.
\end{equation}
We thus see that, as in \cite{Antoniadis:1999xk}, this orientifold involves $\text{O}9_-$ planes together with sixteen $\text{O}5_+$ planes located at the sixteen fixed points of the $g^2$ twisted sector. Differently from \cite{Antoniadis:1999xk}, however, these orientifold planes actually carry a non-trivial twisted R-R charge, as can be deduced from the presence of $\tau_{2,0}$ in $\tilde{\mathscr K}$. Since the $\mathbb{Z}_2$ fixed points, complementary to the $\zeta_i$'s, are identical to those present in \cite{Antoniadis:1999xk}, it follows that only the orientifold planes placed at $\zeta_i$ carry a non-trivial charge with respect the  $g^2$ twisted R-R six-form potentials supported on the four $\mathbb{Z}_4$ fixed points. This will play a crucial role in the D-brane geometry, although it is clearly not a novelty \emph{per se}, since \emph{fractional} orientifold planes have already made their appearance, for instance, in the six-dimensional $\mathbb{Z}_3$ and $\mathbb{Z}_6$ supersymmetric orientifolds \cite{Gimon:1996ay,Dabholkar:1996pc} and in the four-dimensional $Z$ orientifold \cite{Angelantonj:1996uy}.

\subsection{The open-string sector}
\label{Sec:Z4open}

As in the standard BSB, we need to introduce D9 and $\overline{\text{D}5}$ branes to cancel the untwisted R-R charges of the orientifold planes. Since the D9 branes wrap the entire compactification space, the local cancellation of twisted tadpoles requires that fractional $\overline{\text{D}5}$ branes be distributed evenly on the four $\mathbb{Z}_4$ fixed points. For the complete annulus and M\"obius strip amplitudes we refer to Appendix \ref{App:BSBZN}, while the massless contributions to the transverse-channel annulus amplitude read
\begin{equation}
\begin{split}
\tilde{\mathscr A} &= \frac{2^{-5}}{4}\left[ \left( N_0 \sqrt{v} + \frac{1}{\sqrt{v}}\sum_{k=1}^4 D_{(k),0}^2 \right)^2\, \left( \beta_{0,0} - \varphi_{0,1} - \varphi_{0,3} \right) +\left( N_0 \sqrt{v} - \frac{1}{\sqrt{v}}\sum_{k=1}^4 D_{(k),0}^2 \right)^2\, \left( \beta_{0,1} - \beta_{0,3} - \varphi_{0,0} \right)\right]
\\
&+\frac{2^{-5}}{2} \sum_{\alpha =1,3}  \sum_{k=1}^4 \left[ (N_\alpha - 2 D_{(k),\alpha})^2 \beta_{\alpha , 0} - (N_\alpha + 2 D_{(k),\alpha})^2 \varphi_{\alpha , 0} \right]
\\
&+ \frac{2^{-5}}{4} \left[ 12\, N_2^2 (\beta_{2,0}-\varphi_{2,0} ) + \sum_{k=1}^4 \left( (N_2 - 4 D_{(k),2})^2 \beta_{2,0} - (N_2 + 4 D_{(k),2})^2 \varphi_{2,0} \right)\right] \,.
\end{split}
\end{equation}
In this expression we have broken the $\tau_{\alpha, \beta} = \beta_{\alpha, \beta} - \varphi_{\alpha,\beta}$ characters into their bosonic, $\beta_{\alpha,\beta}$, and fermionic, $\varphi_{\alpha , \beta}$, parts since D9 and $\overline{\text{D}5}$ branes couple differently to the closed-string states running in the cylinder. Indeed, from the untwisted tadpoles in the first line, we see immediately that  D9 and $\overline{\text{D}5}$ branes have the same positive tension (coupling to $\beta_{0,0}$) but opposite R-R charge (coupling to $\varphi_{0,0}$). Similarly, for the twisted tadpoles in the second ($g$ and $g^3$ twisted) and third ($g^2$ twisted) lines the branes have different coupling to the twisted forms and scalars in the tensor multiplets, which reflects the breaking of supersymmetry in the open-string sector.  This fact will be relevant when discussing the structure of the anomaly polynomial and the existence of the Kahler $J$ form in these models. Notice also that the D9 branes, which wrap the entire internal space, couple  to the fields associated to both the $\mathbb{Z}_4$ and $\mathbb{Z}_2$ fixed points, while the $\overline{\text{D}5}$ branes are localised only on the $\mathbb{Z}_4$ fixed points.

As usual, the transverse-channel M\"obius strip amplitude is the geometric mean of $\tilde{\mathscr K}$ and $\tilde{\mathscr A}$, and the relevant terms are
\begin{equation}
\begin{split}
\tilde{\mathscr M} &=-\frac{1}{2} \left( \sqrt{v} - \frac{1}{\sqrt{v}}\right) \left( N_0 + \frac{1}{\sqrt{v}}\sum_{k=1}^4 D_{(k),0}\right) \, \hat\beta_{0,0}+
\frac{1}{2} \left( \sqrt{v} - \frac{1}{\sqrt{v}}\right) \left( N_0 - \frac{1}{\sqrt{v}}\sum_{k=1}^4 D_{(k),0}\right) \, \hat\varphi_{0,0}
\\
&-\frac{1}{2} \left( \sqrt{v} + \frac{1}{\sqrt{v}}\right) \left( N_0 - \frac{1}{\sqrt{v}}\sum_{k=1}^4 D_{(k),0}\right) \, (\hat\beta_{0,1}+ \hat\beta_{0,3}) +
\frac{1}{2} \left( \sqrt{v} + \frac{1}{\sqrt{v}}\right) \left( N_0 + \frac{1}{\sqrt{v}}\sum_{k=1}^4 D_{(k),0}\right) \, (\hat\varphi_{0,1}+\hat\varphi_{0,3})
\\
&- \sum_{k=1}^4 \left[ (N_2 - 4 D_{(k),2}) \, \hat\beta_{2,0} - (N_2 + 4 D_{(k),2}) \, \hat\varphi_{2,0} \right] \,.
\end{split}
\end{equation}
The full amplitude can be derived from eqs. \eqref{moebius9} and \eqref{moebius5b} upon a $P$ modular transformation.

The R-R tadpole conditions then read
\begin{equation}
N_0 = 32\,, \qquad \sum_{k=1}^4 D_{(k),0} = 32\,,
\end{equation}
for the untwisted sector,
\begin{equation}
N_1 + 2 D_{(k), 1} =0\,, \qquad N_3 + 2 D_{(k),3}=0\,, \qquad \forall k=1,\ldots , 4
\label{Z4gtadpole}
\end{equation}
for the $g$ and $g^3$ twisted sectors, and
\begin{equation}
\begin{split}
N_2 &=0 & \text{from the $\mathbb{Z}_2$ fixed points}\,,
\\
N_2 + 4 D_{(k),2} &= 32 &\text{from the four $\mathbb{Z}_4$ fixed points}\,,
\end{split}\label{Z4g2tadpole}
\end{equation}
for the $g^2$ twisted sector. 
Given the parametrisations \eqref{CPparam} in terms of the CP charges, these equations admit the three solutions
\begin{equation}
n_0 = 16-4a\,, \quad n_2 = 4a\,, \quad n_1 = n_3 = 8\,, \quad d_{(k),0} = 2a\,, \quad d_{(k),2}=8-2a\,, \quad d_{(k),1}=d_{(k),3}=0\,,
\end{equation}
for $k=1,\dots , 4$, and with $a=0,1,2$. The massless direct channel amplitudes read
\begin{equation}
{\mathscr A}_{99} = (n_0^2 + n_2^2 + 2 n1 n3)\, \tau_{0,0} + 2 (n_0 n_3 + n_2 n_1) \, \tau_{0,1} + 2 (n_0 n_1 + n_2 n_3 ) \tau_{0,3}\,,
\label{A99Z4}
\end{equation}
and 
\begin{equation}
{\mathscr M}_9 = - (n_0 + n_2) \, \hat\tau_{0,0}\,,
\label{M9Z4}
\end{equation}
for open strings stretched between D9 branes,
\begin{equation}
{\mathscr A}_{\bar 5\bar 5} = \sum_{k=1}^4 (d_{(k),0} ^2 + d_{(k),2}^2 )\, \tau_{0,0} \,,
\label{A5b5bZ4}
\end{equation}
and 
\begin{equation}
{\mathscr M}_{\bar 5} = \sum_{k=1}^4 (d_{(k),0} + d_{(k),2} ) \, (\hat\beta_{0,0} + \hat \varphi_{0,0})\,,
\label{M5bZ4}
\end{equation}
for open strings stretched between $\overline{\text{D}5}$ branes, and 
\begin{equation}
{\mathscr A}_{9\bar 5} = \sum_{k=1}^4 2 \left[ (n_0 d_{(k),0}+n_2 d_{(k),2}) \tau^{\hat B}_{2,0} + (n_1 d_{(k),2}+ n_3 d_{(k),0}) \tau^{\hat B}_{2,1}+ (n_1 d_{(k),0}+ n_3 d_{(k),2}) \tau^{\hat B}_{2,3} \right]\,,
\label{A95bZ4}
\end{equation}
for open strings stretched between D9 and $\overline{\text{D}5}$ branes. Notice that, as expected, supersymmetry is broken in the $\overline{\text{D}5}$ brane sector, since the gauge bosons and would-be gauginos undergo a different projection in $({\mathscr A}_{\bar 5\bar 5} + {\mathscr M}_{\bar 5})/2$, while ${\mathscr A}_{9\bar 5}$ involves the GSO projection $\eta^{\hat B}_{a,b}$. As a result, the open-string spectrum comprises gauge bosons in the adjoint representation of
\begin{equation} \label{CPZ4}
G_\text{CP} = \text{SO} (16 -4a) \times \text{SO} (4a) \times \text{U} (8) \Big|_{\text{D}9} \times \prod_{k=1}^4 \text{USp} (8 -2a) \times \text{USp} (2a) \Big|_{\overline{\text{D}5}_k}\,,
\end{equation}
LH MW fermions in the representations
\begin{equation}
(\smalltableau{  \null \\ \null \\ },1,1;1,1)+(1,\smalltableau{  \null \\ \null \\ },1;1,1)+(1,1,\smalltableau{  \null \\}\times \overline{\smalltableau{  \null \\}}
;1,1) + \sum_{k=1}^4 (1,1,1;\smalltableau{  \null \\ \null \\ }_k,1) + (1,1,1;1,\smalltableau{  \null \\ \null \\ }_k)\,,
\end{equation}
four scalars and a RH MW fermion ({\emph{i.e.} a full hypermultiplet) in the representations
\begin{equation}
(\smalltableau{  \null \\ } , 1, \overline{ \smalltableau{  \null \\ }}  ;1,1)+ (1,\smalltableau{  \null \\ } ,\smalltableau{  \null \\ } ;1,1) \,,
\end{equation}
a LH sMW fermion in the representations 
\begin{equation}
\sum_{k=1}^4 (\smalltableau{  \null \\ } , 1, 1; \smalltableau{  \null \\ }_k , 1)+(1, \smalltableau{  \null \\ } ,  1; 1, \smalltableau{  \null \\ }_k )\,,
\end{equation}
and a pair of real scalars in the representations
\begin{equation}
\sum_{k=1}^4 (1,1,\overline{\smalltableau{  \null \\ } }; \smalltableau{  \null \\ } _k,1)+(1,1,\smalltableau{  \null \\ } ;1, \smalltableau{  \null \\ } _k) \,.
\end{equation}
Some comments are in order. First, we notice that there are no fields in the bi-fundamental representation of the symplectic factors living on the $\overline{\text{D}5}$ branes. This means that these branes are stuck on the $\mathbb{Z}_4$ fixed points and cannot be moved away, as in the rigid $\mathbb{Z}_2$ vacuum of Section \ref{SSec:Z2rigid}. Second, the presence of scalars in bi-fundamentals of the D9 gauge group suggests that this brane configuration, instead, is not completely rigid. All this is reflected in the structure of the tadpole conditions. The minimal solution to the twisted tadpoles is, indeed,
\begin{equation}
n_0 = 16-8a\,, \quad n_2 =0\,, \quad n_1=n_3 = 8-4a\,, \quad  d_{(k),0}=2a\,, \quad d_{(k),2}=8-2a\,, \quad d_{(k),1}=d_{(k),3}=0\,,
\end{equation}
for $k=1,\dots , 4$, and with $a=0,1,2$. The remaining $16 a$ D9 branes, necessary to cancel the untwisted tadpole, can then be moved in the bulk, in a $\mathbb{Z}_4$ invariant configuration. The CP gauge group then becomes
\begin{equation}
G_\text{CP} = \text{SO} (4a)\Big|_{\text{D}9\ \text{bulk}} \times \text{SO} (16-8a )\times \text{U} (8-4a) \Big|_{\text{D}9}\times \prod_{k=1}^4 \text{USp} (8 -2a) \times \text{USp} (2a) \Big|_{\overline{\text{D}5}_k}\,. 
\end{equation}
In this case, the open-string spectrum can be extracted from eqs. \eqref{A99Z4}, \eqref{M9Z4}, \eqref{A5b5bZ4}, \eqref{M5bZ4} and \eqref{A95bZ4} upon setting $n_2=0$, with the addition of massless states associated to the bulk branes given by
\begin{equation}
\begin{split}
\text{Hyper in the representation}& \quad (\smalltableau{  \null \& \null \\ } , 1, 1;1,1) \,,
\\
\text{2 scalars and a LH sMW fermion}& \quad \sum_{k=1}^4 (\smalltableau{  \null \\},1,1;\smalltableau{  \null \\}_k,1)+  (\smalltableau{  \null \\},1,1;1, \smalltableau{  \null \\}_k )  \,.
\end{split}
\end{equation}
The hypermultiplet in the bi-fundamental representation $(1, \smalltableau{  \null \\}, \smalltableau{  \null \\};1,1)$ would suggest that the $\text{SO} (16-8a )\times \text{U} (8-4a)$ group on the D9's might be further broken. However, the allowed vacuum expectation values depend on the structure of the (tree-level) $F$ and $D$ terms. In these vacua where supersymmetry is broken at the string scale it is not so obvious which is the scalar potential and therefore a detailed analysis of the vacuum structure requires further investigations along the lines of \cite{Angelantonj:2011hs}. Notice that for $a=2$ all D9 branes are moved in the bulk and the twisted charge of the orientifold planes is entirely compensated by that of the $\overline{\text{D}5}$ branes. Furthermore, unlike the $\mathbb{Z}_2$ case, a recombination of all the branes in the bulk (by turning on the vevs of the scalars between D$9$ and $\overline{\text{D}5}$ branes) would not be possible because of the non-vanishing 
$g^2$-twisted R-R charges of the O-planes.

The universal feature of BSB vacua is the impossibility of cancelling the untwisted NS-NS tadpoles. In this $\mathbb{Z}_4$ vacuum also the twisted NS-NS ones stay un-cancelled, 
\begin{equation} \label{NSNStadpolesZ4}
N_{1,3} - 2 D_{(k),1,3} =32-16a\neq 0\qquad \text{and}\qquad  N_2 - 4\, D_{(k), 2}  - 32 = -64 \neq 0\,, \qquad \forall k=1,\ldots , 4\,,
\end{equation}
so that a potential is generated for the dilaton and the twisted scalars $\xi_{(k),\alpha}$ in the tensor multiplets associated to the $\mathbb{Z}_4$ fixed points, both in the $g^2$, $g$ and $g^3$ twisted sectors. The full potential schematically reads
\begin{equation}
V( \phi, \xi ) = 64\, e^{-\phi} + e^{-\phi} \, \sum_{k=1}^4 \left[ (16 (2-a) (\xi_{(k),1}+ \xi_{(k),3} ) - 64 \, \xi_{(k),2} + O(\xi^2 )\right] \,.
\label{Z4DilPot}
\end{equation}

\section{The $T^4/\mathbb{Z}_6$ orbifold}
\label{Sec:BSBZ6}

The last case to consider is $N=6$, where the orbifold generator $g$ acts on the complex coordinates as $(z_1 , z_2 ) \to (e^{2\pi i/6} z_1 , e^{-2\pi i /6} z_2)$. The complex structures of the two $T^2$'s are again fixed, $U= e^{\pi i /3}$, while only the origin is fixed under all rotations. The elements $g^2$ and $g^4$ form a $\mathbb{Z}_3$ subgroup and, aside from the origin, eight of the nine fixed points arrange in four $\mathbb{Z}_6$ doublets, and the associated Hilbert spaces only support $\mathbb{Z}_3$ projectors. Similarly, $g^3$ generates a $\mathbb{Z}_2$ subgroup and the fifteen fixed points different from the origin, arrange into five $\mathbb{Z}_6$ triplets, with only a $\mathbb{Z}_2$ projection. This structure of fixed points is, as usual, reflected in the various partition functions. 

\subsection{The closed-string sector}

The complete structure of the torus and Klein bottle amplitudes is again given by eqs. \eqref{torusZN} and \eqref{KleinZN}, with $N=6$. Keeping only the massless characters one finds
\begin{equation}
\begin{split}
{\mathscr Z}_\text{IIB} &= |\tau_{0,0}|^2 +\tau_{0,1}\bar\tau_{0,5} + \tau_{0,5}\bar \tau_{0,1}
\\
&+  \, \tau_{1,0} \bar\tau_{5,0}  + \, \tau_{5,0} \bar\tau_{1,0} 
\\
&+ (1+4)\left[ \, \tau_{2,0} \bar\tau_{4,0}  + \, \tau_{4,0} \bar\tau_{2,0} \right]
\\ 
&+(1+5) \, \tau_{3,0} \bar\tau_{3,0} \,,
\end{split}
\end{equation}
and
\begin{equation}
{\mathscr K} = \tau_{0,0} - (1+5) \, \tau_{3,0}\,.
\end{equation}
Again, the minus sign in front of the $\tau_{3,0}$ character in ${\mathscr K}$ comes from the action of the inner automorphism of K3 which marks the BSB vacua. The massless spectrum comprises a gravitational multiplet of ${\mathscr N} =(1,0)$ supersymmetry coupled to thirteen tensor multiplets and eight hypermultiplets. One tensor and two hypers originate from the untwisted sector, one tensor and one hyper from the $g$ and $g^5$ twisted sectors, five tensors and five hypers from the $g^2$ and $g^4$ twisted sectors, while the $g^3$ twisted sector yields six tensor multiplets, which, again, replace the six hypermultiplets of the standard orientifold vacuum \cite{Dabholkar:1996pc, Gimon:1996ay} without the $\sigma$ involution. 

The transverse channel Klein bottle amplitude
\begin{equation}
\tilde{\mathscr K} = \frac{2^{-5}}{6}\, \left(-2^5 \sqrt{v} + \frac{2^5}{\sqrt{v}} \right)^2\, \tau_{0,0} + \frac{2^{-5}}{6}\, \left( - 2^5\, \sqrt{v} - \frac{2^5}{\sqrt{v}} \right)^2\, (\tau_{0,1} + \tau_{0,5})  + \frac{2^5}{3} ( \tau_{2,0} + \tau_{4,0} )
\end{equation}
again shows that this orientifold involves $\text{O}9_-$ and $\text{O}5_+$ planes, which also in this case carry a non-trivial twisted charge. To properly interpret the factor $2^5/3$ one has to understand the geometry of the orientifold planes. In fact, while the O9 planes wrap the entire internal space and therefore couple universally to the twisted six-forms localised on the $\mathbb{Z}_3$ fixed points, the O5 planes,  which are associated to the action of $\varOmega g^3$, are actually located at the position of the sixteen $\mathbb{Z}_2$ fixed points. As a result, only one O5 plane (the one at the origin, which is also a $\mathbb{Z}_6$ fixed point) can carry a twisted charge. Taking into account that, as for the D branes, the twisted charge of an O5 plane is three times\footnote{In general, given $Q_9$ the twisted charge of a D9 brane or O9 plane, that of a D$p$ or O$p$ is given by
\begin{equation} \label{normalisationcharges}
Q_p =  Q_9\sqrt{\frac{\# \text{ fixed points}}{\#\text{ occupied fixed points}}} \,.
\end{equation}
} that of an O9 we are led to interpret
\begin{equation}
\frac{2^5}{3} ( \tau_{2,0} + \tau_{4,0} ) = \frac{2^{-4}}{9} \left[ 8\times Q_t^2 + (Q_t + 3 Q_t )^2 \right] ( \tau_{2,0} + \tau_{4,0} ) \,,
\end{equation}
where $Q_t = 8$ is the twisted R-R charge of the O9 plane.

\subsection{The open-string sector}

Also in this case, the open-string sector involves D9 and $\overline{\text{D}5}$ branes. The previous discussion on the twisted charge of the orientifold planes, suggest now to place all $\overline{\text{D}5}$'s on the unique $\mathbb{Z}_6$ fixed point. As a result, the leading contributions to the transverse-channel annulus amplitude read
\begin{equation}
\begin{split}
\tilde{\mathscr A} &= \frac{2^{-5}}{6} \left( N_0 \sqrt{v} + \frac{D_0}{\sqrt{v}}\right)^2 \, (\beta_{0,0} - \varphi_{0,1}- \varphi_{0,5} ) +
\frac{2^{-5}}{6} \left( N_0 \sqrt{v} - \frac{D_0}{\sqrt{v}}\right)^2 \, (\beta_{0,1} +\beta_{0,5} - \varphi_{0,0} )
\\
&+ \frac{2^{-3}}{6} \, \sum_{\alpha=1,5} \left[ (N_\alpha - D_\alpha)^2 \beta_{\alpha ,0} - (N_\alpha + D_\alpha )^2 \varphi_{\alpha,0}\right]
\\
&+ \frac{2^{-4}}{9} \sum_{\alpha =2,4} \left[ 8\, N_\alpha^2 \, (\beta_{\alpha, 0} - \varphi_{\alpha ,0} ) + (N_\alpha - 3 \, D_\alpha)^2  \beta_{\alpha,0} -  (N_\alpha + 3\, D_\alpha)^2  \varphi_{\alpha,0} \right] 
\\
&+ \frac{2^{-6}}{3} \,\left[ 15\,  N_3^2 \, (\beta_{3,0}-\varphi_{3,0} ) + (N_3 - 4\, D_3 )^2 \, \beta_{3,0} - (N_3 + 4\, D_3 )^2 \, \varphi_{3,0} \right] \,, 
\end{split}
\end{equation}
while the transverse-channel M\"obius amplitude is given by the geometric mean of $\tilde{\mathscr K}$ and $\tilde{\mathscr A}$,
\begin{equation}
\begin{split}
\tilde{\mathscr M} &= - \frac{1}{3} \left[ \left( \sqrt{v} - \frac{1}{\sqrt{v}} \right) \left( N_0 \sqrt{v} + \frac{D_0}{\sqrt{v}}\right) \beta_{0,0} +
\left( \sqrt{v} + \frac{1}{\sqrt{v}} \right)\left( N_0 \sqrt{v} - \frac{D_0}{\sqrt{v}}\right) (\beta_{0,1}+\beta_{0,5} )\right.
\\
&\left. - \left( \sqrt{v} - \frac{1}{\sqrt{v}} \right) \left( N_0 \sqrt{v} - \frac{D_0}{\sqrt{v}}\right) \varphi_{0,0} - \left( \sqrt{v} + \frac{1}{\sqrt{v}} \right)\left( N_0 \sqrt{v} + \frac{D_0}{\sqrt{v}}\right) (\varphi_{0,1}+\varphi_{0,5} )\right]
\\
& - \frac{2^{-2}}{3} \sum_{\alpha=2,4} \left[ 8\,  Q_t\, N_\alpha \, (\beta_{\alpha ,0}-\varphi_{\alpha ,0} ) + 4 Q_t\, (N_\alpha - 3\, D_\alpha)  \beta_{\alpha ,0} - 4 Q_t\, (N_\alpha + 3\, D_\alpha ) \varphi_{\alpha, 0}\right] \,.
\end{split}
\end{equation}
The full amplitudes can be derived from eqs. \eqref{annulus99}, \eqref{annulus5b5b}, \eqref{annulus95b} and eqs. \eqref{moebius9} and \eqref{moebius5b} upon $S$ and $P$ modular transformations, respectively. 
The R-R tadpole conditions then read
\begin{equation}
N_0 = 32\,, \qquad D_0 = 32\,,
\end{equation}
from the untwisted sector, 
\begin{equation}
N_{1,5} + D_{1,5} = 0\,,
\end{equation}
from the $g$ and $g^5$ twisted sectors,
\begin{equation}
N_{2,4} = 8\,, \qquad N_{2,4} + 3 D_{2,4} = 32\,,
\end{equation}
from the $g^2$ and $g^4$ twisted sectors, and, finally,
\begin{equation}
N_3 + 4 D_3 =0\,, \qquad N_3 =0\,,
\end{equation}
from the $g^3$ twisted sector. Using the parametrisations \eqref{CPparam}, the solution is
\begin{equation}
\begin{split}
& n_0 = 8+2a\,, \quad n_1 = n_5 = 4+a\,, \quad n_2 = n_4 = 4-a\,, \quad n_3 =8-2a\,,
\\
& d_0 = 8-2a\,, \quad d_1 = d_5 = 4-a\,, \quad d_2 = d_4 = 4+a\,, \quad d_3 =8+2a\,,
\end{split}
\end{equation}
with $a=0,\ldots, 4$.
Together with the direct channel amplitudes
\begin{equation}
{\mathscr A}_{99} = (n_0^2 + 2 n_1 n_5 + 2 n_2 n_4 + n_3^2 ) \tau_{0,0} + 2 (n_0 n_5 + n_1 n_4 + n_2 n_3 ) \tau_{0,1} + 2  (n_0 n_1 + n_5 n_2 + n_4 n_3 ) \tau_{0,5}\,,
\end{equation}
and
\begin{equation}
{\mathscr M}_9 = - (n_0 + n_3 ) \hat\tau_{0,0}\,,
\end{equation}
for open strings stretched between D9 branes,
\begin{equation}
{\mathscr A}_{\bar 5 \bar 5} = (d_0^2 + 2 d_1 d_5 + 2 d_2 d_4 + d_3^2 ) \tau_{0,0} + 2 (d_0 d_5 + d_1 d_4 + d_2 d_3 ) \tau_{0,1} + 2  (d_0 d_1 + d_5 d_2 + d_4 d_3 ) \tau_{0,5}\,,
\end{equation}
and
\begin{equation}
{\mathscr M}_{\bar 5} = (d_0 + d_3 ) (\hat\beta_{0,0} + \hat\varphi_{0,0})\,,
\end{equation}
for open strings stretched between $\overline{\text{D}9}$ branes, and
\begin{equation}
\begin{split}
{\mathscr A}_{9\bar 5} &= 2 (n_0 d_0 + n_1 d_5 + n_2 d_4 + n_3 d_3 + n_4 d_2 + n_5 d_1 )\tau^{\hat B}_{3,0} 
\\
&+ 2 (n_0 d_5 + n_1 d_4 + n_2 d_3 + n_3 d_2 + n_4 d_1 + n_5 d_0) \tau_{3,1}^{\hat B}
\\
&+2 (n_0 d_1 + n_5 d_2 + n_4 d_3 + n_3 d_4 + n_2 d_5 + n_1 d_0) \tau_{3,5}^{\hat B} \,,
\end{split}
\end{equation}
for open strings stretched between D9's and $\overline{\text{D}5}$'s. Once more, supersymmetry is broken on the $\overline{\text{D}5}$ branes, since the gauge bosons and would-be gauginos come in different representations in $({\mathscr A}_{\bar 5\bar 5} + {\mathscr M}_{\bar 5})/2$, while ${\mathscr A}_{9\bar 5}$ involves the GSO projection $\eta^{\hat B}_{a,b}$. The CP gauge group is then
\begin{equation}
G_\text{CP} = \text{SO} (8+2a ) \times \text{U} (4+a) \times \text{U} (4-a) \times \text{SO} (8-2a ) \Big|_{\text{D}9} \times 
\text{USp} (8-2a ) \times \text{U} (4-a) \times \text{U} (4+a) \times \text{USp} (8+2a ) \Big|_{\overline{\text{D}5}}\,.
\label{CPZ6}
\end{equation}
The charged matter can be readily extracted from the previous partition functions, taking into account that $\tau_{0,0} \sim V_4 - 2 S_4$, $\tau_{0,1}= \tau_{0,5} = 2 O_4 - C_4$, $\tau^{\hat B}_{3,0}= - S_4$ and $\tau^{\hat B}_{3,1} = \tau^{\hat B}_{3,5} = O_4$. It comprises vectors in the adjoint representation of $G_\text{CP}$ together with a LH MW fermions in the representations
\begin{equation}
\smalltableau{  \null \\ \null \\ }_{9_1} + \left( \smalltableau{  \null \\}\times \overline{\smalltableau{  \null \\}} \right)_{9_2} +  \left(\smalltableau{  \null \\}\times \overline{\smalltableau{  \null \\}} \right)_{9_3} +
\smalltableau{  \null \\ \null \\ }_{9_4}  +\smalltableau{  \null \\ \null \\ }_{\bar 5_1} + \left(\smalltableau{  \null \\}\times \overline{\smalltableau{  \null \\}} \right)_{\bar 5_2} +  \left( \smalltableau{  \null \\}\times \overline{\smalltableau{  \null \\}} \right)_{\bar 5_3} + \smalltableau{  \null \\ \null \\ }_{\bar 5_4} \,,
\end{equation}
four scalars and a RH MW fermion ({\emph{i.e.} a full hypermultiplet) in the representations
\begin{equation}
(\smalltableau{  \null \\ }_{9_1} , \overline{ \smalltableau{  \null \\ }}_{9_2}) +
(\smalltableau{  \null \\ }_{9_2} , \overline{ \smalltableau{  \null \\ }}_{9_3}) +
(\smalltableau{  \null \\ }_{9_3} ,  \smalltableau{  \null \\ }_{9_4})  +
(\smalltableau{  \null \\ }_{\bar 5_1} , \overline{ \smalltableau{  \null \\ }}_{\bar 5_2})+
(\smalltableau{  \null \\ }_{\bar 5_2} , \overline{ \smalltableau{  \null \\ }}_{\bar 5_3})+
(\smalltableau{  \null \\ }_{\bar 5_3} ,  \smalltableau{  \null \\ }_{\bar 5_4})
\,,
\end{equation}
a LH  sMW fermion in the representations 
\begin{equation}
(\smalltableau{  \null \\ }_{9_1} ,  \smalltableau{  \null \\ }_{\bar 5_1}) +
(\smalltableau{  \null \\ }_{9_2} , \overline{ \smalltableau{  \null \\ }}_{\bar 5_2}) +
(\smalltableau{  \null \\ }_{9_3} , \overline{ \smalltableau{  \null \\ }}_{\bar 5_3}) +
(\smalltableau{  \null \\ }_{9_4} , \smalltableau{  \null \\ }_{\bar 5_4}) +
(\overline{\smalltableau{  \null \\ }}_{9_2} ,  \smalltableau{  \null \\ }_{\bar 5_2}) +
(\overline{\smalltableau{  \null \\ }}_{9_3} ,  \smalltableau{  \null \\ }_{\bar 5_3})\,,
\end{equation}
and a scalar in the representations
\begin{equation}
(\smalltableau{  \null \\ }_{9_1} ,  \overline{\smalltableau{  \null \\ }}_{\bar 5_2}) +
(\smalltableau{  \null \\ }_{9_2} , \overline{ \smalltableau{  \null \\ }}_{\bar 5_3}) +
(\smalltableau{  \null \\ }_{9_3} ,  \smalltableau{  \null \\ }_{\bar 5_4}) +
(\smalltableau{  \null \\ }_{9_4} , \smalltableau{  \null \\ }_{\bar 5_3}) +
(\overline{\smalltableau{  \null \\ }}_{9_2} ,  \smalltableau{  \null \\ }_{\bar 5_1}) +
(\overline{\smalltableau{  \null \\ }}_{9_3} ,  \smalltableau{  \null \\ }_{\bar 5_2}) + \text{h.c.}
 \,.
\end{equation}
To lighten the notation, the representation $\smalltableau{  \null \\ }_{p_i}$ refers to the fundamental of the $i$-th factor in the gauge group on the $\text{D}p$ (anti-)brane, as ordered in \eqref{CPZ6}, and similarly for the antisymmetric representations.
The presence of scalars in bi-fundamental representations both on the D9 and $\overline{\text{D}5}$ branes, suggests that this vacuum can be deformed by turning on Wilson lines on the D9's and moving (some of) the $\overline{\text{D}5}$'s in the bulk. For instance, one possibility corresponds to the gauge group
\begin{equation}
G_\text{CP} = \text{SO} (4) \Big|_{\text{D}9\ \text{bulk}} \times \text{USp} (4) \Big|_{\overline{\text{D}5}\ \text{bulk}} \times \text{SO} (4)^2 \Big|_{\text{D}9}\times \text{USp} (4)^2 \Big|_{\overline{\text{D}5}}\,,
\end{equation}
with LH MW fermions in the antisymmetric representation of all group factors, four scalars in the $(\boldsymbol{10}, \boldsymbol{1}; \boldsymbol{1}, \boldsymbol{1}; \boldsymbol{1},\boldsymbol{1})+ (\boldsymbol{1}, \boldsymbol{6}; \boldsymbol{1}, \boldsymbol{1}; \boldsymbol{1},\boldsymbol{1})$, a RH MW fermion in the $(\boldsymbol{10}, \boldsymbol{1}; \boldsymbol{1}, \boldsymbol{1}; \boldsymbol{1},\boldsymbol{1})+ (\boldsymbol{1}, \boldsymbol{10}; \boldsymbol{1}, \boldsymbol{1}; \boldsymbol{1},\boldsymbol{1})$ a LH sMW fermion in the $(\boldsymbol{1}, \boldsymbol{1}; \boldsymbol{4}, \boldsymbol{1}; \boldsymbol{4},\boldsymbol{1})+ (\boldsymbol{1}, \boldsymbol{1}; \boldsymbol{1}, \boldsymbol{4}; \boldsymbol{1},\boldsymbol{4})$, and two scalars and a LH sMW fermion in six copies of the $(\boldsymbol{4}, \boldsymbol{4}; \boldsymbol{1}, \boldsymbol{1}; \boldsymbol{1},\boldsymbol{1})$ plus the $(\boldsymbol{4}, \boldsymbol{1}; \boldsymbol{1}, \boldsymbol{1}; \boldsymbol{4},\boldsymbol{1})
+(\boldsymbol{4}, \boldsymbol{1}; \boldsymbol{1}, \boldsymbol{1}; \boldsymbol{1},\boldsymbol{4})
+(\boldsymbol{1}, \boldsymbol{4}; \boldsymbol{4}, \boldsymbol{1}; \boldsymbol{1},\boldsymbol{1})
+(\boldsymbol{1}, \boldsymbol{4}; \boldsymbol{1}, \boldsymbol{4}; \boldsymbol{1},\boldsymbol{1})$ representations.
The branes on the fixed points cannot be further deformed, unless the scalars associated to strings stretching between the bulk branes and those on the fixed points are turned on. This would imply a recombination of the various branes, although one cannot bring everything in the bulk, as in the $\mathbb{Z}_2$ case,  since this would leave the $g^2$ and $g^4$ twisted R-R charge of the O-planes un-matched. The analysis of the allowed Higgsing requires a detailed knowledge of the $F$ and $D$ terms and of the allowed magnetisations of this model, which are beyond the scope of this study. 

Also in this $\mathbb{Z}_6$ vacuum, the NS-NS tadpole conditions cannot be fully satistified. Aside from the (universal) un-cancelled untwisted tadpole, which introduces the dilaton potential $e^{-\phi}$, one finds in general
\begin{equation} \label{NSNStadpolesZ6}
\begin{split}
N_{1,5} - D_{1,5} \neq 0 & \qquad \text{for $\beta_{1,0}$ and $\beta_{5,0}$}\,,
\\
D_{2,4} + 8\neq 0 &\qquad \text{for $\beta_{2,0}$ and $\beta_{4,0}$}\,.
\end{split}
\end{equation}
Thus, they induce new potential terms for the scalars $\xi_\alpha$ in the tensor multiplets from the $\mathbb{Z}_6$ fixed point in the $g$ and $g^2$ twisted sector, and their conjugate. Altogether
\begin{equation}
V (\phi , \xi ) = 64\, e^{-\phi} + e^{-\phi} \left[ 6a\, (\xi_1 + \xi_5 ) + 16\, (\xi_2 + \xi_4 ) + O (\xi^2 )\right]\,.
\label{Z6DilPot}
\end{equation}

\section{Anomaly polynomials and the Kahler $J$ form}
\label{Sec:Anomaly}

\subsection{The anomaly polynomial and the R-R tadpoles} 
\label{Sec:AnomalyRR}

In six dimensions anomalies impose severe constraints on the structure of the low energy effective theories\footnote{In the following, we shall restrict the discussion to perturbative anomalies. It would be interesting to study the role of Dai-Freed anomalies \cite{Dai:1994kq} along the lines of \cite{Debray:2021vob, Debray:2023yrs, Basile:2023knk, Basile:2023zng}.}
 .  Their cancellation is a non-trivial requirement and implies that the irreducible gauge and gravitational anomalies, proportional to $\text{tr} F^4$ and $\text{tr} R^4$, vanish identically, while the reducible ones factorise as
\begin{equation}
I_8 = \tfrac12 \Omega_{\alpha \beta} X_4^\alpha \wedge X_4^\beta \, , 
\label{factor-1}
\end{equation} 
where $\Omega_{\alpha\beta}$ is an $\text{SO} (1,n_\text{T})$ invariant metric with mostly minus signature, $n_\text{T}$ counts the number of tensor multiplets, and 
\begin{equation}
    2 \, X_4^\alpha= a^\alpha \, \text{tr} \, R^2 + \sum_i \frac{b_i^\alpha}{\lambda_i}\, \text{tr} \, F_i^2 \,.
    \label{factor-2}
\end{equation}
Here $a^\alpha$ and $b^\alpha_i$ are model dependent coefficients, while the constant $\lambda_i$ is related to the normalisation of the group generators, and equals one for unitary and symplectic groups, while $\lambda_i = 2$ for orthogonal ones. 
The generalised Green-Schwarz-Sagnotti mechanism \cite{Green:1984bx,Green:1984sg, Sagnotti:1992qw} then requires that the tensor fields $C_2^\alpha$ behave non-trivially under the gauge and Lorentz transformations, so that the reducible anomaly \eqref{factor-1} is cancelled by the Chern-Simons counter-term
\begin{equation}
S_\text{GS} = \int \Omega_{\alpha \beta} \, C_2^\alpha \wedge X_4^\beta \, .
\label{Green-Schwarz}
\end{equation} 

In orientifold vacua, the cancellation of the R-R tadpoles guarantees that the irreducible anomalies vanish identically \cite{Aldazabal:1999nu, Bianchi:2000de}, while their structure, \emph{i.e.} the geometry of D-branes and O-planes, suggests a preferred \emph{stringy} choice \cite{Sagnotti:1992qw,Angelantonj:2020pyr} for the vectors $a$ and $b_i$, which must anyhow span a self-dual lattice\footnote{Clearly, the factorisation of the anomaly polynomial is not unique, and other choices can make manifest the self-duality of the lattice. 
This property  has been argued in \cite{Seiberg:2011dr} to follow from the identification of the dyonic strings as gauge instantons of charges $b_i$ for the gauge group $G_i$ \cite{Duff:1996cf} and a gravitational instanton with charge $a$. This picture is somewhat puzzling in the light of the considerations of Section \ref{Sec:defects}, even though the anomaly lattice preserves the embedding property.} \cite{Seiberg:2011dr}. 

The stringy choice for the $X_4^\alpha$ is justified by the anomalous couplings of D-branes \cite{Douglas:1995bn, Green:1996dd} and O-planes \cite{Morales:1998ux, Stefanski:1998he} to the tensors $C_2^\alpha$ in the Wess-Zumino term. Since the gravitational and tensor multiplets also contain the non-dynamical ten-form and six-form potentials, respectively, these anomalous couplings are then determined by the R-R charges of O-planes, $q_\alpha^O$, and D-branes, $q_\alpha^D$, which can be readily extracted from the R-R tadpoles \cite{Angelantonj:2020pyr}. If the R-R tadpoles are schematically written as
\begin{equation}
\sum_{i=9,5} q^{\text{O}_i}_\alpha + \sum_{i=9,5} \sum_a q^{\text{D}_{i,a}}_\alpha n_{i,a} =0\,, \label{schematictad}
\end{equation} 
where $n_{i,a}$ are the CP labels for $a$-th group factor 
\begin{equation}
G_\text{CP} = \prod_a G_{9,a} \times \prod_a G_{5,a}
\end{equation}
of the D9 and D5 (anti-)branes, the anomaly polynomial takes the form
\begin{equation}
2\, X^\alpha_4 = \frac{\ell_\alpha }{2\, \sqrt{N}} \left[ \sum_{i=9_- ,5_+ } q^{\text{O}_i}_\alpha \, \text{tr} R^2 + \sum_{i=9,\bar 5}\sum_a q^{\text{D}_{i,a}}_\alpha \text{tr} F_{i,a}^2 \right]\,.
\label{factor-3}
\end{equation}
where 
\begin{equation}
\begin{split}
\ell &= \left(1,1; \frac{1}{\sqrt{2} \, \sin (\pi/2)} \boldsymbol{1}_{16} \right)\,, & \text{for}\quad N=2\,,
\\
\ell &= \left(1,1; \frac{1}{\sin (\pi /4 )} \, \boldsymbol{1}_{4} ; \frac{1}{\sqrt{2} \, \sin (2\pi/4)} \boldsymbol{1}_{10} \right)\,, & \text{for}\quad N=4\,,
\\
\ell &= \left(1,1; \frac{1}{\sin ( \pi /6 )} ; \frac{1}{\sin (2\pi/6)} \, \boldsymbol{1}_{5} ; \frac{1}{\sqrt{2} \, \sin (3\pi/6)} \boldsymbol{1}_{6} \right)\,, & \text{for}\quad N=6\,,
\end{split}
\end{equation}
with the coefficients $\sin (a\pi/N)$ clearly related to the number of fixed points in the various twisted sectors, and $\boldsymbol{1}_d$ a 
 $d$-dimensional vector whose entries are all equal to one.
 Therefore, we can read $X^\alpha_4$ from eq. \eqref{schematictad}, so that, up to the overall constant, the net charge of the O-planes determines the coefficient $a^\alpha$ of the Pontryagin class $\text{tr} R^2$, while the charges of the D-branes determine the coefficients $b^\alpha_i$ of the Chern class, and the CP labels $n_{i,a}$ are identified with $\text{tr} F_{i,a}^2$, with the assumption that, in the case of unitary groups, conjugate CP charges yield the same field strength\footnote{In this paper, as in most of the recent literature, we are not interested in Abelian anomalies. Taking these into account requires a straightforward refinement of our identifications of CP labels and Chern classes.}. 

In the  $T^4 /\mathbb{Z}_N$ orientifolds we are interested in, the untwisted and twisted charges of the orientifold planes are 
\begin{equation}
\begin{split}
q^{\text{O}9_-} &= \left( -4,-4; \boldsymbol{0}_{16} \right) \,, \qquad & & q^{\text{O}5_+^i} = \left( \tfrac{1}{4}, -\tfrac{1}{4} ;\boldsymbol{0}_{16} \right)\,, \qquad & \text{for}\quad N=2\,,
\\
q^{\text{O}9_-} &= \left( -4,-4 ; \boldsymbol{0}_4 ; \boldsymbol{0}_{10} \right)\,, \qquad & & q^{\text{O}5_+^i} = \left( \tfrac{1}{4},-\tfrac{1}{4} ; \boldsymbol{0}_4 ; - \boldsymbol{\eta}^i_{10}  \right) \,, \qquad & \text{for}\quad N=4\,,
\\
q^{\text{O}9_-} &= \left( -4,-4 ; 0; - \boldsymbol{1}_5 ; \boldsymbol{0}_{6} \right)\,, \qquad & & q^{\text{O}5_+^i} = \left( \tfrac{1}{4},-\tfrac{1}{4} ; 0 ; -3\, \boldsymbol{\delta}_5^1 ; \boldsymbol{0}_{6}  \right) \,, \qquad &\text{for}\quad N=6\,.
\end{split} \label{Oplanecv}
\end{equation}
In these expressions the semicolon separates the untwisted charges (the first two entries) from the twisted ones. In the $N=2$ case we have one twisted sector with sixteen fixed points, in the $N=4$ case, we have four $\mathbb{Z}_4$ fixed points from the $g$-twisted sector and ten fixed points (four $\mathbb{Z}_4$ and six doublets of $\mathbb{Z}_2$ type) from the $g^2$-twisted sector. Finally, in the $N=6$ case, we have a single $\mathbb{Z}_6$ fixed point from the $g$-twisted sector, five fixed points (one $\mathbb{Z}_6$ and four doublets of $\mathbb{Z}_3$ type) from the $g^2$-twisted sector, and six fixed points (one $\mathbb{Z}_6$ and five triplets of $\mathbb{Z}_2$ type) from the $g^3$-twisted sector. Moreover, $\boldsymbol{0}_d$ denotes a $d$-dimensional null vector,  $\boldsymbol{\delta}_d^i$ is a $d$-dimensional vector whose only non-vanishing component is $\{\boldsymbol{\delta}^i_d\}_j = \delta^i_j$, while $\boldsymbol{\eta}_{10}^i$ is a ten-dimensional vector with components $\{ \boldsymbol{\eta}_{10}^i \}_j = \delta^i_j$, for $i=1,\ldots , 4$, and zero for $i=5,\ldots , 10$. 

Moving to the D-brane charges, these depend on their positions and/or Wilson lines. We have
\begin{equation}
q^{\text{D}9_1} = (1,1; \boldsymbol{1}_{16} ) \,, \qquad q^{\text{D}9_2} = (1,1; - \boldsymbol{1}_{16} ) \,, \qquad q^{\overline{\text{D}5}_i} = (-1, 1; 4\, \boldsymbol{\delta}^i_{16} ) \,,
\label{DCVZ2}
\end{equation}
for the $\mathbb{Z}_2$ vacuum of Section \ref{Sec:BSBZ2} with the $\overline{\text{D}5}$ branes equally distributed among the sixteen fixed points, 
\begin{equation}
\begin{split}
& q^{\text{D}9_1} = (1,1; \boldsymbol{1}_4 ; \boldsymbol{1}_{10} )\,,  & & q^{\text{D}9_2} = (1,1; - \boldsymbol{1}_4 ; \boldsymbol{1}_{10} )\,, \qquad  & & q^{\text{D}9_3} = (2,2;  \boldsymbol{0}_4 ; -2\, \boldsymbol{1}_{10} )\,,
\\
& q^{\overline{\text{D}5}_{k,1}} = (-1,1;  2\, \boldsymbol{\delta}^k_4 ; 4\, \boldsymbol{\eta}^k_{10} )\,,  \quad & &  q^{\overline{\text{D}5}_{k,2}} = (-1,1;  - 2\, \boldsymbol{\delta}^k_4 ; 4\, \boldsymbol{\eta}^k_{10} ) \,,
\end{split} \label{DCVZ4}
\end{equation}
for the $\mathbb{Z}_4$ vacuum of Section \ref{Sec:BSBZ4} with $\overline{\text{D}5}$ on the $\mathbb{Z}_4$ fixed points, and
\begin{equation}
\begin{split}
q^{\text{D}9_1} &= (1,1; 1; \boldsymbol{1}_5 ; \boldsymbol{1}_{6} )\,, \\
q^{\text{D}9_2} &= (2,2; 1 ; - \boldsymbol{1}_5 ; -2\, \boldsymbol{1}_{6} )\,, \\ 
q^{\text{D}9_3} &= (2,2; -1 ;  - \boldsymbol{1}_5 ;  2\, \boldsymbol{1}_{6} )\,, \\
q^{\text{D}9_4} &= (1,1; -1;  \boldsymbol{1}_5 ; - \boldsymbol{1}_{6} ) \,,
\end{split}
\qquad
\begin{split}
q^{\overline{\text{D}5}_{1}} &= (-1,1;1;  3\, \boldsymbol{\delta}_5^1 ; 4\, \boldsymbol{\delta}^1_{6} )\,, \\ 
q^{\overline{\text{D}5}_{2}} &= (-2,2; 1; - 3\, \boldsymbol{\delta}_5^1 ; -8\, \boldsymbol{\delta}^1_{6} ) \,, \\ 
q^{\overline{\text{D}5}_{3}} &= (-2,2;-1;  -3\, \boldsymbol{\delta}_5^1 ; 8\, \boldsymbol{\delta}^1_{6} )\,, \\ 
q^{\overline{\text{D}5}_{4}} &= (-1,1; -1; 3 \, \boldsymbol{\delta}_5^1 ; - 4\, \boldsymbol{\delta}^1_{6} ) \,,
\end{split} \label{DCVZ6}
\end{equation}
for the $\mathbb{Z}_6$ vacuum of Section \ref{Sec:BSBZ6} with all $\overline{\text{D}5}$ branes located on the unique $\mathbb{Z}_6$ fixed point. In these expressions the charges of the various CP factors for the D9 and $\overline{\text{D}5}$ branes are ordered as in eqs. \eqref{CPZ2}, \eqref{CPZ4} and \eqref{CPZ6}. 

Let us discuss in detail some examples. We start from the $\mathbb{Z}_2$ orientifold with almost rigid branes, of Section \ref{SSec:Z2rigid}.  The $\overline{\text{D5}}$ branes are placed on all the fixed points and, using eqs. \eqref{Oplanecv} and \eqref{DCVZ2}, we determine the anomaly coefficients
\begin{equation}\label{eq:anomalyvectorsZ2}
a=\left (0, - 2\sqrt{2};  \boldsymbol{0}_{16} \right ) \, , \quad
b_1= \left ( \tfrac{1}{\sqrt{2}}, \tfrac{1}{\sqrt{2}}; \tfrac{1}{2} \boldsymbol{1}_{16}  \right ) \, ,\quad
b_2= \left ( \tfrac{1}{\sqrt{2}}, \tfrac{1}{\sqrt{2}}; -\tfrac{1}{2} \boldsymbol{1}_{16} \right ) \, ,\quad
b_{2+i}=  \left ( -\tfrac{1}{2 \sqrt{2}}, \tfrac{1}{2 \sqrt{2}};   \boldsymbol{\delta}^i_{16} \right) \,,
\end{equation}
where $i=1, \ldots , 16$ runs over the $\mathbb{Z}_2$ fixed points.
Therefore, the anomaly polynomial
\begin{equation}
\label{poly-interZ2}
\begin{split}
I_8  = \tfrac{1}{64} & \left [ \left(\text{tr} F_{9,1}^2+ \text{tr} F_{9,2}^2 - \sum_{i=1}^{16} \text{tr} F_{\bar 5,i}^2 \right)^2  -  \left(-8\,  \text{tr} R^2+ \text{tr} F_{9,1}^2+ \text{tr} F_{9,2}^2 + \sum_{i=1}^{16} \text{tr} F_{\bar 5,i}^2  \right)^2 \right.  \\
 &  \left. - \frac{1}{2} \sum_{i=1}^{16}  \left(\text{tr} F_{9,1}^2 - \text{tr} F_{9,2}^2 + 4 \, \text{tr} F_{\bar 5,i}^2  \right)^2 \right ] \, .
\end{split}
\end{equation}
clearly reflects the chosen geometry of O-planes and D-branes \cite{Angelantonj:2020pyr}.  As discussed in Section \ref{SSec:Z2rigid}, this vacuum admits an equivalent description in terms of $\text{O}7_-$ and $\text{O}7_+^\prime$ planes and D7 and $\overline{\text{D}7}'$ branes wrapping the cycles \eqref{eq:D9intersectingbranes} and \eqref{eq:D5intersectingbranes}. Consequently, the charge vectors for the O-planes and D-branes change accordingly. In particular, the twisted charges $c^a_{ij}$ and $\tilde c^A_{ij}$ of the D7 and $\overline{\text{D}7}'$ branes, respectively, are given by twice the coefficients of the fractional cycles $\boldsymbol{e}_{ij}$ in eqs.  \eqref{eq:D9intersectingbranes} and \eqref{eq:D5intersectingbranes}, so that the anomaly polynomial now reads
\begin{equation}
\label{poly-interZ2D7}
\begin{split}
I_8  = \tfrac{1}{64} & \left[ \left(\text{tr} F_{7,1}^2+ \text{tr} F_{7,2}^2 - \sum_{A=1}^{16} \text{tr} F_{\bar 7',A}^2 \right)^2  -  \left (-8\,  \text{tr} R^2+ \text{tr} F_{7,1}^2+ \text{tr} F_{7,2}^2 + \sum_{A=1}^{16} \text{tr} F_{\bar 7',A}^2  \right )^2 \right.
\\
 &  \left. - 2 \sum_{i,j=1}^{4}  \left ( \sum_{a=1,2}  c^{a}_{ij} \, \text{tr}  F_{7,a}^2 + \sum_{A=1}^{16} \tilde c^A_{ij} \, \text{tr}  F_{\bar 7',A}^2   \right )^2 \right] \, ,
\end{split}
\end{equation}
which provides a different factorisation of \eqref{poly-interZ2} and reflects the new geometry of O-planes and D-branes.

The $\mathbb{Z}_4$ case is richer, since the O-planes couple to twisted R-R six-forms. From the charges \eqref{Oplanecv} and \eqref{DCVZ4} we can extract the anomaly vectors
\begin{equation}
\begin{split}
a & = \left( 0,- 2; - \tfrac{1}{\sqrt{2}} \boldsymbol{1}_4 ; \boldsymbol{0}_{10} \right) \, , 
\\
b_1 & = \left(\tfrac12,\tfrac12; \tfrac{1}{\sqrt{2}} \boldsymbol{1}_4 ; \tfrac1{2\sqrt{2}} \boldsymbol{1}_4 ; \tfrac12\boldsymbol{1}_6 \right)\,,
\\
b_2 & =  \left(\tfrac12,\tfrac12; -\tfrac{1}{\sqrt{2}} \boldsymbol{1}_4 ;\tfrac1{2\sqrt{2}} \boldsymbol{1}_4 ; \tfrac12\boldsymbol{1}_6 \right)\,,
\\
b_3 & = \left(\tfrac12,\tfrac12;  \boldsymbol{0}_4 ; - \tfrac1{2\sqrt{2}} \boldsymbol{1}_4 ; - \tfrac12\boldsymbol{1}_6 \right)\,,
\\
b_4 & = \left( -\tfrac14, \tfrac14;  \tfrac{1}{\sqrt{2}} \boldsymbol{\delta}^1_4 ; \tfrac1{\sqrt{2}}  \boldsymbol{\eta}^1_{10} \right) \, ,
\\
b_5 & = \left( -\tfrac14, \tfrac14;  \tfrac{1}{\sqrt{2}} \boldsymbol{\delta}^2_4 ; \tfrac1{\sqrt{2}}  \boldsymbol{\eta}^2_{10}  \right)  \, ,
\end{split}
\qquad
\begin{split}
b_6 & = \left( -\tfrac14, \tfrac14;  \tfrac{1}{\sqrt{2}} \boldsymbol{\delta}^3_4 ; \tfrac1{\sqrt{2}}  \boldsymbol{\eta}^3_{10}   \right) \,,
\\
b_7 & = \left( -\tfrac14, \tfrac14;  \tfrac{1}{\sqrt{2}} \boldsymbol{\delta}^4_4 ; \tfrac1{\sqrt{2}}  \boldsymbol{\eta}^4_{10}  \right)  \, , 
\\
b_8 & =  \left( -\tfrac14, \tfrac14;  - \tfrac{1}{\sqrt{2}} \boldsymbol{\delta}^1_4 ; \tfrac1{\sqrt{2}}  \boldsymbol{\eta}^1_{10}  \right)  \,,
\\
b_9 & = \left( -\tfrac14, \tfrac14;  - \tfrac{1}{\sqrt{2}} \boldsymbol{\delta}^2_4 ; \tfrac1{\sqrt{2}}  \boldsymbol{\eta}^2_{10} \right)  \,,
\\
b_{10} & =\left( -\tfrac14, \tfrac14;  - \tfrac{1}{\sqrt{2}} \boldsymbol{\delta}^3_4 ; \tfrac1{\sqrt{2}}  \boldsymbol{\eta}^3_{10}  \right)  \,,
\\
b_{11} & = \left( -\tfrac14, \tfrac14;  - \tfrac{1}{\sqrt{2}} \boldsymbol{\delta}^4_4 ; \tfrac1{\sqrt{2}}  \boldsymbol{\eta}^4_{10} \right)  \,,
\end{split} \label{ab-Z4}
\end{equation}
and the geometric factorisation of the anomaly polynomial
\begin{equation} \label{polz4}
\begin{split}
I_8 &= \tfrac{1}{128}  \Bigg\{ \Bigg[\text{tr} F_{9,1}^2 + \text{tr} F_{9,2}^2 + 2\text{tr} F_{9,3}^2 - \sum_{k=1}^4 \left ( \text{tr} F_{\bar  5_k,1}^2 +  \text{tr} F_{\bar  5_k,2}^2 \right ) \Bigg]^2  
\\ 
&-  \Bigg[ - 8 \, \text{tr} R^2 + \text{tr} F_{9,1}^2 + \text{tr} F_{9,2}^2 +  2\text{tr} F_{9,3}^2 + \sum_{k=1}^4 \left ( \text{tr} F_{\bar 5_k,1}^2 +  \text{tr} F_{\bar 5_k,2}^2 \right ) \Bigg]^2
\\
& -2 \sum_{k=1}^4 \left[ \text{tr} F_{9,1}^2 - \text{tr} F_{9,2}^2+ 2 \left ( \text{tr} F_{\bar 5_k,1}^2 -  \text{tr} F_{\bar 5_k,2}^2 \right ) \right]^2 		
\\
&  -\tfrac{1}{2} \sum_{k=1}^4 \left[ - 4 \text{tr} R^2 + \text{tr} F_{9,1}^2 + \text{tr} F_{9,2}^2 -  2\text{tr} F_{9,3}^2 + 4 \left ( \text{tr} F_{\bar 5_k,1}^2 +  \text{tr} F_{\bar 5_k,2}^2 \right ) \right]^2 
\\
&- 6 \left( \text{tr} F_{9,1}^2+ \text{tr} F_{9,2}^2 -  2\text{tr} F_{9,3}^2 \right)^2  \Bigg\}\,.
\end{split}
\end{equation}
Notice the contribution of the Pontryagin class to the ``twisted'' part (fourth line) of the anomaly polynomial, which reflects the fact that the $\text{O5}_+$-planes are fractional and carry a twisted charge. Moreover, the anomaly vectors \eqref{ab-Z4} and the factorisation \eqref{polz4} applies to the whole family of models in \eqref{CPZ4} parametrised by $a$, which only affects the value of the CP labels but not the charges of D-branes and O-planes.

The $\mathbb{Z}_6$ case and $\mathbb{Z}_2$ vacuum with non-vanishing $B$-field can be described in a similar fashion, and details on the anomaly vectors and on the anomaly polynomial can be found in Appendix \ref{App:Anomaly}.

\subsection{The $J$-vector and the NS-NS tadpoles} 
\label{Sec:JvectNSNS}

In vacua with ${\mathscr N} =(1,0)$ supersymmetry, the anomaly vectors $b_i$ not only determine the Green-Schwarz counter-terms, but also other couplings in the LEEA \cite{Sagnotti:1992qw}, which are indeed related by supersymmetry. In particular, they control the gauge kinetic functions via the coupling\footnote{Following \cite{Kim:2019vuc}, we use the symbol $J_\alpha$ to identify the vector $v^r$ of \cite{Romans:1986er}. $J$ plays the role of the Kahler form determining the geometry of the scalar manifold.}
\begin{equation}
\tfrac{1}{2} \, J_\alpha b_i^\alpha \, \text{tr} \, F_{i} \wedge\star  F_i  \,, \label{GKT}
\end{equation}
between the gauge field strengths and the $n_\text{T}$ scalars $\tau$ in the tensor multiplets, which paramtrise the coset $\text{SO} (1 , n_\text{T})/\text{SO} (n_\text{T} )$. Positivity of the kinetic terms for the scalar fields and the gauge vectors then require \cite{Sagnotti:1992qw}
\begin{equation}
J\cdot J >0\qquad \text{and}\qquad J\cdot b_i >0 \,, \label{JKSV}
\end{equation}
where the inner product is taken with the $\text{SO} (1 , n_\text{T})$ invariant metric $\Omega_{\alpha \beta}$. 

Although the supersymmetry invariance of the LEEA requires that the gauge kinetic function be given in terms of the anomaly vector $b_i$, its origin in string theory is, in principle, different. In orientifold vacua, the gauge multiplets originate from the open-string sector, and thus the coupling \eqref{GKT} should also be extracted from the tadpoles. Indeed, turning on a background magnetic field $H_i$ on the D-branes, and expanding the $J$ function as $J_\alpha \sim \tau_\alpha +  O(\tau^2)$,  eq. \eqref{GKT} becomes schematically
\begin{equation}
\tau_\alpha b_i^\alpha \, H_i^2. \label{GKTtadpole}
\end{equation}
This equation clearly identifies the coefficients $b_i^\alpha$ with the tadpoles of the scalars in the tensor multiplets, \emph{i.e.} the one point functions of the $\tau$ fields with the various D-branes. Now, while the anti-self-dual tensors $C_2^-$ originate from the R-R sector, their scalar superpartners are of NS-NS type, and therefore the $b_i^\alpha$ in \eqref{GKTtadpole} are actually related to the NS-NS tadpoles, and not to the R-R ones. Clearly, supersymmetry relates them, and that's why the same anomaly vectors enter in \eqref{factor-2} and \eqref{GKT}. 

This observation, suggests that some care is needed when dealing with BSB vacua. In fact, in this class of orientifolds, R-R and NS-NS tadpoles are no longer related by supersymmetry, since for the $\overline{\text{D}5}$ branes the tension and charges are not equal. The kinetic terms for the gauge fields should then involve new coefficients $\tilde b_i^\alpha$,
\begin{equation}
\tfrac{1}{2}\, J \cdot \tilde b_i \, \text{tr}\, F_i \wedge \star F_i\,,
\end{equation}
which can be read from the NS-NS tadpoles. They differ from the anomaly vectors for a sign flip for the entries associated to the gauge factors living on the antibranes, thus reflecting the sign difference between their tensions and charges.  This is at all compatible with supersymmetry, since on the $\overline{\text{D}5}$ branes --- or, in general, in the non-supersymmetric sector --- supersymmetry is realised non-linearly, which implies that each term in the corresponding Lagrangian is invariant by itself\footnote{We thank Gianfranco Pradisi and Fabio Riccioni for clarifying this point to us.} \cite{Pradisi:2001yv}. 

For instance, for the $\mathbb{Z}_2$ vacuum of \cite{Antoniadis:1999xk}, the new vectors entering the gauge kinetic functions are
\begin{equation}
\begin{split}
\tilde b_1 &=\left ( \tfrac{1}{\sqrt{2}}, \tfrac{1}{\sqrt{2}}; \tfrac{1}{2} \boldsymbol{1}_{16}  \right ) = b_1 \,,
\\
\tilde b_2 &= \left ( \tfrac{1}{\sqrt{2}}, \tfrac{1}{\sqrt{2}}; -\tfrac{1}{2} \boldsymbol{1}_{16} \right ) = b_2\, ,
\end{split}
\qquad
\begin{split}
\tilde b_{3} &=  \left ( \tfrac{1}{2 \sqrt{2}}, -\tfrac{1}{2 \sqrt{2}};   - \boldsymbol{\delta}^1_{16} \right) = -b_{3} \,,
\\
\tilde b_{4} &=  \left ( \tfrac{1}{2 \sqrt{2}}, -\tfrac{1}{2 \sqrt{2}};   \boldsymbol{\delta}^1_{16} \right) = -b_{4} \,,
\end{split}
\end{equation}
where $\tilde b_{1,2}$ and $\tilde b_{3,4}$ refer to the two orthogonal/symplectic groups on the D9's and  $\overline{\text{D}5}$'s, respectively. With this choice of vectors it is now possible to define a $J$ function, since the conditions 
\begin{equation}
J\cdot J > 0 \qquad \text{and}\qquad J\cdot \tilde b_i >0
\end{equation}
are clearly compatible. If, following \cite{Kim:2019vuc}, we also impose the positivity of the Gauss-Bonnet term, $J\cdot a <0$, the entries of the $J$ vector must satisfy $J_0 > - J_1$ and $J_1 <0$, with the remaining entries chosen for simplicity to vanish,
\begin{equation}
J=  (J_0 , - |J_1 |; \boldsymbol{0}_4 ; \boldsymbol{0}_{10} )\,, \qquad J_0 > | J_1|\,.
\label{eq:Jform}
\end{equation}
This solves the puzzle on the non-existence of the $J$ form in \cite{Angelantonj:2020pyr}, where the na\"\i ve (wrong) choice $\tilde b_i \equiv b_i$ was made.

Similarly, for the $T^4/\mathbb{Z}_2$ almost rigid vacuum with $\overline{\text{D}5}$ branes evenly distributed on the sixteen fixed points, we find
\begin{equation}\label{eq:GKTZ2}
\tilde b_1= \left ( \tfrac{1}{\sqrt{2}}, \tfrac{1}{\sqrt{2}}; \tfrac{1}{2} \boldsymbol{1}_{16}  \right ) = b_1 \, ,\quad
\tilde b_2= \left ( \tfrac{1}{\sqrt{2}}, \tfrac{1}{\sqrt{2}}; -\tfrac{1}{2} \boldsymbol{1}_{16} \right ) = b_2\, ,\quad
\tilde b_{2+i}=  \left (\tfrac{1}{2 \sqrt{2}}, -\tfrac{1}{2 \sqrt{2}};   - \boldsymbol{\delta}^i_{16} \right) = -b_{2+i} \,,
\end{equation}
and the same solution \eqref{eq:Jform} for the $J$ form exists also in this case. 

Finally, for the $\mathbb{Z}_4$ orientifold 
\begin{equation}
\begin{split}
\tilde b_1 & = \left(\tfrac12,\tfrac12; \tfrac{1}{\sqrt{2}} \boldsymbol{1}_4 ; \tfrac1{2\sqrt{2}} \boldsymbol{1}_4 ; \tfrac12\boldsymbol{1}_6 \right) = b_1\,,
\\
\tilde b_2 & =  \left(\tfrac12,\tfrac12; -\tfrac{1}{\sqrt{2}} \boldsymbol{1}_4 ;\tfrac1{2\sqrt{2}} \boldsymbol{1}_4 ; \tfrac12\boldsymbol{1}_6 \right)= b_2 \,,
\\
\tilde b_3 & = \left(\tfrac12,\tfrac12;  \boldsymbol{0}_4 ; - \tfrac1{2\sqrt{2}} \boldsymbol{1}_4 ; - \tfrac12\boldsymbol{1}_6 \right) = b_3\,,
\\
\tilde b_4 & = \left( \tfrac14, -\tfrac14; - \tfrac{1}{\sqrt{2}} \boldsymbol{\delta}^1_4 ; - \tfrac1{\sqrt{2}}  \boldsymbol{\eta}^1_{10} \right) = - b_4\, ,
\\
\tilde b_5 & = \left( \tfrac14, - \tfrac14; - \tfrac{1}{\sqrt{2}} \boldsymbol{\delta}^2_4 ; - \tfrac1{\sqrt{2}}  \boldsymbol{\eta}^2_{10}  \right)  =-b_5 \, ,
\\
\tilde b_6 & = \left( \tfrac14, -\tfrac14;  - \tfrac{1}{\sqrt{2}} \boldsymbol{\delta}^3_4 ; - \tfrac1{\sqrt{2}}  \boldsymbol{\eta}^3_{10}   \right) =- b_6\,,
\end{split}
\qquad
\begin{split}
\tilde b_7 & = \left( \tfrac14, - \tfrac14;  -\tfrac{1}{\sqrt{2}} \boldsymbol{\delta}^4_4 ;- \tfrac1{\sqrt{2}}  \boldsymbol{\eta}^4_{10}  \right) =- b_7  \, , 
\\
\tilde b_8 & =  \left( \tfrac14, - \tfrac14;   \tfrac{1}{\sqrt{2}} \boldsymbol{\delta}^1_4 ; - \tfrac1{\sqrt{2}}  \boldsymbol{\eta}^1_{10}  \right) =- b_8  \,,
\\
\tilde b_9 & = \left( \tfrac14, - \tfrac14;   \tfrac{1}{\sqrt{2}} \boldsymbol{\delta}^2_4 ; - \tfrac1{\sqrt{2}}  \boldsymbol{\eta}^2_{10} \right)  =- b_9 \,,
\\
\tilde b_{10} & =\left(\tfrac14, -\tfrac14;  \tfrac{1}{\sqrt{2}} \boldsymbol{\delta}^3_4 ; - \tfrac1{\sqrt{2}}  \boldsymbol{\eta}^3_{10}  \right)  =- b_{10} \,,
\\
\tilde b_{11} & = \left( \tfrac14, -\tfrac14;  \tfrac{1}{\sqrt{2}} \boldsymbol{\delta}^4_4 ; - \tfrac1{\sqrt{2}}  \boldsymbol{\eta}^4_{10} \right) = - b_{11} \,,
\\
&
\end{split} \label{GKTbt-Z4}
\end{equation}
which are clearly compatible with the same choice \eqref{eq:Jform}. Other cases can be treated in a similar fashion.

To summarise, in full generality if a massless gravitino is present in the spectrum, \emph{i.e.} if ${\mathscr N}=(1,0)$ supersymmetry is exact or non-linearly realised in the open-string sector, the anomaly vectors and the vectors entering the gauge kinetic functions take the form
\begin{equation}
b_i^\alpha = \frac{\ell_\alpha}{2 \sqrt{N}} \, \lambda_i \, q^{D_i}_\alpha\,, \qquad 
\tilde b_i^\alpha = \frac{\ell_\alpha}{2 \sqrt{N}} \, \lambda_i \, t^{D_i}_\alpha\,,
\end{equation}
with $t^{D_i}_\alpha$ and $q^{D_i}_\alpha$ the tensions and charges of the various (anti)branes present in the vacuum, and $\lambda_i = 1,2$ the usual group-theoretical factor associated to the normalisation of the group generators. As for the $a$ vector, it is determined by the orientifold planes,
\begin{equation}
a = \frac{\ell_\alpha}{2 \sqrt{N}} \, \sum_{i=9_- ,5_+ } q^{\text{O}_i}_\alpha \,,
\end{equation}
and it is the same vector determining the coefficient of the Gauss-Bonnet term since a vacuum with massless gravitini must involve just $\text{O}_\pm$ planes, but not $\overline{\text{O}}_\pm$ ones\footnote{A simultaneous presence of O-planes and $\overline{O}$-planes implies that supersymmetry is also broken in the closed-string sector, which requires that gravitini are massive, or projected away.}.

\section{Defects and (new) unitarity constraints}
\label{Sec:defects}

The six-dimensional vacua discussed so far, involve a variable number of tensor multiplets which can naturally be sourced by one-dimensional objects, via the $\text{SO}(1,n_T)$ invariant  coupling
\begin{equation}
S_{2d} = - Q^\alpha \, \Omega_{\alpha\beta} \int C_2^\beta\,.
\end{equation}
The non-trivial transformation of $C_2^\beta$ under local gauge and Lorentz transformations, as required by the Green-Schwarz-Sagnotti mechanism, induces an anomaly inflow on the world-volume of the defect which, in a consistent vacuum, must be cancelled by the anomalous contribution 
\begin{equation}
I_4 = - I_\text{Inflow} = Q\cdot X_4 +\tfrac{1}{2} Q\cdot Q \, \chi (N) = \tfrac{1}{2} \left( Q\cdot a \,\text{tr}\, R^2 +Q\cdot Q\, \chi (N) +\sum_i \frac{Q\cdot b_i}{\lambda_i} \, \text{tr} \, F_i^2 \right)
\end{equation}
of its two-dimensional degrees of freedom (dof's) \cite{Green:1996dd}. Under the decomposition $\text{SO} (1,5) = \text{SO} (1,1) \times \text{SU} (2)_l \times \text{SU} (2)_R$,
\begin{equation}
\text{tr}\, R^2 = -\tfrac{1}{2} p_1 (T_2 ) + c_2 (l) + c_2 (R) \,, \qquad \text{and}\quad \chi (N) = c_2 (R) - c_2 (l)\,,
\end{equation}
so that the anomaly polynomial reads
\begin{equation}
I_4 = - \tfrac{1}{12} 3 Q\cdot a \, p_1 (T_2)  + \tfrac{1}{2} (Q\cdot Q + Q \cdot a )\, c_2 (R)  + \tfrac{1}{2} (Q\cdot Q - Q \cdot a)\, c_2 (l)  +\tfrac{1}{2} \sum_i \frac{Q\cdot b_i}{\lambda_i} \, \text{tr} \, F_i^2\,,
\end{equation}

%where we have removed the contribution of the centre-of-mass (CM) dof's, consisting in four scalars and two right-moving Majorana-Weyl fermions\footnote{Notice that our conventions are different that those of \cite{Kim:2019vuc}, in that the role of $\text{SU} (2)_l$ and $\text{SU} (2)_R$ is exchanged.}, which decouple in the IR. \textbf{Say it better. The CM dof's become non-interacting in the IR and therefore the non-trivial CFT which realised the KM algebras on the defect must have a smaller central charge}

Although the coefficient of $p_1 (T_2)$ can always be expressed in terms of the difference of the central charges from the left and right movers, $c_\text{R} - c_\text{L} =  6 Q\cdot a $, the identification of $c_\text{R}$ with the coefficient of $ c_2 (l)$  \cite{Kim:2019vuc}, is no-longer valid for different reasons\footnote{Note the change of conventions with respect to \cite{Kim:2019vuc}, since for us it is $\text{SU} (2)_l$ which plays the role of the R-symmetry.}
. The simplest and most striking one is that the D1 defects in BSB vacua do not enjoy ${\mathscr N} = (0,4)$ supersymmetry on their world-volume. This is due to the presence of anti-branes which flip the chirality of the excitations of the strings stretched between the D1 and the $\overline{\text{D}5}$ branes, thus explicitly breaking supersymmetry. The second and more subtle reason is that, in general, $\text{SU} (2)_l$ cannot be  identified with the R-symmetry group of the ${\mathscr N} = (0,4)$ algebra, even in the supersymmetric case.  In fact, both left and right moving fields can transform in a non-trivial representations of $\text{SU} (2)_l$, and thus both contribute to the coefficient of $c_2 (l)$, although only right movers should be allowed to transform under the true R-symmetry. As shown in \cite{Angelantonj:2020pyr}, it seems that $\text{SU} (2)_l$ can be identified with the R-symmetry only in the case of a single D1 brane away from the orbifold fixed points. In all other cases, $ c_2 (l)$ gets contributions also from left-moving excitations. 
 We hope to come back to elucidate this point in the near future. For the time being, given the difficulty/impossibility to extract $c_\text{L}$ and $c_\text{R}$ directly from the anomaly inflow,  we shall restrict to a top-down analysis of the consistency of the effective theory on the defects. The main scope of this section is then to accumulate \emph{empirical data} on the two-dimensional CFT on the D1 branes, instrumental for deriving refined consistency conditions in a sought-after bottom-up approach. 

In our class of vacua with D9 and D5 (anti-)branes the defects can be identified with D1 branes localised in the internal space, or with $\text{D}5'$ branes wrapping the entire $T^4/\mathbb{Z}_N$ orbifold\footnote{Clearly, one can also have $\overline{\text{D}1}$, $\overline{\text{D}5}'$ and magnetised $\text{D}5'$ branes. In these cases, the discussion would follow a similar pattern.}.  D1 branes located on a $\mathbb{Z}_N$ fixed point carry the same gauge group of D9 branes, while the gauge group is orthogonal if they are moved in the bulk. The CP group of $\text{D}5'$ branes is equal in structure to that of $\overline{\text{D}5}$ branes sitting on a $\mathbb{Z}_N$ fixed point, although tadpole conditions may force some of the group factors to be absent on the $\overline{\text{D}5}$'s. Open strings stretched between D1 and D9 branes have ND boundary conditions along the eight transverse directions, and therefore the light excitations involve LH fermions which are singlets of the ${\mathscr N} = (0,4)$ superalgebra. Instead, open strings stretched between D1 and $\overline{\text{D}5}$ branes have DD boundary conditions along the internal space, so that the light excitations now comprise a full would-be left-moving (LM) supermultiplet (two LM chiral scalars and two LM MW fermions) and a right-moving (RM) MW fermionic singlet and a pair of RM chiral bosons. These states carry a flipped chirality, incompatible with ${\mathscr N} = (0,4)$ supersymmetry, since D1 and  $\overline{\text{D}5}$ branes would preserve super-charges of opposite chirality. For the interactions of the $\text{D}5 '$ branes with D9 and
$\overline{\text{D}5}$ ones, we have a full supermultiplet and a LM fermionic singlet in the $95'$ sector, and a RM fermionic singlet in the $55'$ sector. The presence of massless scalars in the $51$ and $95'$ strings suggests that $\text{D}5'$ (D1) branes can be interpreted as (anti-) instantons of D9 ($\overline{\text{D}5}$) branes. This identification is further supported by the fact that the twisted charges of the candidate instanton brane/physical brane pairs match. In these cases, the charge vector $Q$ can be expressed in terms of the anomaly vectors $b_i$ associated to the gauge groups of the D9 ($\overline{\text{D}5}$) branes. 

Notice that D1 branes can always be moved in the bulk of the compactification orbifold. In this case, they can only couple to the two-forms $C_2^{0}$ and $C_2^1$ from the untwisted sector. In the decompactification limit, one is expected to recover the ten-dimensional type I superstring with a defect coupled to the (non-chiral) R-R two-form $C_2$ existing in $D=10$. Since, in $D=6$, $C_2^{0}$ and $C_2^1$ are the two chiral components of $C_2$, the D1 brane must then carry the same charge, $Q^0 = \pm Q^1$, thus satisfying the \emph{null-charge} conjecture, $Q\cdot Q =0$ \cite{Angelantonj:2020pyr}. We expect that this is the case for any six-dimensional vacuum with at least one tensor multiplet. The existence of \emph{null-charge} branes allowed us to put in the swampland various models which pass the KSV constraints, but do not admit a string or F-theory construction \cite{Angelantonj:2020pyr}.

Another generic feature of these vacua is that while the gauge group of the D9 branes is realised on the D1 defect as a left-moving Ka$\check{\text{c}}$-Moody algebra, the  $\overline{\text{D}5}$ gauge group can be realised both on the left and right movers. This is reflected in the sign of the Ka$\check{\text{c}}$-Moody level $k_i $, which can be extracted from the coefficient of $\text{tr}\, F_i^2$ in the anomaly polynomial $I_4$,  and in its contribution to the central charge. 
In general, for the gauge group $G_{9,i}$ on the D9 branes with dual Coxeter number $h_i^\vee$, the level $ k_i$ is positive and
\begin{equation}
c_i = \frac{k_i \, \text{dim} \, G_{9,i}}{k_i + h_i^\vee}\,,
\end{equation}
since the Ka$\check{\text{c}}$-Moody algebra is realised by a LM fermion, which is a singlet of ${\mathscr N} = (0,4)$. For the gauge group $G_{\bar 5 ,i}$ on the $\overline{\text{D}5}$ branes, the level of the Ka$\check{\text{c}}$-Moody algebra can be positive or negative. A similar feature also emerges on the $\text{D}5'$ branes, where the levels of the $\overline{\text{D}5}$ gauge group are negative, while those associated to the D9 group can be positive or negative.

Upon removing the contribution of the centre of mass (CM) dof's (four non-chiral scalars and four RM MW fermions) which decouple in the IR, a natural generalisation of the KSV unitarity constraints is 
\begin{equation}
\sum_{i\,|\, k_i>0} \frac{k_i \, \text{dim} \, G_i}{k_i + h_i^\vee} \le c_\text{L}-4\qquad \text{and} \qquad 
\sum_{i\,|\, k_i<0} \frac{|k_i| \, \text{dim} \, G_i}{|k_i| + h_i^\vee} \le c_\text{R}-6\,,
\label{eq:chiralKSV}
\end{equation}
both for the left and right moving sectors. However, in the UV charged (non-chiral) scalar fields are present on the defect, and their role in the realisation of the Ka$\check{\text{c}}$-Moody algebras in the IR is unclear. These scalar fields are non-compact and their dynamics in the IR is difficult to determine. They could be free and generate an Abelian algebra or describe an independent interacting sector of the theory. In both cases, they are expected to contribute to the left and right central charges and, if this happens, the inequalities \eqref{eq:chiralKSV} cannot be saturated. Indeed, this is the case in all string constructions where the defects admit an instanton interpretation.

\subsection{Some examples with D1 and $\text{D}5'$ defects}

We now turn to discuss some specific examples of defects in $T^4/\mathbb{Z}_4$ and $T^4 (B)/\mathbb{Z}_2$ orientifolds. Similar results also hold for the $T^4/\mathbb{Z}_6$ case. Details on the annulus and M\"obius strip amplitudes for D1 and $\text{D}5'$ branes can be found in Appendix \ref{App:ZNDefects}.

\subsubsection{D1 branes in the $T^4 /\mathbb{Z}_4$ orientifold}

We start our analysis by considering D1 branes sitting on a $\mathbb{Z}_4$ fixed point of the $T^4 /\mathbb{Z}_4$ orientifold discussed in Section \ref{Sec:BSBZ4}. This D1 brane supports the CP gauge group $\text{SO}(r_1) \times \text{SO}(r_2) \times \text{U}(r_3)$, which is similar to that of the D9 branes in the same vacuum, and includes states which are charged both under the gauge group of the D9  and that of the $\overline{\text{D}5}$ branes which are on the same fixed point. The light excitations are summarised in table \ref{D1-spectrumZ4}, where the representations of the various fields with respect to the $\text{SO} (4) \sim \text{SU} (2)_l \times \text{SU} (2)_R$ group, transverse to the D1 world-volume, are given.  The second line in the table includes the contribution of the CM dof's which are expected to decouple in the IR. From the microscopic data in the table, we can compute the anomaly polynomial,
\begin{equation} 
\begin{split}
I_4 &= \tfrac12 \left ( -(r_1+r_2) \text{tr}R^2 -  \left ( (r_1-r_3)^2+(r_2-r_3)^2 \right ) \, \chi(N) \right.
\\ 
& \qquad\left. + \tfrac{r_1}{2} \ \text{tr} F_{9,1}^2 + \tfrac{r_2}{2} \ \text{tr} F_{9,2}^2 + r_3 \ \text{tr} F_{9,3}^2 - (r_3-r_1)\ \text{tr} F_{\bar 5_1, 1}^2 
 - (r_3-r_2) \ \text{tr} F_{\bar 5_1,2}^2  \right )\,,
\label{eq:I4D1Z4}
\end{split}
\end{equation}
which indeed cancels the inflow from the bulk for the choice of charge vector
\begin{equation}
Q = -r_1 b_4 - r_2 b_8 + r_3 \left(\tfrac12, \tfrac12; -\sqrt{2}, \boldsymbol{0}_{13} \right) \,,
\label{eq:D1defZ4}
\end{equation}
with $b_4$ and $b_8$ the anomaly vectors given in eq. \eqref{ab-Z4}. Notice that this solution for $Q$ guarantees the positivity of the tension of the defect, $Q\cdot J >0$, with the K\"ahler form $J$ given in eq. \eqref{eq:Jform}. From \eqref{eq:I4D1Z4} we also extract the levels
\begin{equation}
    k_1= r_1 \, , \qquad k_2=r_2 \, , \qquad k_3=r_3 \, , \qquad k_4=r_1-r_3 \, , \qquad k_5= r_2 - r_3 \, , \qquad k_{6,\ldots, 11}=0 \,.
\end{equation}
The vanishing of  $k_{6,\ldots, 11}$ reflects the fact that the associated gauge groups live on a different fixed point than that visited by the D1 brane. We believe that this implies that the world-volume of the defect does not support any dof realising the Ka$\check{\text{c}}$-Moody algebra.

\begin{table}
\centering
\begin{tabular}{| c | c| c|}
\hline
 Representation &   $\text{SO}(1,1)\times \text{SU}(2)_l \times \text{SU}(2)_R $  &  Sector 
 \\
\hline
$\smalltableau{ \null \\ \null \\}_{1_1} + \smalltableau{ \null \\ \null \\}_{1_2} + \left ( \smalltableau{ \null \\} \times \overline{\smalltableau{\null \\}} \right )_{1_3} $  & $(0,1,1) + 2 \times ( \tfrac1 2,2,1)_\text{L}$ & D1-D1
\\
$\smalltableau{ \null \& \null \\}_{1_1} + \smalltableau{ \null \& \null \\}_{1_2} + \left ( \smalltableau{ \null \\} \times \overline{\smalltableau{\null \\}} \right )_{1_3}$  & $ (1,2,2)+2 \times (\tfrac12,1,2)_\text{R}$ &  
\\
$( \,  \smalltableau{\null \\}_{1_1}+ \smalltableau{\null \\}_{1_2},\smalltableau{\null \\}_{1_3}+ \overline{\smalltableau{\null \\}}_{1_3} \, )$  & $(\tfrac12,1,2)_\text{L}$ & 
\\
$(  \,  \smalltableau{\null \\}_{1_1}+ \smalltableau{\null \\}_{1_2},\smalltableau{\null \\}_{1_3}+ \overline{\smalltableau{\null \\}}_{1_3} \, )$  & $2\times (1,1,1)+(\tfrac12,2,1)_\text{R}$ & 
\\ 
\hline
$ ( \,  \smalltableau{\null \\}_{1_1}, \smalltableau{\null \\}_{9_1} \, ) + (\, \smalltableau{\null \\}_{1_2}, \smalltableau{\null \\}_{9_2} \, )+ ( \, \smalltableau{\null \\}_{1_3}, \overline{\smalltableau{\null \\}}_{9_3} \, ) + ( \, \overline{\smalltableau{\null \\}}_{1_3},\smalltableau{\null \\}_{9_3} \, ) $  & $ (\tfrac12,1,1)_\text{L}$ & D1-D9
\\
\hline
$( \,  \smalltableau{\null \\}_{1_1}, \smalltableau{\null \\}_{\bar 5_1} \, ) + (\, \smalltableau{\null \\}_{1_2}, \smalltableau{\null \\}_{\bar 5_2} \, )$  & $ (1,1,2) + 2 \times (\tfrac12,1,1)_\text{L}$ & D1-$\overline{\text{D}5}$ 
\\
$(\, \smalltableau{\null \\}_{1_3}+ \overline{\smalltableau{\null \\}}_{1_3} , \smalltableau{\null \\}_{\bar 5_1} + \smalltableau{\null \\}_{\bar 5_2} \, )$  & $  (\tfrac12,1,1)_\text{R}$ & 
\\
\hline
\end{tabular}\\
\caption{The light spectrum for a probe $\text{SO}(r_1)\times \text{SO}(r_2) \times \text{U}(r_3)$ D1 brane at a $\mathbb{Z}_4$ fixed point in the $T^4/\mathbb{Z}_4$ vacuum with D9/$\overline{\text{D5}}$ branes. The only $\overline{\text{D5}}$ branes contributing to the spectrum are those localised on the same fixed point as the probe D1 branes.}
\label{D1-spectrumZ4}
\end{table}

From the solution \eqref{eq:D1defZ4}, we read that the D1 branes with gauge groups $\text{SO} (r_1 )$ and $\text{SO} (r_2 )$ can be interpreted as instantons of the gauge theories on the $\overline{\text{D}5}$ branes, since $Q \propto b_{4,8}$. This observation is supported by the presence of moduli in the bi-fundamental representations $( \,  \smalltableau{\null \\}_{1_1}, \smalltableau{\null \\}_{\bar 5_1} \, ) + (\, \smalltableau{\null \\}_{1_2}, \smalltableau{\null \\}_{\bar 5_2} \, )$. On the contrary, the contribution of the defects with unitary group $\text{U} (r_3)$ to the charge vector $Q$ is not proportional to any anomaly vector, and therefore cannot be interpreted as instantons of a $\overline{\text{D}5}$ gauge group. Notice that the $\mathbb{Z}_4$ orbifold action would be compatible with an additional unitary group on the $\overline{\text{D}5}$ branes, for which the $r_3$ defects would be instantons. However, tadpole cancellation imposes that this unitary group be absent on the $\overline{\text{D}5}$'s, leaving no role for the $r_3$ D1 branes. This is compatible with the fact that there are no scalars stretched between the $r_3$ branes and the $\overline{\text{D}5}$ ones.

Finally, one can check that the constraints \eqref{eq:chiralKSV} are satisfied with $c_\text{L}$ and $c_\text{R}$ computed from table \ref{D1-spectrumZ4}, once the CM dof's are removed. As expected, this implies that the two-dimensional CFT is unitary and this string construction is consistent. As an example, in the \emph{symmetric} $a=2$ vacuum, where the $\overline{\text{D}5}$ support a $\text{USp} (4)^2$ gauge group on each fixed point, the simple choice $r_1=r_2=0$ and $r_3=1$ gives
\begin{equation}
    \begin{split}
    &\sum_{i \, | \, k_i \geq 0} \frac{k_i \ \text{dim} G_i }{k_i + h_i^{\vee}}=\frac{63}{9}+1= 8 < c_\text{L} -4_\text{CM} = 10 \,, 
    \\
    &\sum_{i \, | \, k_i <0} \frac{|k_i| \ \text{dim} G_i }{|k_i| + h_i^{\vee}}= \frac{10}{4} + \frac{10}{4} = 5 < c_\text{R} -6_\text{CM} = 8  \,,
\end{split}
\end{equation}
where the CM dof's have been removed from the counting of $c_\text{L,R}$. Notice that the in the second line the right-moving Ka$\check{\text{c}}$-Moody algebra does not saturate the available central charge, as one would na\"\i vely expect since in the UV there are no extra RM dof's. However, this is not a problem since the gauge group on the $\overline{\text{D}5}$ branes is symplectic and, in this case, the Sugawara construction is not equivalent to the free-fermion realisation of the algebra and extra dof's are present to account for the full central charge.

\begin{table}
\centering
\begin{tabular}{| c | c | c |}
\hline
	 { Representation} &  $\text{SO}(1,1) \times \text{SU}(2)_l \times \text{SU} (2)_R$ & { Sector} \\
	\hline
	$ \smalltableau{ \null \\ \null \\}   $  & $\left (0, 1, 1 \right ) + 2 \times \left (\tfrac{1}{2}, 2, 1 \right )_\text{L}  $  & D1-D1
	\\
	$ \smalltableau{ \null \& \null \\}   $ & $\left (1, 2, 2 \right ) + 2 \times \left (\tfrac{1}{2}, 1, 2 \right )_\text{R}$ & \ \ 
	\\
	$ \smalltableau{ \null \& \null \\}  $  & $4 \times \left (1, 1, 1 \right ) + 2 \times \left (\frac{1}{2}, 2, 1 \right )_\text{R} $ &  \ \ 
 \\
	 $  \smalltableau{ \null \\ \null \\} $  & $ 2 \times \left (\frac{1}{2},  1,2 \right )_\text{L} $ &  \ \
	\\ 
 \hline
	$  (\, \smalltableau{\null \\}_{1}, \smalltableau{\null \\}_{9_1}+\smalltableau{\null \\}_{9_2}+\smalltableau{\null \\}_{9_3}+\overline{\smalltableau{\null \\}}_{9_3} ) $  & $  \left (\frac{1}{2},  1, 1 \right )_\text{L} $ & D1-D9 
 \\
 \hline
\end{tabular}
\caption{Spectrum for probe D1 branes  in the bulk of the $T^4/\mathbb{Z}_4$ orbifold, supporting an $\text{SO}(r)$ gauge group.}
\label{tab:D1bulkZ4}
\end{table}

One has the option of moving the D1 branes away from fixed points. As in \cite{Angelantonj:2020pyr}, this turns out to be the configuration which satisfies the \emph{minimal} constraints of \cite{Kim:2019vuc}. The gauge group on the defect is $\text{SO} (r)$ and the microscopic dof's are listed in table \ref{tab:D1bulkZ4}, and clearly do not include the D1-$\overline{\text{D}5}$ sector since the open strings with these boundary conditions are massive. Also in this case, the second line in the table contains the CM dof's, which are the singlet in the decomposition of the twofold symmetric representation of orthogonal groups. The anomaly polynomial now reads
\begin{equation}
    I_4= \tfrac{1}{2} \Bigg ( -2 r \, \text{tr} R^2 + \tfrac{r}{2} \text{tr} F_{9,1}^2 + \tfrac{r}{2} \text{tr} F_{9,2}^2 + r \, \text{tr} F_{9,3}^2 \Bigg ) \, ,
\end{equation}
which cancels the inflow of the bulk theory if the charge vector is
\begin{equation}
    Q = \left ( r,-r ; \boldsymbol{0}_{14} \right ) \, .
\end{equation}
As expected, the D1 branes couple only to the two-forms from the untwisted sector, and therefore satisfy the \emph{null-charge} condition \cite{Angelantonj:2020pyr}. For $r=1$, \emph{i.e.} for a single D1 brane, the left and right central charges
\begin{equation}
c_\text{L}= 20 +4_{\text{CM}} \, ,
\qquad
c_\text{R}=  6 +6_{\text{CM}} \, ,
\end{equation}
agree with the those given in \cite{Kim:2019vuc}. We would like to stress, though, that for the case of an arbitrary stack of D1 branes, $r>1$, the expressions of $c_\text{L,R}$ in \cite{Kim:2019vuc} seem no-longer to be correct. 

Reading from $I_4$ that $k_1=k_2=k_3=1$, and noting that the Ka$\check{\text{c}}$-Moody algebra is realised only in the left-moving sector, the unitarity constraint reads
\begin{equation}
\sum_{i \, | \, k_i \geq 0} \frac{k_i \ \text{dim} G_i }{k_i + h_i^{\vee}}= 16 < c_\text{L}-4_\text{CM} =20 \, ,
\end{equation}
and is clearly satisfied. The missing dof's can be clearly identified with the four LM (uncharged) free bosons in the third line of table \ref{tab:D1bulkZ4}.

\subsubsection{$\text{D}5'$ branes in the $T^4/\mathbb{Z}_4$ orientifold}
\label{sssec:D5pZ4}

\begin{table}
\centering
\begin{tabular}{| c | c | c |}
\hline
	 { Representation} &  $\text{SO}(1,1) \times \text{SU}(2)_l \times \text{SU}(2)_R$ & { Sector} 
  \\
	\hline
	$\smalltableau{ \null \& \null \\}_{5'_1} + \smalltableau{ \null \& \null \\}_{5'_2} + \left ( \smalltableau{ \null \\} \times \overline{\smalltableau{\null \\}} \right )_{5'_3} $  & $\left (0, 1, 1 \right ) + 2 \times \left (\tfrac{1}{2}, 2, 1 \right )_\text{L}  $  & $\text{D}5'\text{-D}5'$
	\\
	$\smalltableau{ \null \\ \null \\}_{5'_1} + \smalltableau{ \null \\ \null \\}_{5'_2} + \left ( \smalltableau{ \null \\} \times \overline{\smalltableau{\null \\}} \right )_{5'_3} $ & $\left (1, 2, 2 \right ) + 2 \times \left (\tfrac{1}{2}, 1, 2 \right )_\text{R}$ &
	\\
	$ ( \,  \smalltableau{\null \\}_{5'_1}+ \smalltableau{\null \\}_{5'_2},\smalltableau{\null \\}_{5'_3}+ \overline{\smalltableau{\null \\}}_{5'_3} \, )$  & $2 \times \left (1, 1, 1 \right ) +  \left (\frac{1}{2}, 2, 1 \right )_\text{R} $ &  
	\\
	 $ ( \,  \smalltableau{\null \\}_{5'_1}+ \smalltableau{\null \\}_{5'_2},\smalltableau{\null \\}_{5'_3}+ \overline{\smalltableau{\null \\}}_{5'_3} \, )$  & $\left (\frac{1}{2},  1,2 \right )_\text{L} $ & 
	\\ 
 \hline
	$( \,  \smalltableau{\null \\}_{5'_1}, \smalltableau{\null \\}_{9_1} \, ) + (\, \smalltableau{\null \\}_{5'_2}, \smalltableau{\null \\}_{9_2} \, )+ ( \, \smalltableau{\null \\}_{5'_3}, \overline{\smalltableau{\null \\}}_{9_3} \, ) + ( \, \overline{\smalltableau{\null \\}}_{5'_3},\smalltableau{\null \\}_{9_3} \, )$  & $ \left (1, 2, 1 \right ) + 2 \times \left (\frac{1}{2},  1, 1 \right )_\text{R} $ & $\text{D}5'$-D9 
	\\
	$ ( \,  \smalltableau{\null \\}_{5'_1}, \smalltableau{\null \\}_{9_3} \, ) + (\, \smalltableau{\null \\}_{5'_2}, \overline{\smalltableau{\null \\}}_{9_3} \, )+ ( \, \smalltableau{\null \\}_{5'_3}, \smalltableau{\null \\}_{9_1} \, ) + ( \, \overline{\smalltableau{\null \\}}_{5'_3},\smalltableau{\null \\}_{9_2} \, ) + \text{c.c.}$  & $ \left (\frac{1}{2},  1, 1 \right )_\text{L} $ &  
 \\ 
 \hline
	$ \sum_{k=1}^4 ( \,  \smalltableau{\null \\}_{5'_1}, \smalltableau{\null \\}_{\bar 5_{k,1}} \, ) + (\, \smalltableau{\null \\}_{5'_2}, \smalltableau{\null \\}_{\bar 5_{k,2}} \, ) $  & $\left (\frac{1}{2},  1, 1 \right )_\text{R} $ & $\text{D}5'\text{-}\overline{\text{D}5}$ 
	 \\ 
  \hline
\end{tabular}
\caption{Spectrum for probe $\text{D}5'$ branes wrapping the entire $T^4/\mathbb{Z}_4$. The associated CP group is $\text{USp}(r_1)\times \text{USp}(r_2) \times \text{U}(r_3)$.}
\label{D5-spectrum}
\end{table}

Turning to $\text{D}5'$ defects wrapping the compactification space, the associated CP gauge group is $\text{USp}(r_1)\times \text{USp}(r_2) \times \text{U}(r_3)$ and the spectrum of light excitations is summarised in table \ref{D5-spectrum}. Clearly, the $\text{D}5'$ branes probe all $\mathbb{Z}_4$ fixed points and therefore couple to all the $\overline{\text{D}5}$ branes. The second line contains the contribution of the CM dof's of the defects, associated to the various singlets of the CP group. The anomaly polynomial is 
\begin{equation}
\begin{split}
I_4 &= \tfrac12 \Bigg ( (r_1+r_2) \text{tr} R^2 - (r_1-r_3) \text{tr} F_{9,1}^2 - (r_2-r_3) \text{tr} F_{9,2}^2 + (r_1+r_2-2r_3) \text{tr} F_{9,3}^2 
\\
& - \tfrac{r_1}{2} \sum_{k=1}^4\text{tr} F_{\bar 5_k,1}^2 - \tfrac{r_2}{2} \sum_{k=1}^4 \text{tr} F_{\bar 5_k,2}^2 -  [(r_1-r_3)^2 +(r_2-r_3)^2] \chi(N) \Bigg )
\end{split} 
\label{eq:I4D5'Z4}
\end{equation}
which cancels the inflow for the choice 
\begin{equation}
Q = r_1 \frac{b_1}{2} + r_2 \frac{b_2}{2} + r_3 b_3 \,.
\label{eq:D5'defZ4}
\end{equation}
From this expression it is clear that the $\text{D}5'$ branes can be naturally interpreted as instantons of the D9 ones. Indeed, the light spectrum involves  scalars in the corresponding bi-fundamental representations. 

The solution \eqref{eq:D5'defZ4}, together with the expression \eqref{eq:Jform} for the K\"ahler $J$-form, guarantees that the $\text{D}5'$ branes have positive tension, $Q\cdot J >0$. Moreover, from eq. \eqref{eq:I4D5'Z4} we read
\begin{equation}
    k_1= 2(r_3-r_1) \, , \qquad k_2 = 2(r_3-r_2)\, , \qquad k_3=r_1+r_2-2r_3 \, , \qquad k_{4+2j}=-\frac{r_1}{2} \, , \qquad k_{5+2j}= -\frac{r_2}{2} \, , 
    \label{eq:KacMoodylevelsZ4D5'}
\end{equation}
for $j=0,\ldots, 3$, and the unitarity constraints are satisfied both in the left and right-moving CFT's. For instance, for  the symmetric choice $a=2$, and $r_1=r_2=0$ and $r_3=1$ one finds
\begin{equation}
    \begin{split}
    &\sum_{i \, | \, k_i \geq 0} \frac{k_i \ \text{dim} G_i }{k_i + h_i^{\vee}}=  7 +7 = 14 < c_\text{L}-4_\text{CM} =50 \, ,
    \\
    &\sum_{i \, | \, k_i <0}  \frac{|k_i| \ \text{dim} G_i }{|k_i| + h_i^{\vee}}=  \frac{63}{5}+1 = \frac{68}{5} < c_\text{R} - 6_\text{CM}= 48 \, ,
\end{split}
\end{equation}
where the Ka$\check{\text{c}}$-Moody algebras are realised at level two.

\subsubsection{D1 branes in the $T^4 /\mathbb{Z}_2$ orientifold with a non-trivial $B_{ab}$ background}
\label{sssec:D1T2Z2Bab}

\begin{table}
\centering
\begin{tabular}{| c | c | c |}
\hline
	 { Representation} &  $\text{SO}(1,1) \times \text{SU}(2)_l \times \text{SU}(2)_R$ & { Sector} \\
	\hline
	$ \smalltableau{ \null \\} \times \overline{\smalltableau{\null \\}}  $  & $\left (0, 1, 1 \right ) + 2 \times \left (\tfrac{1}{2}, 2, 1 \right )_\text{L}  $  & D1-D1
	\\
	$  \smalltableau{ \null \\} \times \overline{\smalltableau{\null \\}}  $ & $\left (1, 2, 2 \right ) + 2 \times \left (\tfrac{1}{2}, 1, 2 \right )_\text{R}$ & \ \ 
	\\
	$ \smalltableau{ \null \& \null \\} + \overline{\smalltableau{ \null \& \null \\}} $  & $4 \times \left (1, 1, 1 \right ) + 2 \times \left (\frac{1}{2}, 2, 1 \right )_\text{R} $ &  \ \ 
	\\
	 $  \smalltableau{ \null \\ \null \\} + \overline{\smalltableau{ \null \\ \null \\}} $  & $ 2\times \left (\frac{1}{2},  1,2 \right )_\text{L} $ &  \ \
	\\ 
 \hline
	$2^{b/2}  (\, \smalltableau{\null \\}_{1}, \overline{\smalltableau{\null \\}}_{9} ) + 2^{b/2} (\,   \overline{\smalltableau{\null \\}}_{1} \, , \smalltableau{\null \\}_{9}) $  & $  \left (\frac{1}{2},  1, 1 \right )_\text{L} $ & D1-D9 
 \\
 \hline
\end{tabular}
\caption{The light spectrum for probe D1 branes localised on the $\mathbb{Z}_2$ fixed point supporting an O5$_-$ plane. The defect gauge group is $\text{U}(r)$ and the D1-$\overline{\text{D}5}$ strings are massive since the two sets of branes live on different fixed points.}
\label{tab:D1UxUxUSpZ2}
\end{table}

The next family of defects we want to study is associated to the $T^4/\mathbb{Z}_2$ vacuum with a non-trivial $B$-field background and gauge group \eqref{UxUSpZ2}, introduced in Section \ref{SSec:Z2Bab}. Starting from D1 branes, we have to distinguish the two cases depending on whether the defects are localised on an $\text{O}5_-$ or an $\text{O}5_+$ plane.
The former case is somehow simpler, since there are no interactions between the D1 branes and the $\overline{\text{D}5}$'s, which are clearly localised on a different fixed point supporting an $\text{O}5_+$ plane. The gauge group on the defects is unitary, $\text{U} (r)$, and the light spectrum is listed in table \ref{tab:D1UxUxUSpZ2}.
The anomaly polynomial is given by 
\begin{equation}
    I_4= \tfrac{r}{2} \left ( -2  \, \text{tr} R^2 + 2^{b/2} \,\text{tr} F_9^2 \right ) \,, \label{eq:D1O5manom}
\end{equation}
and cancels the inflow from the bulk theory, if the charge vector for the D1 branes is
\begin{equation}
    Q= \left ( \tfrac{r}{\sqrt{2}}, - \tfrac{r}{\sqrt{2}}; \boldsymbol{0}_{n_T-1} \right ) \, . 
\end{equation}
This D1 brane clearly satisfies the \emph{null-charge} conjecture, $Q\cdot Q=0$, and has a positive tension since $Q \cdot J >0$, with $J$ given in eq. \eqref{eq:Jform}. 
From \eqref{eq:D1O5manom} we can read the level $k=2^{b/2} \, r$ of the Ka$\check{\text{c}}$-Moody algebra, so that the unitarity constraint is satisfied,
\begin{equation}
 \sum_{i \, | \, k_i \geq 0} \frac{k_i \ \text{dim} G_i }{k_i + h_i^{\vee}}= \frac{2^{8-b}-1}{1+2^{4-b}}+1 < c_\text{L} - 4_\text{CM} =26 \,.
 \end{equation}

\begin{table}
\centering
\begin{tabular}{| c | c | c |}
\hline
	 { Representation} &  $\text{SO}(1,1) \times \text{SU} (2)_l \times \text{SU} (2)_R$ & { Sector} \\
	\hline
	$   \smalltableau{ \null \\ \null \\}_{1_1} +  \smalltableau{ \null \\ \null \\}_{1_2} $  & $\left (0, 1, 1 \right ) + 2 \times \left (\tfrac{1}{2}, 2, 1 \right )_\text{L}  $  & D1-D1
	\\
	$   \smalltableau{ \null \& \null \\}_{1_1} +  \smalltableau{ \null \& \null \\}_{1_2}  $ & $\left (1, 2, 2 \right ) + 2 \times \left (\tfrac{1}{2}, 1, 2 \right )_\text{R}$ & \ \ 
	\\
	$ ( \,  \smalltableau{ \null \\}_{1_1} ,  \smalltableau{ \null \\}_{1_2}) $  & $4 \times \left (1, 1, 1 \right ) + 2 \times \left (\frac{1}{2}, 2, 1 \right )_\text{R} $ &  \ \ 
	\\
	 $   ( \,  \smalltableau{ \null \\}_{1_1} ,  \smalltableau{ \null \\}_{1_2})  $  & $ 2 \times \left (\frac{1}{2},  1,2 \right )_\text{L} $ &  \ \
	\\ 
 \hline
	$2^{b/2}  \,  ( \,  \smalltableau{ \null \\}_{1_1} \, + \, \smalltableau{ \null \\}_{1_2} , \smalltableau{ \null \\}_{9}) $  & $  \left (\frac{1}{2},  1, 1 \right )_\text{L} $ & D1-D9 
 \\
 \hline
	$( \,  \smalltableau{ \null \\}_{1_1} ,  \smalltableau{ \null \\}_{\bar 5_1})   + ( \,  \smalltableau{ \null \\}_{1_2} ,  \smalltableau{ \null \\}_{\bar 5_2}) $  & $ \left (1, 1, 2 \right ) + 2 \times \left (\frac{1}{2},  1, 1 \right )_\text{L} $ & D1-$\overline{\text{D5}}$  
 \\
	$( \,  \smalltableau{ \null \\}_{1_1} ,  \smalltableau{ \null \\}_{\bar 5_2})   + ( \,  \smalltableau{ \null \\}_{1_2} ,  \smalltableau{ \null \\}_{\bar 5_1}) $  & $ 2 \times \left (\frac{1}{2},  1, 1 \right )_\text{R} $ & \ \
 \\
 \hline
\end{tabular}
\caption{Spectrum for probe D1 branes on the $\mathbb{Z}_2$ fixed point supporting an O5$_+$ plane, where also the $\overline{\text{D}5}$ branes are localised. The defect gauge group is $\text{SO}(r_1) \times \text{SO}(r_2)$.}
\label{tab:D1SOxUxUSpZ2}
\end{table}

In the case when the defects are localised on an $\text{O}5_+$ plane, the gauge group is $\text{SO}(r_1) \times \text{SO}(r_2)$, and the light spectrum is summarised in table \ref{tab:D1SOxUxUSpZ2}. Clearly, if the D1 branes are separated from the $\overline{\text{D}5}$'s, the open strings are massive and the last line in the table is absent. In the following we shall discuss the situation where they sit on the same fixed point, while the discussion can be straightforwardly adapted to  distant D1 and $\overline{\text{D}5}$ branes. The anomaly polynomial 
\begin{equation}
    I_4= \tfrac{1}{2}  \left ( - (r_1 + r_2) \, \text{tr} R^2 + 2^{b/2} \tfrac{r_1+r_2}{2} \,\text{tr} F_{9}^2 + (r_1-r_2)\,\text{tr} F_{\bar 5,1}^2 + (r_2-r_1)\,\text{tr} F_{\bar 5,2}^2 - (r_1-r_2)^2 \chi(N) \right ) 
\end{equation}
cancels the inflow from the bulk theory if 
\begin{equation} \label{eq:defcharegSO}
    Q= \left ( \tfrac{1}{2\sqrt{2}}(r_1+r_2), - \tfrac{1}{2\sqrt{2}}(r_1+r_2); r_2 - r_1; \boldsymbol{0}_{n_T-2} \right ) \, .
\end{equation}
Notice that the charge vector can be expressed in terms of the anomaly vectors $b_2$ and $b_3$ given in Appendix \ref{App:Anomaly},
\begin{equation}
Q = - r_1 \, b_2 - r_2 \, b_3
\end{equation}
so that D1 branes can indeed be interpreted as (anti-)instantons on the $\overline{\text{D}5}$ branes. This identification is supported by the presence of light scalars in the D1-$\overline{\text{D}5}$ spectrum. Also in this case the tension of the brane is positive, $Q\cdot J >0$, and the unitary constraints are satisfied with the levels
\begin{equation}
    k_1 = 2^{b/2} \cdot  \tfrac{r_1+r_2}{2} \, , \qquad  k_2= r_1-r_2 \, , \qquad \qquad  k_3= r_2-r_1 \, .
\end{equation}
Indeed, for the simple choice $r_1=1, \, r_2=0$, we find\
\begin{equation}
    \begin{split}
    &\sum_{i \, | \, k_i \geq 0}  \frac{k_i \ \text{dim} G_i }{k_i + h_i^{\vee}}= \frac{2^{8-b}-1}{2+2^{5-b}}+1 + \frac{2^{4-b/2}(2^{4-b/2}+1)}{4 + 2^{4-b/2}} 
    \leq  c_\text{L} - 4_\text{CM} = 8 + 48 \cdot 2^{-b/2} \, ,
    \\
    &\sum_{i \, | \, k_i <0}  \frac{|k_i| \ \text{dim} G_i }{|k_i| + h_i^{\vee}}= \frac{2^{4-b/2}(2^{4-b/2}+1)}{4 + 2^{4-b/2}}\leq c_\text{R}- 6_\text{CM} =
    48 \cdot 2^{-b/2} \, .
\end{split}
\end{equation}

\subsubsection{$\text{D}5'$ branes in the $T^4 /\mathbb{Z}_2$ orientifold with a non-trivial $B_{ab}$ background}
\label{sssec:D5pT2Z2Bab}

\begin{table}
\centering
\begin{tabular}{| c | c | c |}
\hline
	 { Representation} &  $\text{SO}(1,1) \times \text{SU} (2)_l \times \text{SU}(2)_R$ & { Sector} \\
	\hline
	$   \smalltableau{ \null \& \null \\}_{5'_1} +  \smalltableau{ \null \& \null \\}_{5'_2} $  & $\left (0, 1, 1 \right ) + 2 \times \left (\tfrac{1}{2}, 2, 1 \right )_\text{L}  $  & D5'-D5'
	\\
	$   \smalltableau{ \null \\ \null \\}_{5'_1} +  \smalltableau{ \null \\ \null \\}_{5'_2}  $ & $\left (1, 2, 2 \right ) + 2 \times \left (\tfrac{1}{2}, 1, 2 \right )_\text{R}$ & \ \ 
	\\
	$ ( \,  \smalltableau{ \null \\}_{5'_1} ,  \smalltableau{ \null \\}_{5'_2}) $  & $4 \times \left (1, 1, 1 \right ) + 2 \times \left (\frac{1}{2}, 2, 1 \right )_\text{R} $ &  \ \ 
	\\
	 $   ( \,  \smalltableau{ \null \\}_{5'_1} ,  \smalltableau{ \null \\}_{5'_2})  $  & $ 2 \times \left (\frac{1}{2},  1,2 \right )_\text{L} $ &  \ \
  \\
 \hline
	$ ( \,  \smalltableau{ \null \\}_{5'_1} ,  \smalltableau{ \null \\}_{9})   +  \, (  \, \smalltableau{ \null \\}_{5'_2} , \smalltableau{ \null \\}_{9}) $  & $ \left (1, 2, 1 \right ) + 2 \times \left (\frac{1}{2},  1, 1 \right )_\text{R} $ & $\text{D}5'$-D9  
 \\
	$( \,  \smalltableau{ \null \\}_{5'_1} ,  \smalltableau{ \null \\}_{9})   +  \, (  \, \smalltableau{ \null \\}_{5'_2} , \smalltableau{ \null \\}_{9}) $  & $ 2 \times \left (\frac{1}{2},  1, 1 \right )_\text{L} $ & \ \
	\\ 
 \hline
	$2^{b/2}  \,  ( \,  \smalltableau{ \null \\}_{5'_1} ,  \smalltableau{ \null \\}_{\bar 5_1})   + 2^{b/2}  ( \,  \smalltableau{ \null \\}_{5'_2} ,  \smalltableau{ \null \\}_{\bar 5_2}) $  & $  \left (\frac{1}{2},  1, 1 \right )_\text{R} $ & $\text{D}5'\text{-}\overline{\text{D5}}$ 
 \\
 \hline
\end{tabular}
\caption{ Spectrum for probe $\text{D}5'$ branes with  a $\text{USp}(r_1) \times \text{USp}(r_2)$ gauge group.}
\label{tab:D5'UxUSpZ2B}
\end{table}

Finally, we can study the consistency of the CFT on the $\text{D}5'$ defect wrapping the entire compact space. We turn on suitable Wilson lines so that the open strings stretched between the $\text{D}5'$ and the $\overline{\text{D}5}$ branes are massless. In this case, the gauge group is $\text{USp}(r_1) \times \text{USp}(r_2) $ and the light excitations are listed in table \ref{tab:D5'UxUSpZ2B}.  The associated anomaly polynomial  reads
\begin{equation}
 \label{eq:I4D5'UxUSpZ2B}
    I_4= \tfrac{1}{2}  \left (  (r_1 + r_2) \, \text{tr} R^2 - 2^{b/2-1}  r_1 \,\text{tr} F_{\bar 5,1}^2 - 2^{b/2-1}  r_2 \,\text{tr} F_{\bar 5,2}^2 - (r_1-r_2)^2 \chi(N) \right ) \, ,
\end{equation}
and cancels the inflow from the bulk theory if
\begin{equation}
    Q = \left ( \tfrac{2^{b/2}}{2\sqrt{2}}(r_1+r_2),  \tfrac{2^{b/2}}{2\sqrt{2}}(r_1+r_2); \tfrac{2^{b/2}}{4}(r_1-r_2) \boldsymbol{1}_{2^{4-b}}; \boldsymbol{0}_{n_T-2^{4-b}-1} \right ) \, .
\end{equation}
Also in this case $Q\cdot J >0$, and the unitarity constraints are satisfied with the Ka$\check{\text{c}}$-Moody levels given by
\begin{equation}
    k_1= 0 \, , \qquad k_2= - 2^{b/2-1} r_1 \, , \qquad k_3=-2^{b/2-1} r_2 \, .
\end{equation}
For the minimal choice $r_1=2 , r_2=0$ we find
\begin{equation}
    \begin{split}
    &\sum_{i \, | \, k_i \geq 0}  \frac{k_i \ \text{dim} G_i }{k_i + h_i^{\vee}}=0 \leq c_\text{L}-4_\text{CM} = 6+ 96 \cdot 2^{-b/2}  \, , 
    \\
    &\sum_{i \, | \, k_i <0}  \frac{|k_i| \ \text{dim} G_i }{|k_i| + h_i^{\vee}}= \frac{2^{4}(2^{4-b/2}+1)}{2 + 2^{4-b/2}+  2^{1+b/2}} \leq 
   c_\text{R} -6_\text{CM} = 16 + 96 \cdot 2^{-b/2}  \, .
\end{split}
\end{equation}

Notice that, although from the anomaly polynomial one reads $k_1=0$, its vanishing is due to the cancellation of equal LM and RM contributions, $k_1= (r_1+r_2)-(r_1+r_2)$, as can be extracted from table \ref{tab:D5'UxUSpZ2B}. This suggests that we have additional algebras realised both on the LM and RM sectors. The new unitarity constraints are
\begin{equation}
    \begin{split}
    &\sum_{i \, | \, k_i \geq 0}  \frac{k_i \ \text{dim} G_i }{k_i + h_i^{\vee}}=\frac{2^{8-b}-1}{1+2^{4-b/2}}+1   < c_\text{L}-4_\text{CM} = 6+ 96 \cdot 2^{-b/2}  \, , 
    \\
    &\sum_{i \, | \, k_i <0}  \frac{|k_i| \ \text{dim} G_i }{|k_i| + h_i^{\vee}}= \frac{2^{4}(2^{4-b/2}+1)}{2 + 2^{4-b/2}+  2^{1+b/2}}  + \frac{2^{8-b}-1}{1+2^{4-b/2}}+1
   < 
   c_\text{R} -6_\text{CM} = 16 + 96 \cdot 2^{-b/2}  \,,
\end{split}
\end{equation}
which are again satisfied.

\section{An example in four dimensions}
\label{Sec:4d}

BSB vacua with almost rigid branes can also be built in four dimensions. The simplest example is based on the $T^6/\mathbb{Z}_2\times \mathbb{Z}_2$ orbifold with discrete torsion. This vacuum does admit a supersymmetry breaking solution since it involves $\text{O}_+$ planes and thus calls for the introduction of anti-branes \cite{Angelantonj:1999ms}.

The three generators of the orbifold group act as
\begin{equation}
g \cdot (z_1,z_2,z_3)= (-z_1,-z_2,z_3) \,,  
\quad   f \cdot (z_1,z_2,z_3)= (-z_1,z_2,-z_3) \,, \quad  h \cdot (z_1,z_2,z_3)= (z_1,-z_2,-z_3) \, ,  
\label{eq:4d1}
\end{equation}
on the complex coordinates $z_i$ of $T^6 = T^2 \times T^2 \times T^2$. Each generator has sixteen fixed torii, each supporting a Hilbert space of twisted strings. Modular invariance of the torus partition function does not fix the action of the $\mathbb{Z}_2\times \mathbb{Z}_2$ group on the twisted states, but leaves the freedom of introducing phases --- in our case a sign, $\epsilon = \pm 1$ --- known as discrete torsion \cite{Vafa:1994rv}. The associated orientifold involves $\text{O}9_-$ planes from the world-sheet parity $\varOmega$, and three types of $\text{O}5_{i, \epsilon_i}$ planes localised at the fixed points of $g\varOmega$, $f\varOmega$ and $h\varOmega$, wrapping the third, second and first $T^2$, respectively. These can have negative or positive tension and charge, $\epsilon_i =\mp 1$, but are constrained by the condition \cite{Angelantonj:1999ms}
\begin{equation}
\epsilon_1 \, \epsilon_2 \, \epsilon_3 = \epsilon\,. \label{eq:4d2} 
\end{equation} 

For simplicity, let us focus on the case $\epsilon =-1$ and $\{ \epsilon_i \} = (1,1,-1)$, with an $\text{O}9_-$ plane and sixteen $\text{O}5_{g,-}$,$\text{O}5_{f,-}$, $\text{O}5_{h,+}$ planes, wrapping the first, second and third $T^2$, respectively. The cancellation of R-R tadpoles thus requires\footnote{More options are possible by allowing internal magnetic fields, including supersymmetric vacua \cite{Dudas:2005jx}.} the introduction of an open-string sector with sixteen sets of D9, $\text{D}5_g$,  $\text{D}5_f$ and  $\overline{\text{D}5}_h$ branes. The orbifold generators act on the CP labels as
\begin{equation}
\begin{split}
N_o &= o+g + {\bar o} +  {\bar g}\,,
\\
N_f &= i (o-g - {\bar o} +  {\bar g})\,,
\\
D_{g_i,o} &= o_1^i+g_1^i + {\bar o}_1^i +  {\bar g}_1^i
\\
D_{g_i,f} &= o_1^i-g_1^i + {\bar o}_1^i -  {\bar g}_1^i
\\
D_{f_i,o} &= o_2^i+g_2^i + {\bar o}_2^i +  {\bar g}_2^i
\\
D_{f_i,f} &= i (o_2^i+g_2^i - {\bar o}_2^i -  {\bar g}_2^i)
\\
D_{h_i,o} &= a^i+b^i + c^i + d^i 
\\
D_{h_i,f} &= a^i-b^i + c^i - d^i   
\end{split}
\qquad
\begin{split}
N_g &= i (o+g - {\bar o} -  {\bar g}) \ , 
\\
N_h &= o-g + {\bar o} -  {\bar g} \ ,
\\
D_{g_i,g} &= i ( o_1^i+g_1^i - {\bar o}_1^i -  {\bar g}_1^i) \ ,
\\
D_{g_i,h} &= -i ( o_1^i-g_1^i - {\bar o}_1^i +  {\bar g}_1^i) \ ,
\\
D_{f_i,g} &=  o_2^i-g_2^i + {\bar o}_2^i -  {\bar g}_2^i\,,
\\
D_{f_i,h} &= i ( o_2^i-g_2^i - {\bar o}_2^i +  {\bar g}_2^i) \ ,
\\
D_{h_i,g} &=   a^i+b^i - c^i - d^i \ ,
\\
D_{h_i,h} &= a^i-b^i - c^i + d^i 
 \, . 
\end{split}
\label{eq:4d3}  
\end{equation}
Here $N_A$ refers to D9 branes which contain two different (complex) gauge group factors $o$ and $g$, and $A=o,g,f,h$ indicates the action of one of the orbifold generators, $o$ being the identity. Similarly for the CP labels of D5 branes where the extra index $i$ means that these branes are localised on the $i$-th fixed point of the corresponding generator.   

The R-R tadpole conditions now read
\begin{equation}
\begin{split}
\text{untwisted} \quad & : \qquad   N_o =  \sum_{i=1}^{16} D_{g_i,o} =  \sum_{i=1}^{16} D_{f_i,o}
= \sum_{i=1}^{16} D_{h_i,o} = 32 \ , 
\\
g\text{-twisted} \quad &:   \qquad N_g - 4 D_{g_i,g} = 0 \,, \quad D_{f_i,g} + D_{h_i,g} = 0 
\,, 
\\
f\text{-twisted} \quad &: \qquad N_f - 4 D_{f_i,f} = 0 \, , \quad D_{g_i,f} + D_{h_i,f} = 0 
\,, 
\\
h\text{-twisted} \quad &: \qquad N_h - 4 D_{h_i,h} = 0 \, , \quad D_{g_i,h} + D_{f_i,h} = 0 
\,, 
\end{split}  \label{eq:4d4} 
\end{equation}
where the twisted tadpole conditions hold independently on each fixed point.

In the vacuum of \cite{Angelantonj:1999ms} all D5 (anti)branes were placed, say, at the origin of the internal space, which corresponds to one of the fixed points, so that the twisted tadpoles were \emph{trivially} cancelled, since each $N_A$ and each $D_{g,f,h,A}$ had to vanish independently, $A=g,f,h$, and pairs of branes with opposite twisted charge could be freely moved to the bulk. This possibility of deforming the brane configurations was reflected in the presence of massless scalars in suitable bi-fundamentals of the gauge group $G_\text{CP} = \text{U} (8)^2 |_9 \times \text{U} (8)^2 |_{5_g} \times \text{U} (8)^2 |_{5_f} \times \text{USp} (8)^4 |_{5_h}$. 

Inspired by the $T^4/\mathbb{Z}_2$ almost rigid vacuum, one can seek for non-trivial solutions of the twisted R-R tadpoles, where the twisted charge of one set of branes is cancelled by that of another type of branes. This possibility exists only if the sixteen D5 branes are distributed democratically among the sixteen fixed points of the corresponding generators, so that the twisted tadpoles become
\begin{equation}
N_h = 8 \,, \qquad   D_{f_i,g}= D_{g_i,f}= -2 \,,  \qquad D_{h_i,g}=D_{h_i,f}=D_{h_i,h}=2 \,, 
\label{eq:4d5}
\end{equation}
with solution
\begin{equation}
o=10 \,, \quad g=6 \,, \qquad g_{1}^{ i} = g_{2}^{ i} = 1 \,, \qquad o_{1}^{i} = o_{2}^{i} = 0 \, , \qquad a_{1}^{i} = 2 \,, \qquad b^i=c^i=d^i=0 \, ,   
\label{eq:4d6}
\end{equation}
for $i=1,\ldots,  16$. The CP gauge group is now
\begin{equation}
G_\text{CP} =  \text{U} (10) \times \text{U} (6) \Big|_{\text{D}9} \times  \text{U}(1)^{16} \Big|_{\text{D}5_g} \times \text{U} (1)^{16}\Big|_{\text{D}5_f}
\times \text{USp}(2)^{16}\Big|_{\overline{\text{D}5}_3}  \, . 
\label{eq:4d7}
\end{equation}
Also in this four-dimensional vacuum, the charged scalars associated to the position moduli of the D5 branes are absent, which reflects the rigidity of this configuration. As in the $T^4/\mathbb{Z}_2$ example, however, the O5 planes do not carry twisted charges, and therefore it is possible to move/recombine all D9 and D5 branes in the bulk, thus connecting this almost rigid vacuum to the one of \cite{Angelantonj:1999ms} via a non-trivial Higgsing of the gauge group.

As a final comment, the non-trivial cancellation \eqref{eq:4d5} of the twisted R-R tadpoles implies that the associated NS-NS ones 
\begin{equation}
\begin{split}
g\text{-twisted}& \qquad  (N_g-4 D_{g_i,g}) \sqrt{v_1} - \frac{2 (D_{f_i,g}-D_{h_i,g})}{\sqrt{v_1}}  \neq 0 \,,
\\
f\text{-twisted} & \qquad (N_f-4 D_{f_i,f}) \sqrt{v_2} + \frac{2 (D_{h_i,f}-D_{g_i,f})}{\sqrt{v_2}}  \neq 0\,, 
\\
h\text{-twisted} &\qquad (N_h+4 D_{h_i,f}) \sqrt{v_3} - \frac{2 (D_{g_i,h}+D_{f_i,h})}{\sqrt{v_3}}  \neq 0 \,, 
\end{split}
\label{eq:4d8}
\end{equation}
stay un-cancelled. Plugging in the solution \eqref{eq:4d6} into eq. \eqref{eq:4d8} we arrive at the non-vanishing terms
\begin{equation}
16\, \sqrt{v_3} \,, \qquad \frac{8}{\sqrt{v_1}} \,, \qquad  \frac{8}{\sqrt{v_2}}\,,\qquad \text{for each}\qquad i=1,\ldots, 16\,,
\end{equation}
which induce new terms to the scalar potentials localised in six ($\sqrt{v_3}$) and four ($1/\sqrt{v_1}$ and $1/\sqrt{v_2}$) dimensions. The full, tree-level, scalar potential, including the contribution of the un-cancelled untwisted tadpole, thus reads
\begin{equation}
\int V (\phi , \xi ) =  \int d^6 x \sqrt{-g_6} \ e^{- \phi} 
 \left( 64 + 16 \sum_{i=1}^{16} \xi_h^i\right) 
 + 8 \sum_{i=1}^{16} \int d^4 x \sqrt{-g_4} \,  e^{- \phi} \, \left(\xi_g^i + \xi_f^i \right) + O(\xi^2) \,, 
\label{eq:4d9}
\end{equation}
where, again, $\phi$ is the ten-dimensional dilaton, $\xi_g^i$ is the twisted NS-NS scalar sitting on the $g_i$ fixed point, etc. The dynamics induced by this potential  deserves a dedicated study. Indeed, in the simplest case  of only untwisted
NS-NS tadpoles plus suitable fluxes, static and time-dependent classical solutions have been derived and their properties,  stability and cosmological implications have been thoroughly studied \cite{Dudas:2000ff, Blumenhagen:2000dc, Dudas:2010gi, Dudas:2012vv, Mourad:2016xbk, Basile:2018irz, Sagnotti:2021mxb, Mourad:2021roa, Raucci:2022jgw,  Raucci:2022bjw, Basile:2022ypo, Mourad:2023loc}.  Moreover, in \cite{Buratti:2021yia, Blumenhagen:2023abk} it was argued that these solutions might play a role in the cobordism conjecture of the Swampland program \cite{McNamara:2019rup, Montero:2020icj}.  Following \cite{Martucci:2022krl}, it would be interesting to  study the consistency of D1 branes in this four-dimensional background.

\section*{Acknowledgements}
We are grateful to Marco Meineri, Gianfranco Pradisi, Fabio Riccioni and Augusto Sagnotti for valuable discussions. C.A. and G.L. would like to thank CPHT of Ecole Polytechnique and the Theory Department at CERN for their kind hospitality during various stages of this work. G.L. would also like to thank the Max-Planck Institute of Munich for its kind hospitality. C.C. was supported by project ``Nucleu" PN 23210101/2024.

\appendix

\section{Partition functions for the $T^4/\mathbb{Z}_N$ orientifolds}
\label{App:BSBZN}

In this appendix we summarise the partition functions for the six-dimensional orientifolds based on the $T^4/\mathbb{Z}_N$ orbifolds. These can be written in a compact way by introducing the building blocks 
\begin{equation}
T_{s} \big[ {\textstyle{\alpha \atop \beta}}\big]  = \text{tr}_\alpha \, \left( g^\beta\, P^s_\text{GSO}\, q^{L_0-c/24} \right)= \sum_{a,b=0,\frac{1}{2}} \tfrac{1}{2}\eta^s_{a,b} \, \frac{\theta^0 \big[{\textstyle{a \atop b}}\big]}{\eta^0}
\prod_{i=1}^4 \frac{\theta \big[{\textstyle{a+\mu_i+\alpha_i \atop b+\beta_i}}\big]}{\theta \big[{\textstyle{1/2+\mu_i+\alpha_i \atop 1/2+\beta_i}}\big]}\, d_{\mu_i+\alpha_i , \beta_i} \,,
\end{equation}
written in terms of the Dedekind eta function and Jacobi theta constants with characteristics, depending on the \emph{nome} $q=e^{2\pi i \tau}$. In this expression the vectors $\boldsymbol{\alpha} = (0,0,\alpha/N , -\alpha/N)$ and $\boldsymbol{\beta} = (0,0,\beta/N , -\beta/N)$ are associated to the orbifold action and label the $\alpha$ twisted sector with the insertion of the group element $g^\beta$ in the trace. The vector ${\boldsymbol \mu} = (\frac{m}{2},-\frac{m}{2}, \frac{n}{2}, -\frac{n}{2})$ is vanishing identically for the closed strings, while it takes into account the boundary conditions for open strings along the space-time ($m$) and the internal ($n$) coordinates: $m,n=0$ if one imposes NN or DD boundary conditions, while $m,n=1$ for ND or DN boundary conditions. The correct values for $m$ and $n$ are specified by the labels of the annulus amplitudes. The phase $\eta^s_{a,b}$ defines the GSO projection, with
\begin{equation}
\eta^A_{a,b} = (-1)^{2a+2b}\,, 
\quad
\eta^B_{a,b} = (-1)^{2a+2b+4ab}\,, 
\quad
\eta^{\hat A}_{a,b} = (-1)^{2a}\,,
\quad
\eta^{\hat B}_{a,b} = (-1)^{2a +4ab}\,,
\end{equation}
where the first two choices preserve supersymmetry, while the last two are associated to open strings stretched between branes and anti-branes. We shall also use the convention that $T_{AB}$ and $T_{\hat A \hat B}$ involve the averaged of the GSO phases $(\eta^A_{a,b} + \eta^B_{a,b})/2$ and $(\eta^{\hat A}_{a,b} + \eta^{\hat B}_{a,b})/2$, respectively. Moreover, the coefficients $d_{p ,q} = 2\, \sin  (\pi q)$ if $p =0$ or equal to one if $p \neq 0$, remove the numerical degeneracy from the theta functions associated to the compact bosons. Finally, the contribution $\theta^0 / \eta^0$ associated to the light-cone coordinates is trivially equal to one, but it is essential to pin point  the correct excitations living on the world-volume of the defects once expressed in terms of the (trivial) characters $V_0$, $O_0$, $S_0$ and $C_0$.

The $T$'s can be conveniently diagonalised 
\begin{equation}
T_s \big[ {\textstyle{\alpha \atop \beta}}\big] = \sum_{\gamma =0}^{N-1} e^{2 i \pi \beta \gamma/N} \, \tau^s_{\alpha , \gamma} (m,n)
\end{equation}
in terms of the \emph{characters} $\tau_{\alpha\, \beta}^s (m,n)$ which are eigenstates of the orbifold action. Clearly, we have different characters for different choices of GSO projection and for different boundary conditions of open strings, labeled by $m,n$. When dealing with closed strings, {\em i.e.} in the torus and Klein bottle amplitudes, we shall suppress most of the labels and simply refer to them as $\tau_{\alpha ,\beta}$, since the vector ${\boldsymbol \mu}\equiv 0$ and only the GSO phase $\eta^B_{a,b}$ is present. These characters are clearly representations of the ${\mathscr N}=(1,0)$ supersymmetry algebra. The massless ones are
\begin{equation}
\tau_{0,0} \sim V_4 - 2 S_4\,, \qquad \tau_{0,1}\,,\ \tau_{0,N-1}\,,\ \tau_{\alpha ,0} \sim 2O_4 - C_4 \qquad (\alpha\neq 0)\,,
\end{equation}
all other characters being massive. The $N=2$ case is special in that the untwisted character $\tau_{0,1}$ yields a full hypermultiplet. 

The torus partition function for the type IIB superstring on the $T^4/\mathbb{Z}_N$ orbifold thus reads
\begin{equation}
{\mathscr Z}_\text{IIB} = \tfrac{1}{N} \sum_{\alpha, \beta =0}^{N-1} n_{\alpha , \beta}\, T_B\big[ {\textstyle{\alpha \atop \beta}}\big] \, \bar T_B \big[{\textstyle{-\alpha \atop -\beta}}\big]\,\varLambda \big[{\textstyle{\alpha \atop \beta}}\big]\,,
\label{torusZN}
\end{equation}
with
\begin{equation}
\varLambda \big[{\textstyle{\alpha \atop \beta}}\big] = \begin{cases} \varLambda_{(4,4)} & \text{for} \quad \alpha = \beta = 0\,,
\\
1 & \text{otherwise}\,.
\end{cases}
\end{equation}
$\varLambda_{(4,4)}$ being the Narain lattice associated to the $T^4$, while $n_{0 , \beta} =1$ and $n_{\alpha , \beta}$ count the number of fixed points as given by the Lefschetz theorem. In all cases, the massless spectrum comprises the gravitational multiplet of ${\mathscr N}=(2,0)$ supersymmetry coupled to twenty-one tensor multiplets. 

For $N$ even, the Klein bottle amplitude reads\footnote{Actually, this expression for ${\mathscr K}$ is valid also for $N=3$ with the convention that terms associated to $N/2$ are set to zero. We shall not dwell with this case here since it only involves O9 planes and thus does not induces brane supersymmetry breaking.},
\begin{equation}
{\mathscr K}= \frac{1}{N} \sum_{\beta=0}^{N-1} T_B \big[ {\textstyle{0 \atop 2\beta}}\big] \, \varGamma_\beta + \frac{\epsilon}{N} \sum_{\beta=0}^{N-1} \, n_{N/2, \beta} \, T_B \big[ {\textstyle{N/2 \atop 2\beta}}\big]\,,
\label{KleinZN}
\end{equation}
where $\varGamma_0 \equiv P$ encodes the sum over the KK momenta, $\varGamma_{N/2} \equiv W$ takes into account the contribution of the winding states, while the remaining $\varGamma$'s are equal to one. The sign $\epsilon = \pm 1$ defines the action of the world-sheet parity $\varOmega$ on the twisted two-cycles and determines the type of orientifold planes involved in the construction. The choice $\epsilon =+1$ is associated to O9 and O5 planes of the same type, while for $\epsilon =-1$ the orientifold projection introduces $\text{O}9_\mp$ and $\text{O}5_\pm$ planes. In the following we shall restrict to this choice for $\epsilon$, with $\text{O}9_{-}$ and $\text{O}5_{+}$, as in the original BSB vacuum \cite{Antoniadis:1999xk}. 

Cancellation of the untwisted R-R charges than calls for the introduction of D9 and $\overline{\text{D}5}$ branes, so that the annulus partition functions read
\begin{equation}
{\mathscr A}_{99} = \frac{1}{N} \sum_{\beta=0}^{N-1} N_\beta^2 \, T_B \big[ {\textstyle{0 \atop \beta}}\big] \, P^{\delta_{\beta ,0}} 
\label{annulus99}
\end{equation}
for open strings stretched between a pair of D9 branes, 
\begin{equation}
{\mathscr A}_{\bar 5 \bar 5} =\frac{1}{N}\sum_{k,\ell} D_{(k)\, , \,0 } \, D_{(\ell)\, , \,0 } \, T_B \big[ {\textstyle{0 \atop 0}}\big] W (y_k - y_\ell ) + \frac{1}{N} \sum_{\beta=1}^{N-1} \sum_{k} D_{(k)\, , \,\beta}^2 \, T_B \big[ {\textstyle{0 \atop \beta}}\big] 
\label{annulus5b5b}
\end{equation}
for open strings stretched between a pair of $\overline{\text{D}5}$ branes sitting at fixed points $y_k$ and $y_\ell$, with 
\begin{equation}
N_\beta = \sum_{\gamma=0}^{N-1} e^{2 \pi i \beta \gamma/N}\, n_\gamma\,,
\qquad
D_{(k)\,,\, \beta} = \sum_{\gamma=0}^{N-1} e^{2 \pi i \beta \gamma/N}\, d_{(k)\,,\,\gamma}\,,
\label{CPparam}
\end{equation}
denoting the action of $g^\beta$ on the CP multiplicities $n_\gamma$ ($d_{(k)\,,\, \gamma}$) associated to the D9 ($\overline{\text{D}5}$) branes. The CP labels $n_0$ and $n_{N/2}$ are real and are associated to orthogonal gauge groups, while $n_\gamma$ and $n_{N-\gamma}$, with $\gamma \neq 0, N/2$, form conjugate pairs associated to the fundamental and anti-fundamental representations of a unitary gauge group. Similarly for the CP labels associated to the $\overline{\text{D}5}$ branes proviso that orthogonal groups are replaced by symplectic ones.   
In $W$ the winding contributions are shifted by the relative distance $y_k - y_\ell$ of the $\overline{\text{D}5}$ branes sitting at different fixed points. Similarly one could have introduced suitable Wilson lines on the D9 branes, which shift the KK momenta in $P$. The massless characters in there amplitudes coincide with those employed by the closed strings.

Open strings stretched between a D9 and an $\overline{\text{D}5}$ brane contribute with
\begin{equation}
{\mathscr A}_{9 \bar 5} = \frac{2}{N} \sum_{\beta =0}^{N-1} \sum_k N_\beta \, D_{(k)\,,\,\beta} \, T_{\hat B} \big[ {\textstyle{N/2 \atop \beta}}\big]
\label{annulus95b}
\end{equation}
and involves the complementary GSO projection $\eta^{\hat B}_{a,b}$. Following our prescription $(m,n)=(0,0)$ in ${\mathscr A}_{99}$ and ${\mathscr A}_{\bar 5 \bar 5}$, while $(m,n)=(0,1)$ in ${\mathscr A}_{9 \bar 5}$. In this amplitude, the massless characters are $\tau^{\hat B}_{N/2,0} (0,1)$ contributing a symplectic-Majorana-Weyl left-fermion, $S_4$, and $\tau^{\hat B}_{N/2,1} (0,1)$ and $\tau^{\hat B}_{N/2,N-1} (0,1)$ each contribute with a massless scalar, $O_4$. For $N=2$ we assist at the standard phenomenon where the characters $\tau^{\hat B}_{N/2,1} (0,1)$ and $\tau^{\hat B}_{N/2,N-1} (0,1)$ merge and yield a pair of scalars. 

Finally, the M\"obius strip amplitude for the D9 branes is standard 
\begin{equation}
{\mathscr M}_9 = - \frac{1}{N} \sum_{\beta=0}^{N-1} N_{2\beta}\, T_B \big[ {\textstyle{0 \atop \beta}}\big]\, P^{\delta_{0,\beta}}\,, 
\label{moebius9}
\end{equation}
while
\begin{equation}
{\mathscr M}_{\bar 5} = \frac{1}{N} \sum_{\beta=0}^{N-1} \sum_k D_{(k)\,,\, 2\beta}\, \left(B_B\big[ {\textstyle{0 \atop \beta}}\big] + F_B \big[ {\textstyle{0 \atop \beta}}\big] \right)\, W^{\delta_{0,\beta}}
\label{moebius5b}
\end{equation}
symmetrises the NS sector and anti-symmetrises the R sector, where we have decomposed $T = B - F$ in its bosonic  ($B$) and fermionic ($F$) parts. 

\subsection{Adding a background $B$-field}

The $T^4 / \mathbb{Z}_2$ orientifold  allows for a quantised $B$-field background of rank $b$ \cite{Bianchi:1990yu, Bianchi:1991eu, Angelantonj:1999jh}. Its presence implies some  changes in the Klein bottle, annulus and M\"obius strip amplitudes which are explicitly given in \cite{Angelantonj:1999jh} both for the supersymmetric and the BSB vacua. For clarity of presentation, we report here the various amplitudes of \cite{Angelantonj:1999jh}, and discuss their modification for the new vacua with unitary and symplectic/orthogonal gauge groups. The Klein bottle amplitude
\begin{equation}
{\mathscr K}^{(b)}= \tfrac{1}{2} T_B \big[ {\textstyle{0 \atop 0}}\big] \,  \sum_{\beta=0}^1 \varGamma_\beta^{(b)} - \, 2^{4-b/2} \, T_B \big[ {\textstyle{1 \atop 0}}\big]\,,
\label{KleinZ2B}
\end{equation}
involves the standard sum with $\beta =0$ over KK modes
\begin{equation}
    \varGamma_0^{(b)}= \sum_{\{m \} } q^{\frac{\alpha'}{2} m^\intercal \, g^{-1} \, m  } \equiv P(0,0) \, ,
\end{equation}
while for $\beta=1$ it enforces the projection on the windings
\begin{equation}
    \varGamma_1^{(b)}= \sum_{\{n \} }  \sum_{\{ \zeta \in \{ 0,1 \} \} }  q^{\frac{1}{2\alpha'} n^\intercal \, g \, n} \,  e^{\frac{2 \pi i}{\alpha'} n^\intercal B \zeta} \equiv W(0,B) \, .
\end{equation}
Open strings stretching between pairs of D9 branes are described by the annulus amplitude
\begin{equation}
{\mathscr A}_{99}^{(b)} = \tfrac{1}{2} \sum_{\beta=0}^{1} N_\beta^2 \, T_B \big[ {\textstyle{0 \atop \beta}}\big] \, \left ( 2^{b} \, P(B,0) \right )^{\delta_{\beta ,0}} \,,
\label{annulus99B}
\end{equation}
which depends on $B$ via the shifted momenta
\begin{equation}
   P(B,0)=  2^{-4} \sum_{\{ m \} }  \sum_{\{ \zeta \in \{ 0,1 \} \} }  q^{\frac{\alpha'}{2} (m+ \frac{1}{\alpha'} B \zeta )^\intercal \, g^{-1} \, (m+ \frac{1}{\alpha'} B \zeta ) } \, .
\label{eq:KKmomentaB}
\end{equation}
On the contrary open strings stretched between pairs of $\overline{\text{D5}}$ branes are unaffected, and the associated annulus partition function is simply given by eq.  \eqref{annulus5b5b}. 
Finally, open strings stretched between D9 and $\overline{\text{D5}}$ branes involve, as usual, the non-supersymmetric GSO phase $\eta_{\hat B}$ and come in $2^{b/2}$ families,
\begin{equation}
{\mathscr A}_{9 \bar 5}^{(b)} = 2^{b/2} \sum_{\beta =0}^{1} \sum_{k=1}^{2^{4-b}} N_\beta \, D_{(k)\,,\,\beta} \, T_{\hat B} \big[ {\textstyle{1 \atop \beta}}\big] \, .
\label{annulus95bB}
\end{equation}

The M\"obius strip amplitudes involve the signs $\Tilde{\gamma}$ and $\gamma$ \cite{Bianchi:1991eu,Angelantonj:1999jh} which reflect the different interaction of $\text{O}_\pm$ planes with D-branes. One finds
\begin{equation}
\begin{split}
{\mathscr M}_9^{(b)} &= - \frac{1}{2} \sum_{\beta=0}^{1} N_{2\beta}\, ( T_B \big[ {\textstyle{0 \atop \beta}}\big]\, \left ( 2^{(b-4)/2} P^{(\gamma)}(B,0) \right)^{\delta_{0,\beta}}\,, 
\\
{\mathscr M}_{\bar 5}^{(b)} &= \frac{1}{2} \sum_{\beta=0}^{1} \sum_k D_{(k)\,,\, 2\beta} \,\left(B_B\big[ {\textstyle{0 \atop \beta}}\big] + F_B \big[ {\textstyle{0 \atop \beta}}\big] \right)\, \left ( W^{(\tilde{\gamma})}(0,B) \right)^{\delta_{0,\beta}} \, , 
\end{split}
\end{equation}
with
\begin{equation}
\begin{split}
   P^{(\gamma)}(B,0) &=  2^{-2} \sum_{\{ m \} }  \sum_{\{ \zeta \in \{ 0,1 \} \} }  \gamma_{\zeta} \, q^{\frac{\alpha'}{2} (m+ \frac{1}{\alpha'} B \zeta )^\intercal \, g^{-1} \, (m+ \frac{1}{\alpha'} B \zeta ) } \, ,
\\
W^{(\tilde{\gamma})}(0,B) &= 2^{-2} \sum_{\{n \} }  \sum_{\{ \zeta \in \{ 0,1 \} \} }  \tilde\gamma_{\zeta}\, q^{\frac{1}{2\alpha'} n^\intercal \, g \, n} \,  e^{\frac{2 \pi i}{\alpha'} n^\intercal B \zeta} \, .
\end{split}
\end{equation}
Compatibility among the transverse channel Klein bottle, annulus and M\"obius strip amplitudes requires that the signs $\gamma$ and $\tilde \gamma$ satisfy \cite{Bianchi:1991eu,Angelantonj:1999jh}
\begin{equation}
    \sum_{\{ \zeta \}} \gamma_\zeta = 2^{4/2} \, , \qquad \sum_{\{ \zeta \in \text{ker}(B) \}} \tilde \gamma_\zeta = 2^{(4-b)/2} \, ,
\end{equation}
and
\begin{equation}
     \sum_{\{ \zeta  \in \text{ker}(B) \}} \gamma_\zeta = \xi \cdot 2^{(4-b)/2} \, , \qquad \sum_{\{ \zeta\}} \tilde \gamma_\zeta = \tilde{\xi} \cdot 2^{4/2} \, .
\end{equation}
The signs $\xi,\tilde \xi=\pm 1$ are {\em a priori} unconstrained and determine the type of  gauge group living on the D-branes. The cases $\xi=\tilde \xi=\pm 1$ have already been discussed in \cite{Angelantonj:1999jh} and are quickly reviewed in Section \ref{SSec:Z2Bab}. The new solution $\xi=-\tilde \xi =\pm 1$ is also possible, contrary to what stated in \cite{Angelantonj:1999jh}, and involves unitary and symplectic/orthogonal groups. Consistency of the construction requires that now open strings stretched between D9 and  $\overline{\text{D}5}$ branes do not feel the $\mathbb{Z}_2$ projection, and thus  the $\beta=1$ term is absent in eq. \eqref{annulus95bB}. In Section \ref{SSec:Z2Bab} we have discussed the vacuum with $\xi=-\tilde \xi=1$ and gauge group \eqref{UxUSpZ2}. For the sake of completeness, we shall present here the open-string spectrum associated to the choice $\xi=-\tilde \xi=-1$. The CP gauge group is
\begin{equation}
G_\text{CP} = \text{SO} (2^{4-b/2}) \times \text{SO} (2^{4-b/2}) \Big|_{\text{D}9} \times \text{U} (2^{4-b/2} ) \, \Big|_{\overline{\text{D}5}}\,,
\label{SOxUZ2}
\end{equation}
while matter comprises LH MW fermions in the  $ ( \smalltableau{  \null \\ \null \\ }\, 1;,1)+  ( 1, \smalltableau{  \null \\ \null \\ }\, ; 1 ) + (1,1;\smalltableau{  \null \\}\times \overline{\smalltableau{  \null \\}}\,)$ representation, four scalars in the $(\smalltableau{  \null \\} , \smalltableau{  \null \\}\, ;1) + (1,1 ;  \smalltableau{  \null \\ \null \\ }\, + \overline{ \smalltableau{  \null \\ \null \\ }\,} )$ representation, a RH MW fermion in the $( \smalltableau{  \null \\} , \smalltableau{  \null \\}\, ;1) + (1,1 ;\smalltableau{  \null \& \null \\} + \overline{\smalltableau{  \null \& \null \\}}) $ representation, together with $2^{b/2-1}$ copies of two scalars and a LH sMW fermion in the representation $ ( \smalltableau{  \null \\ }\, , 1; \smalltableau{  \null \\ }\, + \overline{\smalltableau{  \null \\ }}\, ) +  ( 1, \smalltableau{  \null \\ }\, ; \smalltableau{  \null \\ }\, + \overline{\smalltableau{  \null \\ }}\, )$. In these vacua with a unitary group on D9 or $\overline{\text{D}5}$ branes, the twisted NS-NS tadpoles vanish together with the R-R ones. As a result, the disk scalar potential only depends on the dilaton, as in \eqref{DilPot}.

\section{Partitions functions for defects in the $T^4/\mathbb{Z}_N$ orientifolds}
\label{App:ZNDefects}

The $T^4/\mathbb{Z}_N$ orientifold vacua have a varying number of tensor multiplets which naturally couple to one-dimensional defects. These can be D1 branes fully localised in the internal space, or $\text{D}5'$ branes which, on the contrary, wrap the full $T^4/\mathbb{Z}_N$. Following \cite{Dudas:2001wd,Angelantonj:2020pyr}, the associated partition functions read schematically\footnote{Here we have neglected the contribution of the KK and/or winding modes. Clearly, these do not affect the massless spectrum of open strings stretched between the defects and the bulk branes. We have also omitted the sectors associated to D1 and $\overline{\text{D}5}$ branes sitting at different fixed points since, in these cases, the open strings are necessarily stretched and, thus, massive.}
\begin{equation}
\begin{split}
{\mathscr A}_{11} &= \frac{1}{N} \sum_{\beta=0}^{N-1} R_\beta^2\, T_{AB} \big[ {\textstyle{0 \atop \beta}}\big] \,,
\\
{\mathscr A}_{19} &=\frac{2}{N} \sum_{\beta=0}^{N-1} R_\beta\, N_\beta \, T_{AB} \big[ {\textstyle{0 \atop \beta}}\big] \,,
\end{split}
\qquad\qquad
\begin{split}
{\mathscr M}_{1} &= - \frac{1}{N} \sum_{\beta=0}^{N-1} \, R_{2\beta}\, \hat T_{AB} \big[ {\textstyle{0 \atop \beta}}\big]\,, 
\\
{\mathscr A}_{1\bar 5_k} &= \frac{2}{N} \sum_{\beta=0}^{N-1} R_\beta\, D_{(k)\,,\, \beta}\, T_{\hat A \hat B} \big[ {\textstyle{0 \atop \beta}}\big]\,,
\end{split}\label{eq:D1partfunc}
\end{equation}
which involve two GSO projections and $(m,n) = (0,0)$ in ${\mathscr A}_{11}$, $(m,n) = (1,1)$ in ${\mathscr A}_{19}$, and $(m,n) = (1,0)$ in ${\mathscr A}_{1\bar 5}$. As usual, 
\begin{equation}
R_\beta = \sum_{\gamma=0}^{N-1} e^{2\pi i \beta\gamma/N}\, r_\gamma
\end{equation} 
encodes the action of the $\mathbb{Z}_N$ orbifold on the CP labels $r_\gamma$ of the D1 branes, which are assumed to sit at the  $y_k$ fixed point where some $\overline{\text{D}5}$ are also localised. Clearly, if the D1 branes sit at an empty fixed point (with no $\overline{\text{D}5}$'s) the ${\mathscr A}_{1\bar 5_k}$ only involves massive states, while if the D1 brane in moved in the bulk only $\beta =0$ contributes. The M\"obius amplitude involves new objects 
\begin{equation}
\hat T_{AB} \big[ {\textstyle{\alpha \atop \beta}}\big]  = \sum_{a,b=0,\frac{1}{2}} \tfrac{1}{4}\left( \eta^\text{A}_{a,b}+\eta^\text{B}_{a,b} \right) \, \frac{\theta^0 \big[{\textstyle{a \atop b}}\big]}{\eta^0}
\prod_{i=1}^4 \frac{\theta \big[{\textstyle{a+\alpha_i \atop b+\nu_i +\beta_i}}\big]}{\theta \big[{\textstyle{1/2+\alpha_i \atop 1/2+\nu_i + \beta_i}}\big]}\, d_{\alpha_i , \nu_i + \beta_i} \,,
\end{equation}
which have ${\boldsymbol \mu}=0$ but the lower characteristic is shifted by the vector ${\boldsymbol \nu} = (\frac{1}{2}, -\frac{1}{2}, \frac{1}{2}, -\frac{1}{2})$. 

Similarly, the partition functions relevant for the $\text{D}5'$ defect read
\begin{equation}
\begin{split}
{\mathscr A}_{5'5'} &= \frac{1}{N} \sum_{\beta=0}^{N-1} S_\beta^2\, T_{AB} \big[ {\textstyle{0 \atop \beta}}\big]\,,
\\
{\mathscr A}_{5'9} &=\frac{2}{N} \sum_{\beta=0}^{N-1} S_\beta\, N_\beta \, T_{AB} \big[ {\textstyle{0 \atop \beta}}\big] \,,
\end{split}
\qquad\qquad
\begin{split}
{\mathscr M}_{5'} &= \frac{1}{N} \sum_{\beta=0}^{N-1} \, S_{2\beta}\, \hat T_{AB} \big[ {\textstyle{0 \atop \beta}}\big]\,, 
\\
{\mathscr A}_{5'\bar 5} &= \frac{2}{N} \sum_{\beta=0}^{N-1} \sum_k S_\beta\, D_{(k)\,,\, \beta}\, T_{\hat A \hat B} \big[ {\textstyle{0 \atop \beta}}\big]\,,
\end{split} \label{eq:D5'partfunc}
\end{equation}
with $(m,n) = (0,0)$ in ${\mathscr A}_{5'5'}$, $(m,n) = (1,0)$ in ${\mathscr A}_{5'9}$, and $(m,n) = (1,1)$ in ${\mathscr A}_{5' \bar 5}$ while 
\begin{equation}
S_\beta = \sum_{\gamma=0}^{N-1} e^{2\pi i \beta\gamma/N}\, s_\gamma
\end{equation} 
refers to the $\mathbb{Z}_N$ action on the CP multiplicities $s_\gamma$ of $\text{D}5'$ branes. In the M\"obius amplitude $\hat T_{AB}$ now involves the vector ${\boldsymbol \nu} = (\frac{1}{2}, -\frac{1}{2}, 0, 0)$. 

The CP labels for the D1 ($\text{D}5'$) defects identify orthogonal (symplectic) and unitary gauge groups as in the D9 ($\overline{\text{D}5}$) case.

The $T$'s admit the usual decomposition in terms of characters. The massless ones are
\begin{equation}
\begin{split}
\tau^{AB}_{0,0} (0,0) &= V_0 O_4 + O_0 V_4 - 2 S_0 S_4 - 2 C_0 C_4\,,
\\
\tau^{AB}_{0,1} (0,0)&= \tau^{AB}_{0,N-1} (0,0) = 2 O_0 O_4 - S_0 C_4 - C_0 S_4\,,
\\
\tau^{AB}_{0,0} (1,0) &=O_0 S_4 -2 C_0 O_4\,,
\\
\tau^{AB}_{0,1} (1,0) &=\tau^{(10)}_{0,N-1} = -  S_0 O_4\,,
\\
\tau^{AB}_{0,0} (1,1) &= - S_0 O_4\,,
\end{split}
\end{equation}
and 
\begin{equation}
\begin{split}
\tau^{\hat A \hat B}_{0,0} (1,0) &= O_0 C_4-2S_0 O_4\,,
\\
\tau^{\hat A \hat B}_{0,1} (1,0) &= \tau^{\hat A \hat B}_{0,N-1} (0,0) = - C_0 O_4\,,
\\
\tau^{\hat A \hat B}_{0,0} (1,1) &= - C_0 O_4\,.
\end{split}
\end{equation}
where $V_0$, $O_0$, $S_0$ and $C_0$ encode the nature of the excitations on the two-dimensional world-volume of the defect, while the characters in the second position indicate the representation with respect to the SO(4) group associated to the transverse (non-compact) space-time directions, which also yield the corresponding multiplicities. 

Finally, 
\begin{equation}
\begin{split}
\hat\tau^{AB}_{0,0} (0,0) &= \hat V_0 \hat O_4 - \hat O_0 \hat V_4 - 2 \hat S_0 \hat S_4 + 2 \hat C_0 \hat C_4\,,
\\
\hat\tau^{AB}_{0,1} (0,0)&= \tau^{AB}_{0,N-1} (0,0) = - 2 \hat O_0 \hat O_4 - \hat S_0 \hat C_4 + \hat C_0 \hat S_4\,,
\end{split}
\end{equation}
in ${\mathscr M}_1$, and
\begin{equation}
\begin{split}
\hat\tau^{AB}_{0,0} (0,0) &= \hat V_0 \hat O_4 - \hat O_0 \hat V_4 - 2 \hat S_0 \hat S_4 + 2 \hat C_0 \hat C_4\,,
\\
\hat\tau^{AB}_{0,1} (0,0)&= \tau^{AB}_{0,N-1} (0,0) =  2 \hat O_0 \hat O_4 + \hat S_0 \hat C_4 - \hat C_0 \hat S_4\,,
\end{split}
\end{equation}
in ${\mathscr M}_{5'}$.

Adding a background value for the Kalb-Ramond field provides a richer scenario. Indeed, since both $\text{O5}_+$ and $\text{O5}_-$ planes are present in the construction, different physical situations arise according to the choice of the fixed point on which D1 branes are placed. Indeed, whenever they are placed on a $\text{O5}_-$  the CP group is unitary, $\text{U}(r_0)$, while if D1 branes are placed on a $\text{O5}_+$ plane they carry an $\text{SO}(r_0) \times \text{SO}(r_1)$ gauge group. Furthermore, since strings stretched between D1 branes have DD boundary conditions in the compact directions, the D1/D9 spectrum is $2^{b/2}$-fold degenerate. The other amplitudes in \eqref{eq:D1partfunc} are left unchanged.  

For $\text{D}5'$ branes the situation is slightly different since strings stretched among them carry NN boundary conditions in the compact directions.  This means that the KK momenta in \eqref{eq:D5'partfunc} are modified as in \eqref{eq:KKmomentaB} and discrete Wilson lines can be turned on to select unitary $\text{U}(r_0)$ or symplectic gauge groups $\text{USp}(r_0) \times \text{USp}(r_1)$. In addition, the $2^{b/2}$ degeneracy is now present for the  $\overline{\text{D5}}/\text{D}5'$ strings, leaving the $\text{D}9/\text{D}5'$ ones unaffected.

\section{Anomaly polynomials}
\label{App:Anomaly}

All the models described so far arise from a K3 orientifold and, as shown in Section \ref{Sec:AnomalyRR}, the anomaly polynomial admits a writing which reflects the geometry of the orbifold $T^4/\mathbb{Z}_N$, and the charges and positions of O-planes and  D-branes. In the following, we collect the main formul\ae\ for computing anomaly polynomials and provide further examples from $\mathbb{Z}_2$ orbifolds with non-trivial $B$-field background, and from the $\mathbb{Z}_6$ one.  As in \cite{Angelantonj:2020pyr}, our conventions are such that a LH MW fermion in the representation $\mathbf{R}$ of the gauge group, a LH MW gravitino and self-dual tensor contribute to the anomaly as
\begin{equation}
\begin{aligned}
    &I_8^{1/2}= - \text{dim}_{\mathbf{R}} \left ( \tfrac{1}{360}  \text{tr} R^4 + \tfrac{1}{288} \left ( \text{tr} R^2 \right )^2 \right ) + \tfrac{1}{24} \text{tr}_{\mathbf{R}} F^2 \text{tr} R^2 - \tfrac{1}{24} \text{tr}_{\mathbf{R}} F^4 \, ,
    \\
    & I_8^{3/2}= -  \tfrac{245}{360}  \text{tr} R^4 + \tfrac{43}{288} \left ( \text{tr} R^2 \right )^2 \, ,
    \\
    &I_8^{A}= - \tfrac{28}{360}  \text{tr} R^4 + \tfrac{7}{288} \left ( \text{tr} R^2 \right )^2  \, .
\end{aligned}
\end{equation}
Standard group theoretical facts imply the following decompositions of the traces
\begin{equation}
\begin{split}
\text{tr}_{\pm} \, F^4 &= (N\pm 8)\, \text{tr} \, F^4 + 3\, (\text{tr} \, F^2)^2 \,,
\\
\text{tr}_{(n,m)} \, F^4 &= n\, \text{tr} \, F^4_m + m\, \text{tr} \, F^4_n + 6\, \text{tr} \, F^2_n \, \text{tr} \, F^2_m\,,
\\
\text{tr}_{\pm} \, F^2 &= (N\pm 2)\, \text{tr} \, F^2\,,
\\
\text{tr}_{(n,m)} \, F^2 &= n\, \text{tr} \, F^2_m + m\, \text{tr} \, F^2_n\,,
\end{split}
\end{equation}
where $\pm$ indicates the symmetric/antisymmetric representation, while $(n,m)$ refers to the bi-funda\-mental. Here and in the whole manuscript $\text{tr} $ without a suffix denotes the trace in the fundamental representation, and we have omitted terms with odd powers of the field strength since they are non-vanishing only for Abelian groups, and we are not interested here in $\text{U} (1)$ anomalies.

Using the general expression of the reducible anomaly polynomial \eqref{factor-1} and the connection between the R-R charges of O-planes and D-branes \eqref{factor-3} and the anomaly vectors \eqref{factor-2}, we find the following expression for the anomaly polynomial 
\begin{equation}
\begin{split}
I_8  = \tfrac{1}{64} & \Bigg \{ \left( 2^{\frac{b}{2}} \text{tr} F_{9,1}^2+ 2^{\frac{b}{2}} \text{tr} F_{9,2}^2 - \sum_{i=1}^{2^{4-b}} 
 \text{tr} {F_{\bar 5_i }}^2 - 4\left(1-2^{-\frac{b}{2}}\right) \,  \text{tr} R^2 \right)^2  
\\
& -  \left( 2^{\frac{b}{2}} \text{tr} F_{9,1}^2+ 2^{\frac{b}{2}} \text{tr} F_{9,2}^2 + \sum_{i=1}^{2^{4-b}}  \text{tr} {F_{\bar 5_i}}^2 - 4\left(1+2^{-\frac{b}{2}}\right) \,  \text{tr} R^2 \right)^2  
\\
 &  - 2^{b-1} \sum_{i=1}^{2^{4-b}}  \left(\text{tr} F_{9,1}^2 - \text{tr} F_{9,2}^2 - 2^{2- \frac{b}{2}} \text{tr} {F_{\bar 5_i}}^2 \right)^2 \Bigg  \} \, ,
\end{split}
\end{equation}
and the anomaly vectors
\begin{equation} \label{eq:anomalyvectorsZ2USpUSp}
	\begin{split}
		a &=\left ( - \tfrac{2(1-2^{-\frac{b}{2}})}{\sqrt{2}}, - \tfrac{2(1+2^{-\frac{b}{2}})}{\sqrt{2}}; \boldsymbol{0}_{n_T-1} \right ) \, ,
		\\
		b_1&=  \left ( \tfrac{2^{b/2}}{\sqrt{2}}, \tfrac{2^{b/2}}{\sqrt{2}}; \tfrac{2^{b/2} }{2} \boldsymbol{1}_{2^{4-b}} , \boldsymbol{0}_{}  \right ) \, ,
		\\
		b_2 &=  \left ( \tfrac{2^{b/2}}{\sqrt{2}}, \tfrac{2^{b/2}}{\sqrt{2}}; -\tfrac{2^{b/2} }{2} \boldsymbol{1}_{2^{4-b}}, \boldsymbol{0}_{}  \right ) \, ,
		\\
		b_{2+i} &=  \left ( - \tfrac{1}{2\sqrt{2}}, \tfrac{1}{2\sqrt{2}}; - \boldsymbol{\delta}^i_{2^{4-b}} , \boldsymbol{0}_{} \right) \, ,
	\end{split}
\end{equation}
with $i=1,\ldots, 2^{4-b}$, for the vacuum with gauge group \eqref{CPBgauge}, originating from the $T^4/\mathbb{Z}_2$ orientifold with a rank-$b$ $B$-field background. 
Similarly, the anomaly polynomial for the new vacuum with gauge group $ \text{U}(2^{4-\frac{b}{2}}) \times \text{USp}(2^{4-\frac{b}{2}}) \times \text{USp}(2^{4-\frac{b}{2}}) $ reads
\begin{equation}
   \begin{split}
I_8  = \tfrac{1}{64} & \Bigg \{ \left( 2^{1+\frac{b}{2}} \text{tr} F_9^2  -  \text{tr} {F_{\bar 5, 1}}^2 - \text{tr} {F_{\bar 5 ,2}}^2 -4 \left(1-2^{-\frac{b}{2}}\right) \,  \text{tr} R^2 \right)^2  
\\
& -  \left(  2^{1+\frac{b}{2}} \text{tr} F_9^2 +  \text{tr} {F_{\bar 5, 1}}^2 + \text{tr} {F_{\bar 5 ,2}}^2 - 4\left(1+2^{-\frac{b}{2}}\right) \,  \text{tr} R^2 \right)^2  
\\
 &  - 8  \left( \text{tr} {F_{\bar 5 ,1}}^2 - \text{tr} {F_{\bar 5 ,2}}^2  \right)^2 \Bigg  \} \, ,
\end{split}
\end{equation}
and the associated anomaly vectors are
\begin{equation} \label{eq:anomalyvectorsZ2UU}
	\begin{split}
		a&=\left ( - \tfrac{2(1-2^{-\frac{b}{2}})}{\sqrt{2}}, - \tfrac{2(1+2^{-\frac{b}{2}})}{\sqrt{2}}; \boldsymbol{0}_{n_T-1} \right ) \, ,
		\\
		b_1&=  \left ( \tfrac{2^{b/2}}{\sqrt{2}}, \tfrac{2^{b/2}}{\sqrt{2}}; \boldsymbol{0}_{n_T-1} \right ) \, ,
		\\
		b_{2} &=  \left ( - \tfrac{1}{2\sqrt{2}}, \tfrac{1}{2\sqrt{2}}; 1, \boldsymbol{0}_{n_T-2} \right ) \,  ,
  \\
		b_{3} &=  \left ( - \tfrac{1}{2\sqrt{2}}, \tfrac{1}{2\sqrt{2}}; -1, \boldsymbol{0}_{n_T-2} \right ) \, .
	\end{split}
\end{equation}

The last example we have discussed in Section \ref{Sec:BSBZ6}, based on the $\mathbb{Z}_6$ orbifold, has gauge group \eqref{CPZ6}, and anomaly polynomial
\begin{equation}
\begin{split}
    I_8=\tfrac{1}{192} &\left \{ \left (  \text{tr} F_{9,1}^2 + 2\text{tr} F_{9,2}^2 + 2 \text{tr} F_{9,3}^2 + \text{tr} F_{9,4}^2 - \text{tr} F_{\bar 5,1}^2  - 2\text{tr} F_{\bar 5,2}^2 - 2 \text{tr} F_{\bar 5,3}^2 - \text{tr} F_{\bar 5,4}^2 \right )^2 \right. 
    \\
    &- \left ( -8 \text{tr} R^2 + \text{tr} F_{9,1}^2 + 2 \text{tr} F_{9,2}^2 + 2 \text{tr} F_{9,3}^2 +  \text{tr} F_{9,4}^2 + \text{tr} F_{\bar 5,1}^2 + 2 \text{tr} F_{\bar 5,2}^2 + 2 \text{tr} F_{\bar 5,3}^2 +  \text{tr} F_{\bar 5,4}^2 \right )^2  
     \\
    &  -4 \left (   \text{tr} F_{9,1}^2 + \text{tr} F_{9,2}^2 -  \text{tr} F_{9,3}^2 -  \text{tr} F_{9,4}^2 + \text{tr} F_{\bar 5,1}^2 + \text{tr} F_{\bar 5,2}^2 - \text{tr} F_{\bar 5,3}^2-  \text{tr} F_{\bar 5,4}^2    \right )^2
    \\
    & - \tfrac{4}{3} \left [ \left ( - 4 \text{tr} R^2 + \text{tr} F_{9,1}^2 - \text{tr} F_{9,2}^2 -  \text{tr} F_{9,3}^2 + \text{tr} F_{9,4}^2 + 3 \left ( \text{tr} F_{\bar 5,1}^2 - \text{tr} F_{\bar 5,2}^2 -  \text{tr} F_{\bar 5,3}^2 +  \text{tr} F_{\bar 5,4}^2 \right ) \right )^2  \right. 
    \\
    & \left. \qquad \qquad + 8 \left (  -\text{tr} R^2 + \text{tr} F_{9,1}^2 - \text{tr} F_{9,2}^2 - \text{tr} F_{9,3}^2 + \text{tr} F_{9,4}^2 \right )^2 \right ]  
    \\
    & -\tfrac{1}{2} \left [ \left ( \text{tr} F_{9,1}^2 - 2 \text{tr} F_{9,2}^2 + 2\text{tr} F_{9,3}^2 - \text{tr} F_{9,4}^2 + 4 \left ( \text{tr} F_{\bar 5,1}^2 - 2 \text{tr} F_{\bar 5,2}^2 + 2\text{tr} F_{\bar 5,3}^2 -  \text{tr} F_{\bar 5,4}^2 \right ) \right )^2 \right. 
    \\
    & \left.  \left. \qquad \qquad + 15  \left ( \text{tr} F_{9,1}^2 - 2 \text{tr} F_{9,2}^2 + 2\text{tr} F_{9,3}^2 - \text{tr} F_{9,4}^2 \right )^2 \right ] \right \} \, . 
 \end{split}
\end{equation}
The anomaly lattice is now spanned by
\begin{equation} \label{eq:anomalyvectorsZ6}
	\begin{split}
		a&=\left ( 0; -\tfrac{4}{\sqrt{6}}; 0; - \tfrac{2 \sqrt{2}}{3}; - \tfrac13, - \tfrac13, - \tfrac13, -\tfrac13; 0; 0, 0, 0, 0, 0 \right ) \, ,
		\\
		b_1&= \left ( \tfrac{1}{\sqrt{6}}; \tfrac{1}{\sqrt{6}};  \sqrt{\tfrac{2}{3}}; \tfrac{\sqrt{2} }{3}; \tfrac23, \tfrac23, \tfrac23, \tfrac23; \tfrac{1}{2\sqrt{3}}; \tfrac12, \tfrac12, \tfrac12, \tfrac12, \tfrac12 \right ) \, ,
		\\
		b_2 &= \left ( \tfrac{1}{\sqrt{6}}; \tfrac{1}{\sqrt{6}};\tfrac{1}{\sqrt{6}}; -\tfrac{1 }{3\sqrt{2}}; -\tfrac13, -\tfrac13, -\tfrac13, -\tfrac13; -\tfrac{1}{2\sqrt{3}}; -\tfrac12,- \tfrac12, -\tfrac12, -\tfrac12, -\tfrac12 \right ) \, ,
		\\
		b_3 &= \left ( \tfrac{1}{\sqrt{6}}; \tfrac{1}{\sqrt{6}};-\tfrac{1}{\sqrt{6}}; -\tfrac{1 }{3\sqrt{2}};- \tfrac13, -\tfrac13, -\tfrac13, -\tfrac13; \tfrac{1}{2\sqrt{3}}; \tfrac12, \tfrac12, \tfrac12, \tfrac12, \tfrac12 \right ) \, ,
        \\
        b_4&= \left ( \tfrac{1}{\sqrt{6}}; \tfrac{1}{\sqrt{6}};  -\sqrt{\tfrac{2}{3}}; \tfrac{\sqrt{2} }{3}; \tfrac23, \tfrac23, \tfrac23, \tfrac23; -\tfrac{1}{2\sqrt{3}}; -\tfrac12, -\tfrac12, -\tfrac12, -\tfrac12, -\tfrac12 \right ) \, ,
		\\
		b_5 &= \left (- \tfrac{1}{2\sqrt{6}}; \tfrac{1}{2\sqrt{6}};  \tfrac{1}{\sqrt{6}}; \tfrac{1}{ \sqrt{2}}; 0,0,0,0; \tfrac{1}{\sqrt{3}}; 0,0,0,0,0 \right ) \, ,
		\\
		b_6 &= \left (- \tfrac{1}{\sqrt{6}}; \tfrac{1}{\sqrt{6}};  \tfrac{1}{\sqrt{6}}; - \tfrac{1}{ \sqrt{2}}; 0,0,0,0; -\tfrac{2}{\sqrt{3}}; 0,0,0,0,0 \right ) \, ,
		\\
		b_7 &= \left (- \tfrac{1}{\sqrt{6}}; \tfrac{1}{\sqrt{6}};  -\tfrac{1}{\sqrt{6}}; -\tfrac{1}{ \sqrt{2}}; 0,0,0,0; \tfrac{2}{\sqrt{3}}; 0,0,0,0,0\right ) \, ,
  \\
		b_8 &= \left (- \tfrac{1}{2\sqrt{6}}; \tfrac{1}{2\sqrt{6}}; - \tfrac{1}{\sqrt{6}};   \tfrac{1}{ \sqrt{2}}; 0,0,0,0; -\tfrac{1}{\sqrt{3}}; 0,0,0,0,0 \right ) \, .
	\end{split}
\end{equation}

\newpage

\bibliographystyle{utphys}

\end{document}